\documentclass[a4paper,titlepage,hypertex,12pt]{article}
\usepackage{graphicx}
\usepackage[ansinew]{inputenc}
\usepackage{hyperref}
\hypersetup{colorlinks,linkcolor=black,citecolor=black}

\begin{document}

\begin{titlepage}
\begin{minipage}[][\textheight][c]{153mm}
\centering
\LARGE
ToF-SIMS Investigations on Dental \\ Implant Materials and Adsorbed Protein Films \\[4ex]

Diploma Thesis\\[10ex]

\Large \mdseries Falk Bernsmann \\[10ex]

\large Group for Interfaces, Nano Materials and Biophysics \\ Department of Physics \\ Technische Universit\"at Kaiserslautern \\[10ex]

Supervised by \\ Prof. Dr. Christiane Ziegler \\ and \\ apl. Prof. Dr. habil. Hubert Gnaser \\[10ex]

July 2007
\end{minipage}
\end{titlepage}

\thispagestyle{empty}
\tableofcontents
\newpage

\section{Introduction}
Biofilms play an important role in the health sector, in bioanalytics, in the food industry and in engineering science \cite{mondon:2002} because the adsorption of organic molecules can alter the physical, biological and chemical properties of a surface. This work deals with the formation of biofilms on dental implant materials.

When a dental implant is placed in the oral cavity, within seconds its surface is covered by a biofilm called pellicle consisting mainly of proteins and other macromolecules \cite{hannig:2006}. Since the adsorption of proteins is a highly selective process, the proportions of proteins found in the pellicle differ significantly from the ones found in saliva \cite{yao:2003}. The pellicle is of great physiological importance because it serves as lubricant, as diffusion barrier to demineralising agents and as reservoir for remineralising electrolytes \cite{hannig:2006}. Furthermore, proteins in the pellicle play an important role in the colonisation of the surface by bacteria and thus in the formation of dental plaque. On the one hand there are proteins, like amylase, that exhibit specific binding sites for bacterial adsorption \cite{hannig:2004}. On the other hand enzymes, like lysozyme, immobilized in the pellicle have anti-bacterial properties \cite{hoch:2005}. Since the adsorption process of proteins is a subject not yet fully understood, this work shall further investigate the adsorption of the proteins amylase, lysozyme and serum albumin on two experimental dental implant materials.

The chosen method is Time-of-Flight Secondary Ion Mass Spectrometry (ToF-SIMS) because 
it offers the following advantages \cite{belu:2003}: The mass distribution of molecules adsorbed on a sample's surface can be measured with a high mass resolution and a high surface sensitivity. It
is possible to create depth profiles with a depth resolution of below one nanometre.
And the analysis of non-conducting samples is possible without further preparation steps. ToF-SIMS has already been used to analyse adsorbed protein films on different substrates (see section \ref{sec:literature}) but up to now there are no ToF-SIMS studies of protein films on dental implant materials.  

The difficulty in interpreting mass spectra of proteins is their complexity \cite{graham:2006}. Every protein consists of a combination of the same twenty amino acids which dissociate within the ToF-SIMS analysis to numerous fragments. Hence one has to take into account the intensities of many different masses for analysis. To reduce the number of variables (i.e. masses), Principal Component Analysis (PCA) is used. This multivariate technique concentrates the variance of the spectra onto only a few variables, called Principal Components (PC).

This work is subdivided into the following sections: In this first section an introduction to the subject is given. The second section deals with previous work on the analysis of adsorbed protein films by ToF-SIMS and multivariate data analysis. Theoretical aspects of the examined systems and the applied techniques are detailed in the third section. The experiments are described in the fourth section. The fifth section contains the following results: First, mass spectra of dental implant materials are examined to determine their elemental surface composition. Then mass spectra of proteins adsorbed to silane coated silicon substrates are analysed to develop the methods and programmes necessary for distinguishing different proteins by their mass spectra. Since this does not work very well, the examined system is simplified to proteins adsorbed directly to silicon substrates. Here the different proteins can be recognized by their mass spectra and the developed statistical models perform well in evaluation tests. The adsorption conditions are varied to obtain the best results. Additionally the mutual influences of two proteins adsorbed at the same time or consecutively to the same silicon substrate are studied. The results obtained are confirmed by enzymatic activity measurements. Finally the mass spectra of proteins adsorbed to dental implant materials are examined. With only little modifications in data pre-treatment, the programmes developed to analyse the spectra of proteins on silicon can be used to distinguish between different proteins adsorbed to dental implant materials. Again, information on the mutual influence of the proteins upon adsorption is obtained. The sixth section gives a summary and an outlook on possible future investigations.
\pagebreak
\section{Previous work on protein analysis by ToF-SIMS}
\label{sec:literature}
In this section a brief overview of the literature available on multivariate analysis (MVA) of time-of-flight secondary ion mass spectra (ToF-SIMS) of adsorbed protein films is given. 

In 2001 Wagner and Castner published their article ``Characterization of Adsorbed Protein Films by Time-of-Flight Secondary Ion Mass Spectrometry with Principal Component Analysis'' \cite{wagner:2001}. For analysis of single component protein films of various proteins adsorbed to poly(tetrafluorethylene) (PTFE), mica or silicon substrates with principal component analysis (PCA), several peaks of the mass spectra of positively charged ions were selected. The selection was based upon the work of Mantus and others \cite{mantus:1993} who had developed a spectral interpretation protocol for protein spectra based on strong peaks in the spectra of amino acid homopolymers adsorbed to glass substrates. Wagner and Castner were able to distinguish between several proteins by their scores on the first two principle components. Furthermore, the PCA model developed with the mass spectra of single component protein films allowed qualitative insight into the composition of a complex adsorbed protein film from bovine plasma. 

In another article \cite{wagner:2003} published in 2002, Wagner and others described their attempt to quantitatively characterise multicomponent adsorbed protein films by ToF-SIMS. As before, only peaks related to amino acid fragments were selected from the mass spectra of positively charged ions for analysis. For binary protein films composed of fibrinogen and immunglobulin G adsorbed on mica or PTFE, a good agreement between surface concentrations predicted from the mass spectra by a partial least squares regression (PLSR) model and radio labelling experiments was observed. PLSR is a method of multivariate analysis closely related to PCA. It is described for example in \cite{geladi:1986}. For ternary films composed of fibrinogen, immunglobulin G and bovine serum albumin, only major trends in the surface composition could be traced by a PLSR model. As in the preceding article, qualitative information about the composition of complex protein films adsorbed from bovine serum or bovine plasma was obtained with a PCA model developed with the mass spectra of single component adsorbed protein films. 

Still in 2002, Wagner and others compared the interpretation of static ToF-SIMS mass spectra of adsorbed protein films on mica or PTFE by different methods of multivariate pattern recognition \cite{wagner:2002}. The unsupervised technique PCA and the two supervised techniques discriminant principal component analysis (DPCA) and linear discriminant analysis (LDA) were used. An improved discrimination between the mass spectra of different proteins was observed when comparing the supervised techniques to the unsupervised one. Furthermore, Wagner and others introduced a method to classify unknown spectra to the previously examined proteins using a PCA model developed by the mass spectra of these proteins. A successful classification was possible using PCA but it could be improved by the use of DPCA and especially LDA. Yet the very good classification results of LDA went along with a high risk of spurious discrimination.

In their article ``Classification of adsorbed protein static ToF-SIMS spectra by principal component analysis and neural networks'' \cite{sanni:2002}, Sanni and others compared the performance of PCA and the artificial neural network (ANN) ``NeuroSpectraNet'' applied to the mass spectra of proteins adsorbed to silicon substrates. An introduction to neural networks is given for example by Kriesel \cite{kriesel:2006}. Sanni and others concluded that a discrimination of different proteins using PCA with peak selection was possible but the classification of unknown spectra to the known proteins was difficult due to numerous outliers. On the other hand ``NeuroSpectraNet'' allowed a full classification of unknown spectra using the complete mass spectra of positively and negatively charged ions. According to the authors the major drawbacks of neural networks lie in the complexity of their algorithms and of data interpretation. 

Xia and Castner published the article ``Preserving the structure of adsorbed protein films for time-of-flight secondary ion mass spectrometry analysis'' \cite{xia:2003} in 2003. They wanted to preserve the structure of fibrinogen layers on gold coated silicon substrates upon dehydration. The samples were fixated with trehalose or glutardialdehyde and ToF-SIMS spectra of positively charged ions were analysed with PCA. It was found that unfolding and exposure of hydrophobic domains induced by drying could be prevented by both methods. 

In 2003 Belu and others published a review on techniques and applications for characterisation of biomaterial surfaces by ToF-SIMS \cite{belu:2003}. They discuss the ToF-SIMS technique with regard to biomaterial samples and give examples of applications and data interpretation. 

Michel and Castner reviewed the ``Advances in time-of-flight secondary ion mass spectrometry analysis of protein films'' \cite{michel:2006} in 2006. The article deals mainly with characterisation and classification as well as conformation and orientation of proteins, quantitative studies, spatial distribution of proteins, cluster ion sources and matrix-assisted desorption techniques.

Also in 2006 Graham and others gave an overview of current techniques and future needs in ToF-SIMS data interpretation by multivariate analysis in the article ``Information from complexity: Challenges of ToF-SIMS data interpretation'' \cite{graham:2006}.
\pagebreak
\section{Theoretical aspects}

\subsection{Formation of biological films in the oral cavity}
\subsubsection{Saliva}

In the oral cavity, saliva fulfils the following tasks:
\begin{itemize}
\item The regeneration of dental enamel is enabled by ions solved in the saliva.
\item Enzymes like amylase allow the pre-digestion of food.
\item The oral cavity is cleaned by removal of nutrition residues.
\item Saliva buffers acids either supplied by food or produced by bacteria.
\item Mucines form a lubricating film on the tooth surfaces to reduce the mutual abrasion. 
\end{itemize} 

Human saliva consists to over 99 \% of water. The residue is composed to two thirds by organic
and to one third by inorganic compounds  \cite{schmitt:2006}. 
The most abundant inorganic materials are the anions hydrogen \nolinebreak[1] carbonate, chloride and
phosphates as well as the cations of potassium, sodium and calcium.
A large amount of the organic material is formed by proteins. The most frequent of these are 
albumin, amylase and lysozyme  \cite{schmitt:2006}. These proteins analysed in this work are described in
detail in section \ref{sec:proteins}.

\subsubsection{Acquired enamel pellicle}
On the surface of teeth as well as dental implants, proteins and other macromolecules are selectively adsorbed to form a film called pellicle. Salivary proteins form an initial layer of 10 to 20 nanometres thickness within a couple of minutes \cite{hannig:2006}. According to the work of Hannig and Joiner \cite{hannig:2006}, the adsorption is governed by ionic interactions between the proteins' charged groups and calcium and phosphate ions of the enamel surface assisted by van der Waals interactions. In a second phase, proteins and protein aggregates are continuously adsorbed from the saliva. Its thickness reaches a plateau after 30 to 90 minutes and increases further within 60 minutes to reach 100 to 1 000 nanometres. Afterwards the pellicle attains a dynamic equilibrium of adsorption and desorption.   

The major salivary components of in-vivo formed pellicle are proteins and glycoproteins \cite{hannig:2006}. Their proportions are not the same as in whole saliva indicating that the adsorption is a selective process \cite{yao:2003}. The three proteins albumin, amylase and lysozyme studied in this work are abundantly found in pellicle.

The main functions of pellicle formed on enamel surfaces are:
\begin{itemize}
	\item Lubrification of the tooth surface.
	\item Formation of a diffusion barrier for acidic agents to protect the enamel from erosion.
	\item Inhibition of mineral precipitation from the tooth surface.
	\item Modulation of bacterial adherence onto the surface.
\end{itemize}
Proteins in the pellicle can influence the adhesion of bacteria in different ways. Some of them, like amylase, offer specific binding sites for bacteria \cite{hannig:2004} while others, like lysozyme, can decompose bacteria by enzymatic processes \cite{hoch:2005}.  

Bacteria and other micro-organisms colonise the pellicle and form dental plaque which is clearly distinguished from pellicle \cite{hannig:2006}. It is a whitish layer difficult to wipe off the dental surface and it can lead to caries and gingivitis. By formation of phosphate crystals the plaque can mineralise to form dental calculus \cite{schmitt:2006}. 

\subsection{Proteins}
\label{sec:proteins}
Proteins are the most complex known molecules. Due to their manifold structure they can fulfil various tasks. They serve the immune system as antibodies, they transport and store metabolic materials (as haemoglobin does), they allow signal transmission as hormones, they catalyse metabolic reactions in form of enzymes, they form supporting structures (e.g. collagen) and allow the movement of muscles (actin and myosin). An introduction to function and structure of proteins can be found for example in the book by Light \cite{light:1974} and in the diploma thesis of Schmitt \cite{schmitt:2006} which have served as main sources for this section.  

\begin{figure}
\centering
\includegraphics{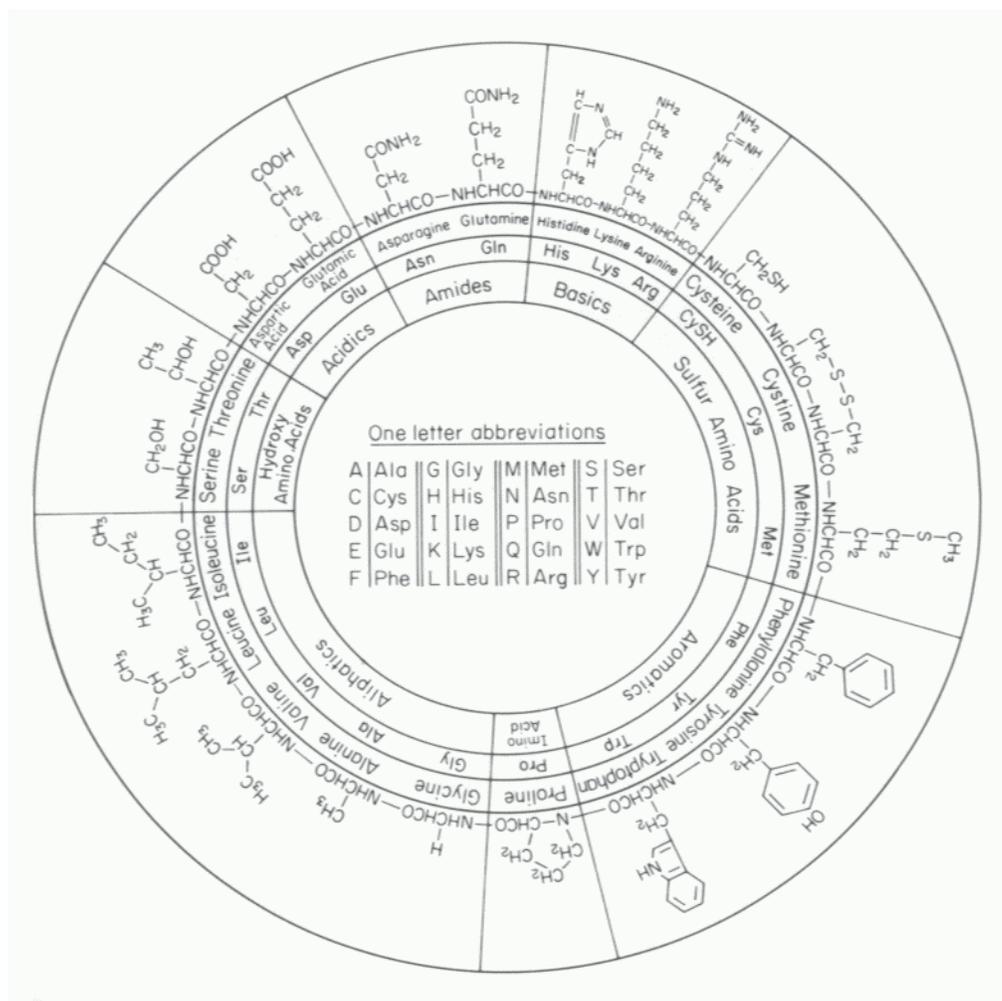}
\caption[The amino acid wheel]{Structures, names and abbreviations of the amino acid residues found in proteins grouped by the properties of their side chains \cite{light:1974}}
\label{fig:aminoacids}
\end{figure}

Every protein is composed of a combination of the twenty proteinogenic amino acids.
An amino acid consists of a carboxylic acid ($\mathsf{-COOH}$) with an amino group
($\mathsf{-NH_2}$), which is usually bound to the $\alpha$ carbon atom of the carboxylic acid,
and a side chain bound to the same carbon atom. Different amino acids are discriminated by the composition of their side chain which will be denoted $\mathsf{R}$ in the following. The structures of the amino acid residues as they are found in proteins are shown in figure \ref{fig:aminoacids}.

\begin{figure}
	\centering
		\includegraphics{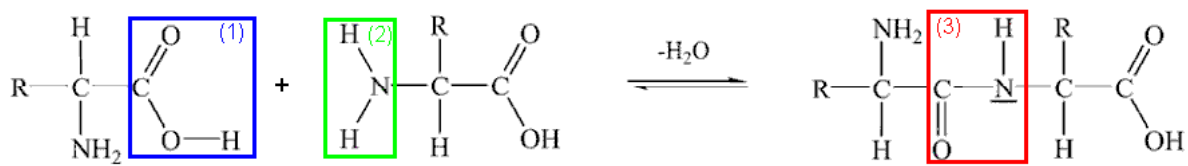}
	\caption[Formation of an amide bond]{Formation of an amide bond: (1) = carboxylic group; (2) = amino group; (3) = amide bond \cite{schmitt:2006}}
	\label{fig:AmideBond}
\end{figure}

The carboxylic group of one amino acid can react with the amino group 
of another by dissection of water. Thereby an amide bond ($\mathsf{O=C-NH}$) is created between the two amino acids (see figure \ref{fig:AmideBond}). A protein is a chain of many amino acids linked by amide bonds. It is unambiguously determined by its amino acid sequence called \emph{primary structure}.

Each amide bond contains a partially negatively charged group ($\mathsf{CO}$)
and a partially positively charged one ($\mathsf{NH}$). This favours the building of hydrogen bonds ($\mathsf{C=O \cdots H-N}$) which stabilise the polypeptide chain in certain conformations. The most common of these \emph{secondary structures} are the $\alpha$ helix and the $\beta$ sheet. 

\begin{figure}
	\centering
	\includegraphics{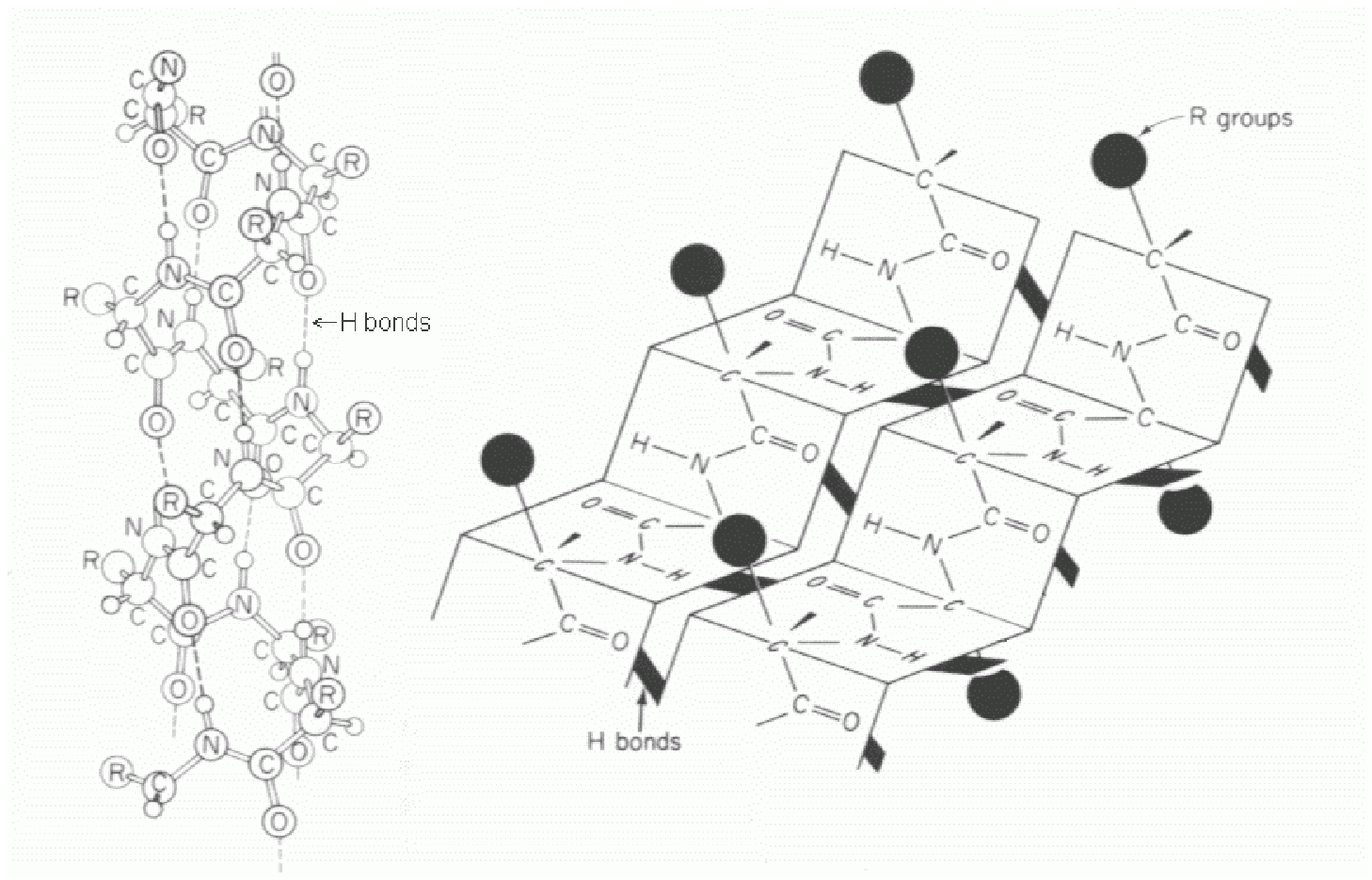}
	\caption[Models of an $\alpha$ helix and a $\beta$ sheet]{Models of an $\alpha$ helix (left) and a $\beta$ sheet (right) (modified from \cite{light:1974})}
	\label{fig:secondarystructures}
\end{figure}

In an $\alpha$ helix the polypeptide chain winds around a central axis with 3.6 amino acid residues per turn and a translational distance along the axis of 5.4 \AA ngstr\"oms per turn \cite{light:1974} (see figure \ref{fig:secondarystructures}). This way every carbonyl oxygen is hydrogen bonded to the amide hydrogen of the fourth peptide further along the chain. The side chains of the amino acid residues mostly point away from the helical axis.

In a $\beta$ sheet some polypeptide chains are closely aligned side by side. To allow the maximum number of hydrogen bonds between the chains, these have to be shorter than a fully extended chain resulting in a conformation that resembles a pleated sheet (see figure \ref{fig:secondarystructures}). The side chains of the amino acid residues are located alternately above or below the plane of the sheet. Regions of the polypeptide chain showing an extended form  without of one of the secondary structures are called \emph{random coil} regions.    

The overall spatial structure (\emph{tertiary structure}) of a protein is built by interactions between the side chains of amino acid residues. These can be salt bridges, hydrogen bonds, van der Waals interactions or disulfide bonds ($\mathsf{S-S}$). The tertiary structure is vital for the functioning of a protein.
If this conformation is changed for example by dehydration or changes in temperature or pH, the protein usually can not fulfil its tasks any more. One speaks of a \emph{denatured} protein in this case.

A protein contains acidic as well as basic amino acid residues that are partly dissociated depending on the pH of the surrounding medium. The dissociation processes create electrically charged residues. Since the numbers of positive and negative charges are usually not equal, the protein carries a non zero net charge.  Anyway, there exists for a given protein a pH at which its net electrical charge is zero. This pH is called the proteins \emph{isoelectric point} (pI). 
 
\subsubsection{Serum albumin}
Serum albumin is the most abundant plasma protein in human blood. It accounts for roughly $60 \%$ of the protein mass in the plasma \cite{schmitt:2006}. Its major task is to maintain the colloidal osmotic pressure. Since the concentration of albumin in the blood vessels is higher than in the surrounding tissue, a leakage of water from the vessel is prevented by the osmotic pressure.  
Furthermore, albumin serves as a transport molecule and buffers the pH-value \cite{friedli:1996}.
Besides in the blood plasma, serum albumin can also be found in the skin, in muscles, in the saliva and in the cerebrospinal fluid. 

In this work bovine serum albumin (BSA) is used because its amino acid sequence is to $76 \%$ identical to the one of human serum albumin (HSA) and it is significantly cheaper \cite{schmitt:2006}.  

\begin{figure}
	\centering
		\includegraphics{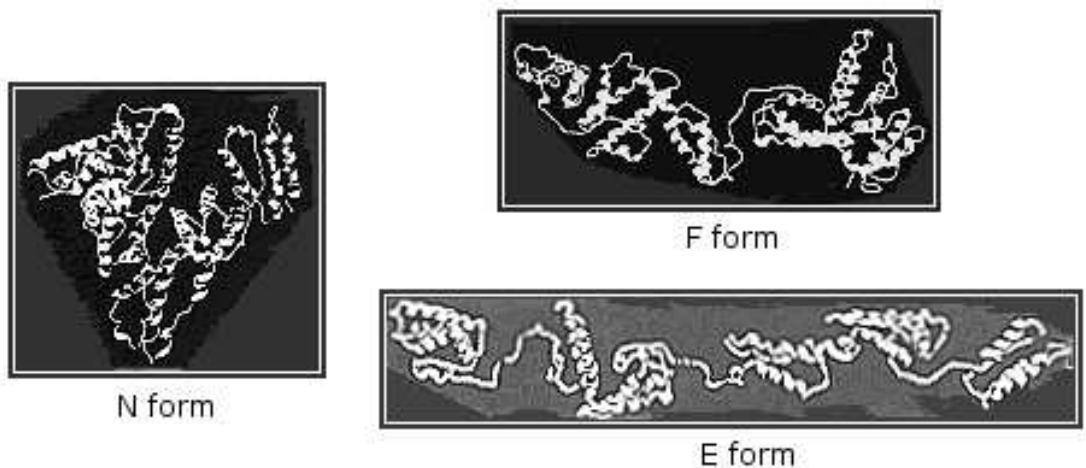}
	\caption[Models of different conformations of BSA]{Models of different conformations of BSA. $\alpha$ helices are represented by helices and random coil regions by cords \cite{friedli:1996}}
	\label{fig:BSAConformation}
\end{figure}

BSA consists of 607 amino acids and has a molecular weight of 69 kg/mol \cite{proteindatabank}. 
In physiological conditions it shows a heart-shaped tertiary structure called normal form (N form) with a size of roughly 11 nm by 8 nm by 8 nm \cite{proteindatabank}.
Its isoelectric point is at pH 4.7 \cite{yao:2003}.
Dependent of the pH-value the shape of BSA is reversibly changed. Above pH 8 one finds the
basic form (B form), between pH 2.7 and pH 4.3 the fast migrating form (F form) and below pH 2.7 the extended form (E form) \cite{friedli:1996}. Figure \ref{fig:BSAConformation} shows simplified models of different spatial conformations of BSA. Instead of the atoms of the polypeptide chain only the different secondary structures and their relative position are drawn.   

\subsubsection{Lysozyme}
Lysozyme can be found abundantly in most body liquids like saliva, plasma and tears. It is the main pellicle-bound bacteriolytical component \cite{hoch:2005}. Its antibacterial effect is caused by the ability to enzymatically dissolve the cell wall of bacteria and to activate bacterial autolysis \cite{hoch:2005}.

The lysozyme used in this work is extracted from chicken egg white. It consists of 129 amino acids and has a molecular weight of approximately 14.3 kg/mol \cite{schmitt:2006}. Its shape is globular with a size of roughly 4 nm by 3 nm by 2.5 nm \cite{proteindatabank} as visualised by the model shown in figure \ref{fig:AmyLysConformation}. Lysozyme is a basic protein with its isoelectric point at pH 9.3 \cite{yao:2003}. It shows optimum efficiency at pH 4.5 \cite{schmitt:2006}.

\subsubsection{Amylase}
Amylase is a digestive enzyme that breaks down starch. It is produced in the salivary glands and in the pancreas. In human saliva it is the most abundant enzyme and it is a major component of pellicle as well \cite{hannig:2005}. Like lysozyme it keeps its enzymatic activity in the pellicle \cite{hannig:2004}. Amylase complexes in the pellicle are binding sites for pioneer bacteria and play thus an important role in plaque formation \cite{hannig:2004}.
Salivary amylase also called pytalin can break polysaccharides down into maltose and glucose. It works only at a pH of about 7 and is inactivated in the stomach by gastric acid \cite{schmitt:2006}. The isoelectric point of amylase is at pH 6.3 \cite{yao:2003}.

\begin{figure}
	\centering
		\includegraphics{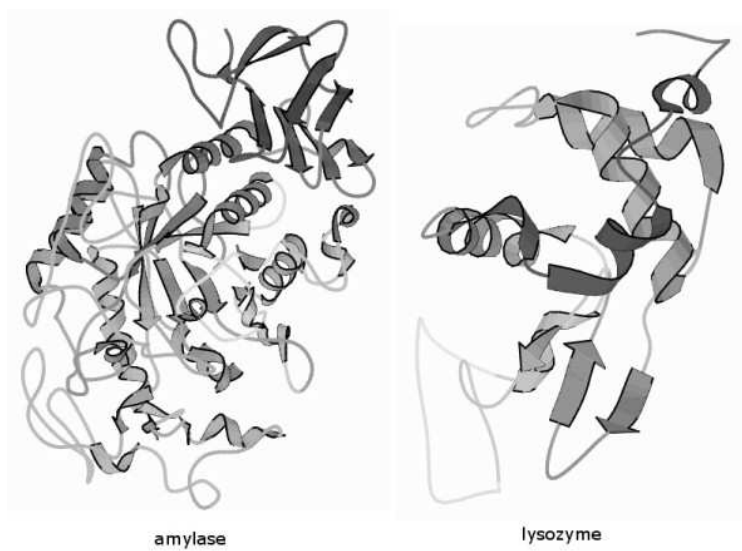}
	\caption[Conformation models of amylase and lysozyme]{Conformation models of amylase and lysozyme. $\alpha$ helices are represented by helices, $\beta$ sheets by flat arrows and random coil regions by cords  \cite{proteindatabank}}
	\label{fig:AmyLysConformation}
\end{figure}

Human salivary amylase consists of 496 amino acids and has a molecular weight of 56 kg/mol. Its size is approximately 7.5 nm by 4.5 nm by 4.5 nm \cite{proteindatabank}. A model of the conformation is show in figure \ref{fig:AmyLysConformation}.

\subsection{Time-of-flight secondary ion mass spectrometry}
In secondary ion mass spectrometry (SIMS), a beam of fast primary ions forms secondary particles on a bombarded surface from which they are ejected. Only few of the secondary atoms and molecules are ionized by the collision and can be examined by a mass analyser. In the case of time-of-flight secondary ion mass spectrometry (ToF-SIMS), the ions are accelerated to a certain energy. Then ions of different masses are separated by the time they need to fly a given distance. 

The most important condition for the examined samples is their ability to tolerate ultra high vacuum conditions with pressures below $10^{-8}$ mbar. A vacuum is necessary to prevent collisions of the primary and secondary ions on their drift trajectories with air molecules. Furthermore, it prevents a contamination of the sample surface during analysis.
Electric charging of the sample must be prevented because it may degrade or even suppress the secondary ion signal. To examine electric insulators they are flooded with slow electrons between the primary ion pulses to compensate charging of the sample.

\subsubsection{Creation of primary ions}
\begin{figure}
	\centering
		\includegraphics{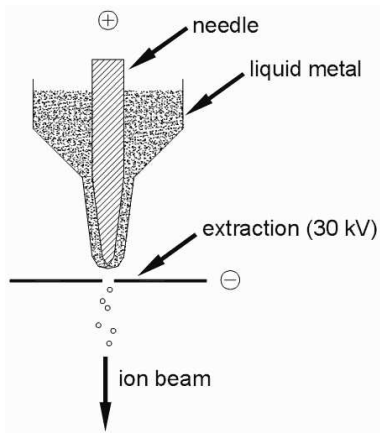}
	\caption[Sketch of a liquid metal ion source]{Sketch of a liquid metal ion source \cite{schnieders:2006}}
	\label{fig:LMIG}
\end{figure}

The primary ions can be created for example by electron collision ionisation, plasma ionisation, surface ionisation or in a liquid metal ion source (LMIS). The latter is employed in the ToF-SIMS apparatus used for this work. The source is a metal tip in a reservoir of liquid metal that forms a film on the surface of the tip. Between the tip and an aperture lying in front of it, a strong electric field (for a gallium source about $10^{10}$ V/m \cite{gnaser:2007}) is applied to create ions from the metal film by field ionisation and accelerate them towards the sample (see figure \ref{fig:LMIG}). Compared to other sources \cite{gnaser:2007} a liquid metal source generates a very intense beam and has a small lateral source size of about 50 nanometres. This creates large space charge effects near the tip of the source causing the emitted ions to have a relatively high energy spread of more than five electron volts. Thus the spot size on the surface is limited by chromatic aberrations. Anyway, liquid metal ion sources are the ones that offer smallest spot size compared to other ion sources \cite{gnaser:2007}.

\subsubsection{Sample interaction}
Before their arrival on the sample surface, the primary ions are accelerated to energies of several kilo-electron volts. Since these energies are significantly larger than typical binding energies, most of the molecular bonds near the impact site are broken and mainly atomic secondary particles are emitted. A part of the primary ion's energy propagates into the sample in form of a collision cascade. Thus particles can be emitted from sites further apart from the impact, too. Since the available energy is smaller at these sites, molecular bonds are not necessarily broken and molecules can be emitted in larger fragments or as an integrated whole. The diameter of the collision cascade is typically smaller than five nanometres in organic compounds and smaller than twenty nanometres in metals \cite{schnieders:2006}. Only particles from the uppermost monolayers have enough energy to overcome the surface binding energy and to escape from the sample. A small fraction of them 
($10^{-6} - 10^{-1}$ \cite{belu:2003}) is charged and can be mass analysed.

To limit the analysis to the uppermost monolayers and to minimize sample damage, an experiment has to be done in static mode (Static Secondary Ion Mass Spectrometry: SSIMS).
This is the case if the probability of any sample site to be hit by more than one primary ion is negligible. The mean number $\bar n$ of primary ions hitting an area $A$ is related to the ion dose $D$ by  
\begin{equation}
D A= \bar n.
\end{equation}
Assuming for static mode that $\bar n < 0.01$ and $A=10 \ \mathrm{nm^2}$, one can calculate the maximum ion dose in SSIMS as
\begin{equation}
D<10^{13} \ \mathrm{cm^{-2}}.
\label{eq:StaticLimit}
\end{equation}

The secondary ion yield $Y$ is defined as the number of detected secondary ions per primary ion.
It is strongly dependent on the chemical environment at the emission site. This \emph{matrix effect} complicates the analysis of ToF-SIMS spectra because the varying yield for different secondary ions as well as for different sample sites leads to an intensity distribution in the mass spectrum that does not necessarily reflect the sample surface composition. 

\subsubsection{Time-of-flight mass analysis}
The principle of time-of-flight (ToF) mass analysis is as follows: All secondary ions are accelerated to the same kinetic energy and drift over a field free distance. At the end of it, the ions are detected and one can discriminate different masses by their differing times-of-flight. To measure a time interval, one needs a well-defined starting time. Therefore the primary ion beam is pulsed and it is blanked after each pulse until all secondary ions have arrived at the detector. The advantage over other mass analysers like quadrupole or magnetic sector field systems lies in the parallel acquisition of signals from all masses. In addition, ToF mass analysers offer a greater transmission \cite{belu:2003} which allows a better exploitation of the limited number of available secondary ions.

Usually the secondary ions are accelerated to kinetic energies of several kilo-electron volts. Thus their kinetic energy is negligible against their rest energy, which amounts already to nearly one giga-electron volt for a single proton. Hence the following calculations are done in non-relativistic approximation.  

To calculate the time-of-flight, the following variables are defined: From the sample surface secondary ions of mass $m$ and electric charge $q$ are accelerated on a distance $d$ by an electric field of strength $E$. Thus they traverse a voltage $V=Ed$. Afterwards they drift on a field free distance $D$ and are detected at the time $t$. Within the acceleration the force $F$ on the ions and their acceleration $a$ are related by
\begin{eqnarray}
F=Eq \quad &\mbox{and}& \quad F=ma \\
\Rightarrow a&=&Eq/m.
\end{eqnarray}
The final velocity $u$ and the time $t_a$ to traverse the acceleration range are thus
\begin{eqnarray}
u=\int_{t_0} ^{t_a} a dt  = u_0 + Et_aq/m \\
\Rightarrow t_a=\frac{(u-u_0)m}{Eq}.
\end{eqnarray}
$u_0$ is the initial velocity of the ions at the starting time $t_0$. From the kinetic energy $T$ at the end of the acceleration one can calculate the drift velocity $u_d$
\begin{eqnarray}
T=qV=qEd=1/2mu_d^2 \\
\Rightarrow u_d=u-u_0=\sqrt{\frac{2qV}{m}}.
\end{eqnarray}
Assuming a small initial velocity ($u_0 \ll u$), it follows that
\begin{equation}
u \approx u_d = \sqrt{\frac{2qV}{m}}.
\end{equation}
Hence the drift time $t_D$ is
\begin{equation}
t_D = D/u=\frac{D}{\sqrt{2qV/m}}.
\label{eq:ToFApproximation}
\end{equation}
The observed time-of-flight $t$ is the sum of the preceding times and the reacting time $t_R$ of the detector
\begin{equation}
t=t_0+t_a+t_D+t_R.
\label{eq:ToF}
\end{equation}

Since the acceleration occurs on a distance of typically some millimetres and the drift distance measures about one metre, the drift time is with some 100 $\mu$s much longer than all the other terms and one can use equation (\ref{eq:ToFApproximation}) as good approximation of the time-of-flight. 

The charge of the secondary ions amounts to $q=ze$. $z$ is the charge number and $e$ the elementary charge. Thus it follows from equation (\ref{eq:ToFApproximation}) that
\begin{equation}
t^2 = \frac{D^2}{2V}\frac{m}{q} \propto \frac{m}{z}.
\end{equation}
To derive the mass to charge ratio $m/z$ from the time-of-flight, one has to determine by calibration the proportionality constant $n$ in 
\begin{equation}
\frac{m}{z} = nt^2
\end{equation}

The mass of an ion of given charge is determined by $m = c t^2$ with a proportionality constant $c$. By deriving, one gets $dm = 2 c dt$. Hence the mass resolution
$m / \Delta m$ is inversely proportional to the relative uncertainty in the measurement of the time-of-flight $\Delta t/t$:
\begin{equation}
\frac{m}{\Delta m}=\frac{t}{2 \Delta t}.
\end{equation}
Reasons for this uncertainty are the spread of initial velocities, energies and positions as well as the spread in the formation times of the secondary ions. Furthermore non-ideal accelerating fields and jitter of the detection system contribute to the uncertainty. 

In case of a solid sample all secondary ions are created at its surface and traverse the same acceleration distance. With an initial energy $T_0$, their drift energy $T_D$ and their drift time $t_D$ are  
\begin{eqnarray}
T_D&=&T_0+qV \\
t_D&=&D\sqrt{\frac{m}{2(T_0+qV)}}.
\end{eqnarray}
Assuming a long drift distance ($t = t_D$),
\begin{eqnarray}
dt &=& dt_D = \frac{\frac{D}{2}\sqrt{\frac{m}{2}}}{(T_0+qV)\sqrt{T_0+qV}}dT_0 \\
\Rightarrow \frac{dm}{m}&=&\frac{2dt}{t}=\frac{dT_0}{T_0+qV}
\end{eqnarray}
holds.

\begin{figure}
	\centering
		\includegraphics{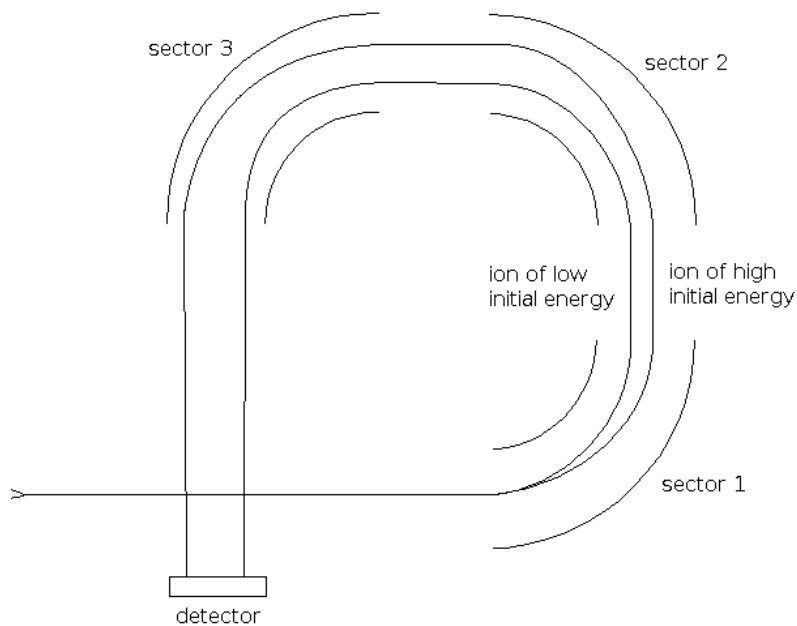}
	\caption{Effect of the initial energy on the trajectories of ions of equal mass}
	\label{fig:SectorFields}
\end{figure}

Since the acceleration energy usually is much larger than the initial kinetic energy
($qV\gg T_0$), the mass resolution is
\begin{equation}
\frac{m}{\Delta m}=\frac{T_0+qV}{\Delta T_0}\approx \frac{qV}{\Delta T_0}.
\end{equation}
To achieve a high mass resolution, the spread in the initial energy has to be compensated. This is usually done by letting high energy ions traverse a longer distance than low energy ions. In the mass spectrometer used for this work, this is realised by a system of three electric 90° sector fields. Therein the trajectories of high energy ions have larger radii than the ones of low energy ions as sketched in figure \ref{fig:SectorFields}. 

The uncertainty in the time of ion creation $t_0$ is mainly determined by the length of the primary ion pulse. For the apparatus employed for this work, it amounts to roughly ten nanoseconds.

Because of their high sensitivity, electron multipliers and micro channel plates are used as detectors.
In the ideal case, all electrons created by the impact of one secondary ion should arrive at the same time at the detector anode. Furthermore an amplifier with a short time constant and an exact determination of the starting time $t_0$ are needed to prevent signal broadening. Actually none of these conditions is perfectly met, so that a single secondary ion usually creates a signal pulse of up to ten nanoseconds width \cite{guilhaus:1995}. 

\begin{figure}
	\centering
		\includegraphics{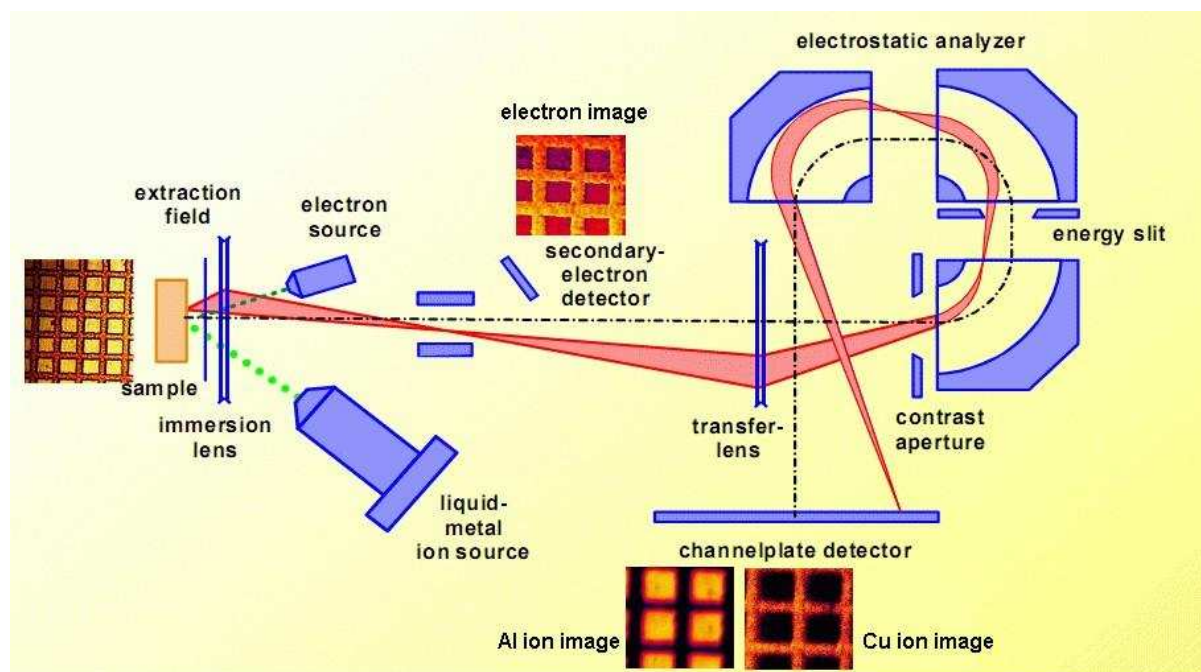}
	\caption[Layout of the TRIFT mass spectrometer]{Layout of the TRIFT mass spectrometer (modified from \cite{gnaser:2007})}
	\label{fig:TRIFT}
\end{figure}
Figure \ref{fig:TRIFT} shows the layout of the triple focusing time-of-flight (TRIFT) mass spectrometer used in this work. Due to a system of electrostatic lenses, it offers not only energy focusing but also direction focusing. Hence on the channel plate detector an image of the ion distribution on the sample's surface is created. In the figure, the ion images of aluminium and copper from an aluminium copper grid are shown to illustrate this.

\subsection{Fluorescence microscopy}
Fluorescence microscopy is a kind of optical microscopy allowing the visualization of objects containing a fluorescing molecule, a so-called fluorophore. 

\begin{figure}
	\centering
		\includegraphics{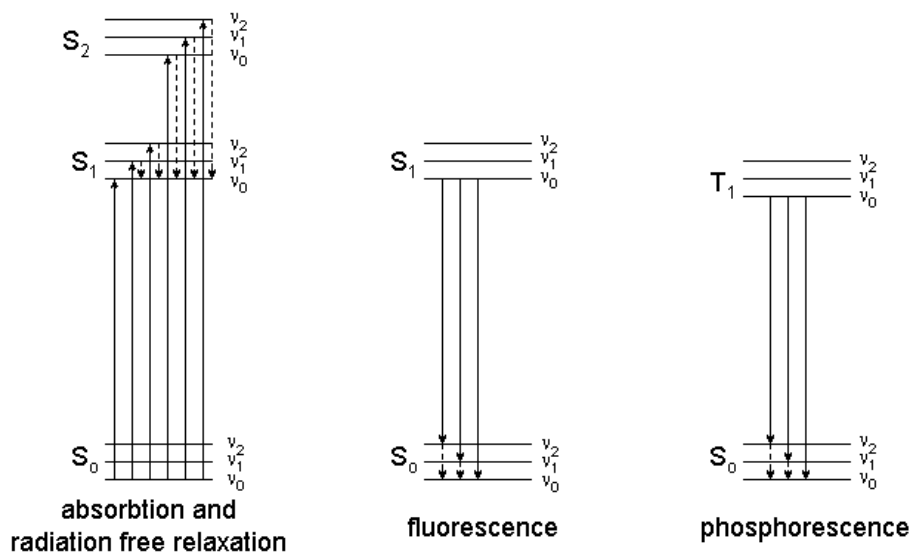}
	\caption[Scheme of excitation and relaxation in a fluorophore]{Scheme of excitation and relaxation in a fluorophore. Radiation transitions are symbolised by solid lines, radiation free transitions by dashed lines}
	\label{fig:FluorescenceScheme}
\end{figure}

Absorption of light in the visible or ultraviolet range by a molecule can lead to an electronically excited state. Afterwards there exist several different ways for the molecule to return to its ground state. If this relaxation is accompanied by the emission of radiation, one speaks of luminescence.
Due to electron pairing, the electronic ground state of a molecule with an even number of electrons is usually a singlet state called $S_0$ with vanishing total spin. This electronic state can be subdivided into several vibrational states (denoted $\nu_0,\nu_1,\nu_2,...$), but at room temperature mainly the vibrational ground state is occupied. By absorption of a photon with the matching energy, the molecule can be excited into a vibrational state of the first ($S_1$) or higher ($S_2$, ...) excited electronic singlet states. The principles of excitation and relaxation in a fluorophore are sketched in figure \ref{fig:FluorescenceScheme}. An excitation into a triplet state is forbidden by quantum selection rules because it would require a spin-flip which cannot be caused by a photon \cite{demtroeder:2005}.

Then the molecule relaxes by radiation free transitions to the vibrational ground state of the first excited electronic state. The energy is dispersed to inner vibrational states of the molecule or to vibrational states of neighbouring molecules \cite{galla:1988}. Anyway, the molecule does not relax by radiation free transitions from the first excited electronic state to the electronic ground state. Since the energetic difference among the states $S_1$ and $S_0$ is larger than between the others, more vibrations would have to be excited simultaneously reducing the probability of a radiation free transition to occur. Thus the molecule relaxes by emission of a photon.

Once the molecule has arrived in the $S_1$ vibrational ground state \emph{intersystem crossing} can occur with little probability. In this case the transition from the singlet system to the triplet system takes place. Another possible way to leave the $S_1$ state is relaxation by fluorescence \emph{quenching} if the excitation energy is transferred to a so-called quencher molecule.

\emph{Fluorescence} occurs upon the radiating transition from the $S_1$ vibrational ground state into a vibrational state of the electronic ground state $S_0$. Typical transition rates are $10^7$ to $10^8$ per second \cite{galla:1988}. 

\emph{Phosphorescence} is generated by the radiating transition from the vibrational ground state of the first excited electronic triplet state $T_1$ to a vibrational state of the electronic ground state $S_0$. Its transition rate is much smaller with typical values ranging from less than one up to $10^4$ per second \cite{galla:1988}. 

The setup of a fluorescence microscope resembles the one of an ordinary light microscope. Typically a mercury lamp or a laser is used to illuminate the samples. In the first case the wavelength for exciting the fluorophore is selected by a set of filters. A dichriotic mirror reflects the light through the observation optics onto the sample surface. The fluorescing molecules emit light which is collected by the observation optics. Since the wavelength of the emitted light is longer then the one of the exciting light, it can pass the dichriotic mirror to be guided to the ocular pieces, a camera or a photomultiplier where it is detected. 

The advantages of fluorescence microscopy over normal optical microscopy are the following:
\begin{itemize}
	\item Objects showing little contrast with respect to one another can be distinguished by marking them with different fluorophores emitting at different wavelengths.
	\item Since they act as light sources, fluorescing objects much smaller than the optical resolution of the microscope can be detected.	 
\end{itemize}
A drawback of the method is the possible modification of the sample's chemical or physical properties by the introduction of a fluorophore. Additionally, the properties of the fluorophore, especially its fluorescence yield, can be strongly dependent on its chemical environment making comparisons of different samples difficult. 

Using normal light microscopy the protein layers dealt with in this work are not visible. So it is necessary to modify them with fluorescence markers for imaging.   

\subsection{Scanning force microscopy}
Scanning force microscopy (SFM) is a scanning probe microscopy method. It allows imaging of the topography of non conducting samples in air as well as in liquid with a typical lateral resolution of some nanometres and a height resolution of less than one \AA ngstr\"om \cite{binnig:1986}. Since it is possible to achieve a resolution of atomic scale, this method is also called atomic force microscopy (AFM).

\begin{figure}
	\centering
		\includegraphics{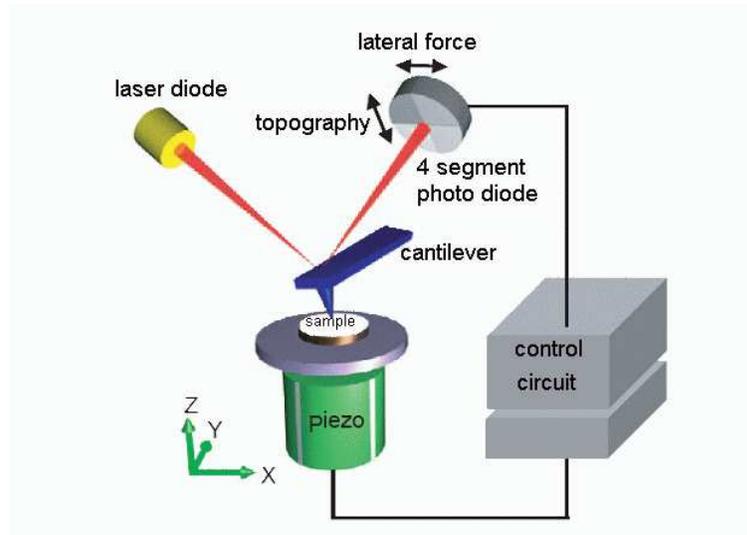}
	\caption[Layout of a scanning force microscope]{Layout of a scanning force microscope (modified from \cite{schmitt:2006})}
	\label{fig:AFMSchema}
\end{figure}
Figure \ref{fig:AFMSchema} shows the layout of a scanning force microscope. The very fine tip of a cantilever serves as probe to explore the surface of a sample placed on a stage which can be moved in three dimensions by piezoelectric actuators. A laser spot is reflected from the cantilever onto the centre of a four segment photo diode. By measuring the light intensities on the four segments and calculating the differences between the intensity arriving on the upper and lower half and between the right and left half of the photo diode, movements of the cantilever can be detected. If the cantilever is bent, the first difference is non-zero; if it is twisted, the second difference is non-zero.

In the so-called \emph{contact mode} the sample is approached to the cantilever until its tip feels the repulsive force caused by the sample surface. This force causes an upwards bending of the cantilever and can thus be detected via the photo diode. There are two possibilities to scan the sample surface:

In the \emph{constant height mode} the z position of the sample is held constant while scanning in x and y directions. Here x, y and z are the axes of a Cartesian coordinate system with the z axis perpendicular to the sample surface and the x and y axes in the sample surface plane (see figure \ref{fig:AFMSchema}). Any height differences of the sample surface cause a change in the force acting on the cantilever tip which is recorded to calculate an image of the sample surface. Due to the limited flexibility of the cantilever, this mode can only be used on very flat surfaces.

In the \emph{constant force mode} the z position of the sample is modified by a control circuit to maintain the force acting on the cantilever tip constant while scanning. This way the distance between cantilever tip and sample surface remains constant. Hence the variations of the position of the z piezoelectric drive correspond directly to the height variations on the sample surface and can be used to create an image. This mode can be used on flat as well as on rough surfaces but it is slower than the constant height mode. 

Another possibility for imaging the sample surface is the \emph{dynamic mode} (also called tapping mode or AC mode). In this case the cantilever is set to oscillation by a piezoelectric element at a frequency near its resonance frequency. The amplitude of the cantilever oscillation is detected via the photo diode. If the cantilever approaches the sample surface and forces act, the amplitude changes. Similar to the constant force mode, the z position of the sample is modified while scanning in the x and y directions to maintain the amplitude change constant. Hence the distance between the cantilever tip and the surface remains constant and the position variations of the z piezoelectric drive can be used to create an image of the surface topography. Since in dynamic mode the force acting on the sample surface is smaller than in contact mode, the former is especially useful for imaging soft and delicate samples. The advantage of the contact mode lies in its higher lateral resolution \cite{schmitt:2006}.  

\subsection{Scanning electron microscopy}
\begin{figure}
\centering
\includegraphics{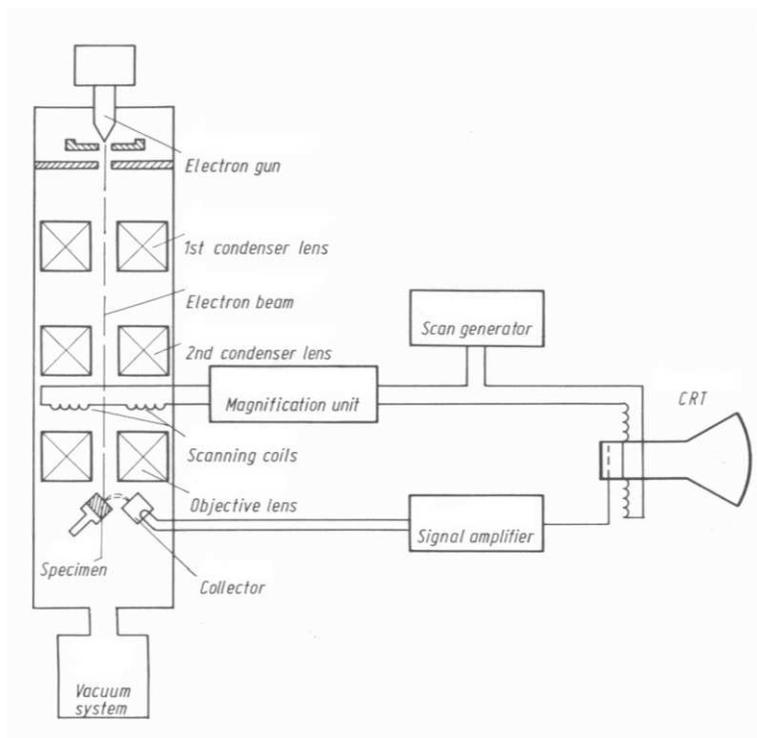}
\caption{Block circuit diagram of a scanning electron microscope (modified from \cite{ohnsorge:1978})}
\label{fig:SEMscheme}
\end{figure}

The principle of a scanning electron microscope (SEM) is represented by figure \ref{fig:SEMscheme}. 
The whole system has to be evacuated to below $10^{-5}$ millibar \cite{ohnsorge:1978} to achieve long enough mean free pathways for the electrons. In an electron gun a beam of so-called \emph{primary electrons} (PE) with typical energies of about $10^4$ electron volts \cite{ohnsorge:1978} is produced by thermal emission or field emission. By a column of lenses it is focused on the investigated specimen. Scanning coils scan the primary beam over the surface of the specimen. The scan generator synchronizes the scanning of the primary beam to the formation of an image on a computer screen (in former times on a cathode ray tube (CRT)). This way the scanning image is faithfully reproduced on the screen. When the primary electrons hit the surface of a specimen, several interactions can occur. For imaging only the produced secondary electrons of low energies (0 - 50 eV \cite{ohnsorge:1978}) and backscattered electrons of high energies (about $10^4$ eV \cite{ohnsorge:1978}) are used. A laterally placed, positively biased detector collects the electrons. The intensity of its output signal is used to modulate the brightness of the point on the screen corresponding to the point on the sample surface being scanned. Due to their small energy, secondary electrons can only be emitted from a specimen if they are produced very close to the surface. Backscattered electrons can also leave the specimen from deeper layers. They contribute to image formation mainly by releasing more secondary electrons on their way back to the surface. The main cause of contrast in the created image is the number of secondary electrons created close to the surface. It is influenced by the following properties of the specimen:
\begin{itemize}
\item Tilting: Surfaces non-perpendicular to the primary beam are brighter than surfaces perpendicular to it.
\item Topography: Spikes and edges appear brighter than plane surfaces.
\item Electric charge: Negatively charged regions are brighter than positively charged ones.
\item Chemical composition: Heavy elements appear brighter than light ones. 
\end{itemize}
If the latter two effects are negligible, the image creates a three dimensional impression of the specimen. 

The main advantages of SEM imaging over optical microscopy imaging are an about 300 times greater depth of focus \cite{ohnsorge:1978} due to the small illumination aperture (0.06° to 0.6°) of the primary beam and a point resolution of down to two nanometres \cite{ohnsorge:1978}.

\subsection{Principal component analysis}
Mass spectra of proteins are very complex, because every protein can decompose into several different fragments visible as peaks in the mass spectra. Considering only the masses from 1 amu/z (atomic mass unit per charge number) to 200 amu/z, every measured spectrum can be interpreted as a point in a two hundred dimensional space spanned by unit vectors corresponding to the different masses. The intensity of a mass peak determines its projection onto the corresponding vector. In this space, spectra of different proteins should be grouped at different sites due to their similarity. Unfortunately it is not possible to visualize such a high dimensional space.  

Principal component analysis (PCA) is a method of linear algebra with the goal to concentrate the information contained in a data set onto a small number of variables called principal components (PC). 

Comprehensive information on PCA can be found in the books by Jackson \cite{jackson:1991} and Jolliffe \cite{jolliffe:2002}. The following explanation is based upon the tutorial by Shlens \cite{shlens:2005}.

\subsubsection{Change of basis}
Let a mass spectrum be represented by a column vector $\vec x$ of length $m$ containing as elements the intensities at the masses from 1 amu/z to $m$ amu/z. Out of $n$ spectra one can construct an $m \times n$ data matrix $X$:
\begin{equation}
X=[\vec x_1 \cdots \vec x_n].
\end{equation}
PCA searches a basis to represent the data set in a way that concentrates the information onto just a few of the new basis vectors. The new basis vectors should be linear combinations of the old ones, to allow the use of the well established techniques of linear algebra. Thus a transformation matrix $P$ is needed to transform the data matrix into a new 
$m \times n$ matrix $Y$:
\begin{equation}
Y=PX.
\end{equation}
The row vectors $p_1 \cdots p_m$ of $P$ are the new basis vectors and by definition the principal components of $X$. 

\subsubsection{Variance and covariance}
A well transformed matrix $Y$ should fulfil mainly two criteria:
Firstly, it should contain as much relevant information as possible. Assuming the measured data are not too noisy, directions in the data space showing a large variance are the interesting ones. By assumption, the effects of interest show stronger dynamics than the measurement noise.
Secondly, there should not be any redundancy. Since the number of relevant variables should be as small as possible, mutual dependencies between them have to be eliminated.

The mutual dependency of two variables is characterised by their covariance.
Let $a$ and $b$ be two mean centred row vectors of the data matrix $X$. Their elements are thus the intensities at one mass in different measurements minus the mean intensity at the respective mass.
Their covariance $\sigma^2_{ab}$ is defined as  
\begin{equation}
\sigma^2_{ab}:=\frac{1}{n-1}a b^T.
\end{equation}
$b^T$ is the transposed vector of $b$.
The covariance fulfils the following:
\begin{itemize}
	\item $\sigma^2_{ab} \geq 0$. The covariance is zero if and only if the variables are uncorrelated.
	\item $a=b \Rightarrow \sigma^2_{ab} = \sigma^2_a$. The covariance of a variable with itself is its variance.
	\item $\sigma^2_{ab} = \sigma^2_{ba}$. The covariance is symmetric with respect to the variables.
\end{itemize}
Further explanations concerning the concept of covariance can be found in the tutorial on PCA by Smith \cite{smith:2002}.

Since the $x_i$ are the rows of the data matrix $X$,
one can generalise the definition of the covariance to a covariance matrix $C_X$:
\begin{equation}
C_X:=\frac{1}{n-1}X X^T.
\end{equation}
It has the following properties:
\begin{itemize}
\item $C_X$ is a symmetric $m \times m$ matrix.
\item The diagonal elements are the variances of the variables contained in $X$.
\item The off-diagonal elements are the covariances between the variables contained in $X$.
\end{itemize}

To minimize covariance, the covariance matrix of the data in the transformed basis  $C_Y$ has to be diagonal. This way all covariances are zero which is their smallest possible value.

\subsubsection{Diagonalisation of the covariance matrix}
A matrix $P$ of basis vectors $p_1 \cdots p_m$ is searched to diagonalise the covariance matrix
$C_Y$ of $Y=PX$. The basis vectors are ordered by their importance by assuming that vectors with the highest variance are the most important ones. As a further assumption, PCA postulates orthonormality of the new basis vectors. This is to make the solution achievable by methods of linear algebra. By this assumption, $P$ is an orthogonal matrix ($P^T=P^{-1}$ : The transposed matrix of $P$ is its inverse).
One can express the new covariance matrix $C_Y$ as a function of the searched matrix $P$:
\begin{eqnarray}
C_Y&=&\frac{1}{n-1}Y Y^T \quad // \quad Y=PX \\
&=&\frac{1}{n-1}PXX^TP^T \quad // \quad C_X=\frac{1}{n-1}X X^T \\
&=& PC_XP^T. \label{eq:CovarianceDiagonalisation}
\end{eqnarray}
The covariance matrix $C_X$ is symmetric by its construction:
\begin{equation}
C_X^T=(\frac{1}{n-1}X X^T)^T=\frac{1}{n-1}X^{T^T}X^T=\frac{1}{n-1}X X^T=C_X.
\end{equation}

Every symmetric matrix can be diagonalised by an orthogonal matrix $O$. The columns of $O$ are the eigenvectors of $C_X$. The elements of the resulting diagonal matrix $D$ are the eigenvalues of $C_X$ (a proof can be found in \cite{shlens:2005}):
\begin{equation}
C_X=ODO^T.
\label{eq:Diagonalisation}
\end{equation}

A matrix $C_X$ of rank $r \leq m$ has exactly $r$ orthonormal eigenvectors. If $C_X$ is degenerate (i. e. $r<m$), one can freely choose $m-r$ additional orthonormal vectors to fill up $O$.
These do not influence the solution because their eigenvalues are zero. 

Let the row vectors $p_i$ of $P$ be the eigenvectors of $C_X$. Consequently $P=O^T$ holds. Hence from equation(\ref{eq:Diagonalisation}) it follows that
\begin{equation}
C_X=P^TDP.
\end{equation}
With this, one can substitute $C_X$ in equation (\ref{eq:CovarianceDiagonalisation}):
\begin{eqnarray}
C_Y&=&PP^TDPP^T \quad // \quad P^T=P^{-1} \\
&=&D
\end{eqnarray}
Obviously the covariance matrix is diagonal by employing this $P$.
Its diagonal elements are the variances of the respective variables.
The solution of PCA can be summarised as follows: 
\begin{itemize}
\item The mean centred $m \times n$ data matrix $X$ is transformed by the orthogonal matrix $P$ to a new $m \times n$ matrix $Y$ ($Y=PX$). Its elements are the so-called \emph{scores} of the $n$ measurements on the $m$ new basis vectors, the principal components. The scores link the measurements to the principal components.
\item To calculate the principal components, the covariance matrix $C_X$ of $X$ is used: $C_X= \frac{1}{n-1}  XX^T$.
\item The eigenvectors of $C_X$ are the principal components of $X$ and the row vectors of $P$. The elements of the PC, called \emph{loadings}, show how strongly the initial variables (masses) influence the PC. Loadings link the initial variables to the principal components.
\item The eigenvalues of $C_X$ are the diagonal elements of $C_Y$. Their values are the variances of the respective principal components.
\item The eigenvectors and -values are ordered by descending variance, that is by descending eigenvalues.
\item This way, the relevant information is concentrated onto the first principal components.
\end{itemize}

\subsubsection{Graphical representation}
To visualise the results of principal component analysis, one usually plots the first two or three scores vectors and loading vectors. In the scores plot, the positions of the measurements in the space spanned by the first principal components are visible. Under favourable circumstances the measurements are grouped according to their sample type in this space.
The corresponding loadings plot shows which of the initial variables are responsible for the positioning of the measurements in the scores plot. 

\begin{figure}
	\centering
		\includegraphics{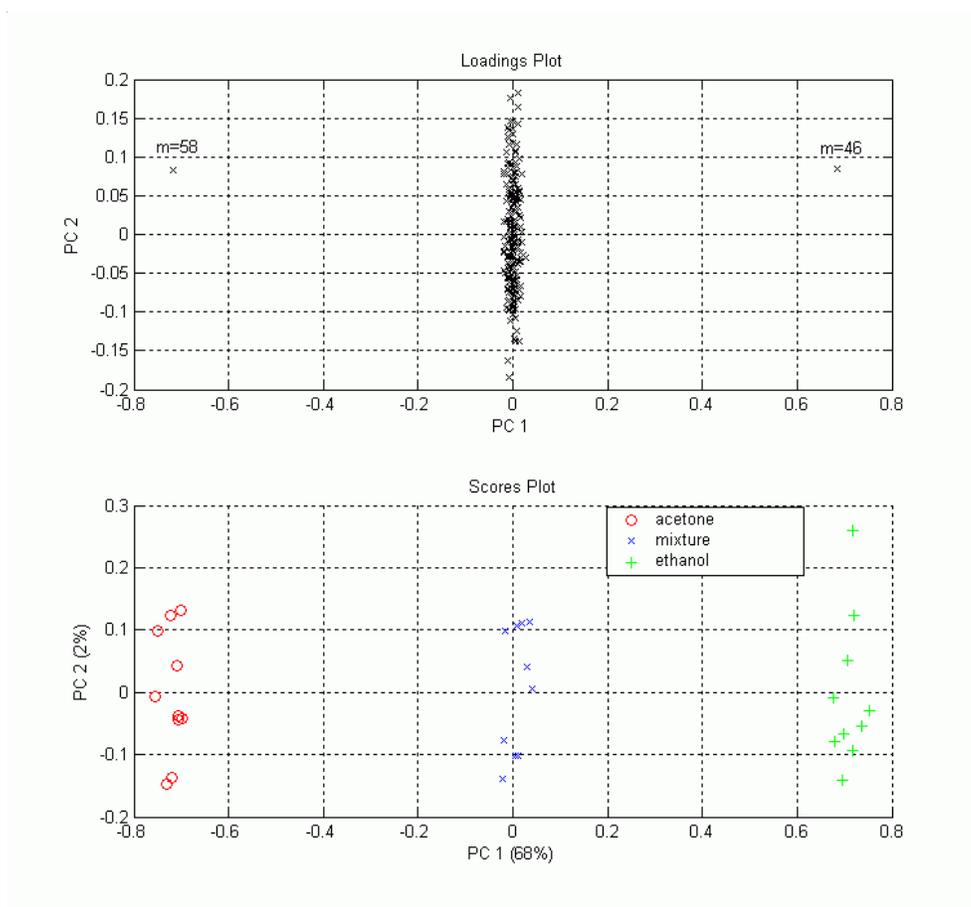}
	\caption{Results of PCA on simulated data}
	\label{fig:PcaSimulation}
\end{figure}

This shall be clarified by an example. Here, the ``measurements'' are ten simulated mass spectra each of acetone, ethanol and a 1:1 mixture of the two. The scores and loadings on the first two PC are given in figure \ref{fig:PcaSimulation}. The percent values at the axes are the relative variances explained by an axis.    

The scores plot shows that the first principal component separates the three sample types. The spread on the second PC is caused by the equally simulated noise of the measurements. In the loadings plot one can see that the first PC is dominated by the masses 58 amu/z (acetone) in the negative direction and by 46 amu/z (ethanol) in the positive direction. Accordingly, in the scores plot acetone shows negative values on the first PC and ethanol shows positive values.

\begin{figure}
\centering
\includegraphics{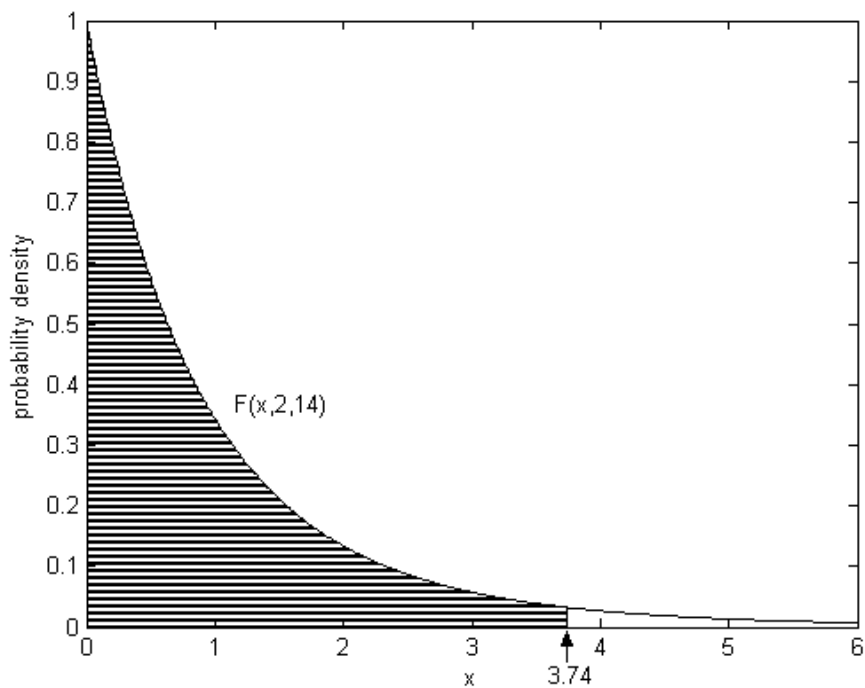}
\caption[Example of an $F$ distribution]{Example of an $F$ distribution with the parameters $a=2$ and $b=14$ and its critical value $F_{2,14,0.05}=3.74$ for a significance level of $\alpha=0.05$}
\label{fig:fdistribution}
\end{figure}

\label{TsquareEllipse}
In the representation of a two dimensional scores plot, it is possible to draw a probability ellipse around a group of data points. Therefore one has to do once again PCA on these data represented in the space of the first two  principal components of the prior analysis.
The centre of the ellipse is given by the mean values of the data points on the two axes. 
The two new principal components define the directions of the axes of the ellipse. The first principal component gives the direction of highest variance, the major axis of the ellipse. The second principal component is by definition orthonormal to the first one and gives the direction of the minor axis. The eigenvalues $\lambda_1$ and $\lambda_2$ associated to the principal components are used to calculate the lengths of the axes. The length of the semimajor axis is given by 
\begin{equation}
l_1=\sqrt{\lambda_1 T^2_\alpha}.
\end{equation} 
The length of the semiminor axis is 
\begin{equation}
l_2=\sqrt{\lambda_2 T^2_\alpha}.
\end{equation}
Here $T_\alpha^2$ is the critical value of the $T^2$ distribution which is related to the critical value $F_{p,n-p,\alpha}$ of the $F$ distribution by
\begin{equation}
T^2_{\alpha,n,p}=\frac{p(n-1)}{n-p}F_{p,n-p,\alpha}.
\end{equation}
$p$ is the number of variables (here $p=2$) and $n$ is the number of measurements in the data set. The $T^2$ statistic predicts that a data point of the considered sample can be found with a probability of $P=1-\alpha$ in the so defined ellipse \cite{jackson:1991}. In this work a level of significance $\alpha=0.05$ is used. The $F$ distribution is a probability density function of a continuous positive random variable $x$ with two independent degrees of freedom $a$ and $b$ as parameters. In the case of the probability ellipses the parameters $a$ and $b$ correspond to the number of variables $p$ and the reduced number of measurements $n-p$. The $F$ distribution is described by the following formula \cite{hartung:1982}:
\begin{equation}
F(x|a,b)=\frac{\Gamma(\frac{a}{2}+\frac{b}{2})}{\Gamma(\frac{a}{2})\Gamma(\frac{b}{2})}\frac{x^{\frac{a}{2}-1}}{(1+\frac{a}{b}x)^{\frac{a+b}{2}}}.
\end{equation}
$\Gamma(x)$ is the gamma function given by
\begin{equation}
\Gamma(x)=\int^\infty_0 t^{x-1}\exp(-t)dt.
\end{equation}
Further properties of the $F$ distribution can be found in the book ``Statistik'' by Hartung \cite{hartung:1982}. The critical value of the $F$ distribution $F_{a,b,\alpha}$ is defined by
\begin{equation}
\int^{F_{a,b,\alpha}}_0 F(x|a,b)dx = 1-\alpha.
\end{equation}
Therefore the random variable has a probability of $1-\alpha$ to be smaller than the critical value of the $F$ distribution. Figure \ref{fig:fdistribution} illustrates this for the case $a=2$ and $b=14$. The hatched area represents the probability of $95\%$ 
for $x$ to be smaller than the critical value $F_{2,14,0.05}=3.74$. 
Critical values of the F distribution are tabulated for example in the ``User's guide to principal components'' by Jackson \cite{jackson:1991}.

\subsubsection{Discriminating principal component analysis}
The goal of this work is to recognize different adsorbed proteins by their mass spectra. Therefore a representation of the data stressing the differences between the proteins is necessary. Yet, PCA searches the components with largest variance regardless of whether this is the variance between spectra of one protein (``internal'' variance) or between the spectra of different proteins (``external'' variance).  
 
To achieve a good discrimination between the different sample types, it would be better to maximise the relation of external variance to internal variance. This is the goal of discriminant principal component analysis (DPCA) as described by Yendle \cite{yendle:1989}.

A simple approach to DPCA would be to replace the spectra of one protein by the mean spectrum of this protein. Thus the internal variance would be zero and normal PCA would find the axes of highest external variance. With the resulting principal components one could transform the initial data matrix.  

However, in the initial data the axes of highest external variance could be the ones of highest internal variance as well. Hence the simple approach would not maximise just the distance between different proteins but also the spread within the protein groups. Thus the internal variance has to be taken into account to calculate the principal components. 

Therefore the initial variables (masses) are scaled by their total internal standard deviation. That is, first one calculates the internal variances of each variable for all the different proteins. These are summed to form the total internal variance of each variable. By root extraction one obtains the total internal standard deviations to scale the values of the respective variables. This way variables with a small internal variance are scaled up and vice versa.    

In the resulting scaled data set all the spectra of a protein are replaced by its mean spectrum. Afterwards a normal PCA is done. The resulting principal components are the discriminating principal components (DPC) for transforming the scaled data set. The transformed data can be used to create scores plots that maximise the external variance while minimising the internal variance, to allow the best possible discrimination between the sample types.   

The general disadvantage of DPCA is the need of a training data set with known discrimination of the measurements into groups to calculate the principal components. 
Therefore DPCA is called a \emph{supervised} technique of multivariate data analysis while PCA is an \emph{unsupervised} technique. After having calculated the DPC, one can use these for transforming new measurement data, to see to which of the known groups these belong. The only condition is that the differences between training data and unknown data must not be too large. 

Since the goal of this work is not the recognition of unknown patterns in the data but the attribution of measurements (mass spectra) to known groups (proteins), a training phase is also necessary in normal PCA to know, where in the scores plot the proteins can be found. Hence in this case the named disadvantage of DPCA does not matter.

\subsubsection{Evaluation}
In this work the \emph{Leave-One-Out-Technique (LOO)} shall be used to evaluate how well a (D)PCA projection fits the data. As suggested by the name, one measurement is left out of the data set before performing (D)PCA. Afterwards this measurement is projected into the space spanned by the principal components using the loadings matrix. 

Now the measurement can be assigned to one of the sample groups. This can be done by calculating the euclidean distances on the first principal components to the centres of gravity of the sample groups and assigning the left out measurement to the closest group. The great advantage of this method is its simplicity but on the other hand it cannot deal with outliers. Even samples that do not belong to any group are assigned to one. Another possibility is to draw probability ellipses around the sample groups on the first two principal components and to assign the left out measurement to a group if its projection lies in the corresponding ellipse. With this method outliers can be identified because they should not lie in any probability ellipse. The problem lies in the fact that projections can be found in the overlap of more than one ellipse. In this case no unambiguous assignment is possible.    

The procedure is repeated with leaving out different measurements until all of them have been left out once. If most of the measurements are assigned to the correct group, the (D)PCA projection fits well the data and can be used to project unknown data and assign them to the known sample groups. If on the other hand many measurements are assigned to the wrong groups, the (D)PCA projection cannot be used to represent the data.  
\pagebreak
\section{Experimental aspects}

\subsection{Chemicals}
\begin{itemize}
\item Bovine serum albumin fraction V, approximately $99 \%$, Sigma-Aldrich, Germany.
\item Lysozyme from chicken egg white, approximately $95 \%$, Sigma-Aldrich, Germany.
\item $\alpha$-Amylase from human saliva, Fluka BioChemika, USA.
\item Bovine serum albumin fluorescein conjugate, Invitrogen, Germany
\item Glutardialdehyde solution, $50 \%$ in water, Merck-Schuchardt, Germany.
\item 3-Aminopropyl-tri(ethoxy)silane, minimum $98 \%$, Sigma-Aldrich, Germany.
\item Sodium di(hydrogen)phosphate monohydrate, minimum $99.5 \%$, Fluka BioChemika, Switzerland.
\item Disodium hydrogenphosphate, minimum $99 \%$, Riedel-de Ha\"en, Germany.
\item Bidistilled water with a resistivity of $182 \; \mathrm{k\Omega m}$ from a ``Milli-Q A10'' water purification system, Millipore, USA.
\item Dehydrated, denatured ethanol, Department of Chemistry, Technische Universit\"at Kai\-sers\-lau\-tern, Germany.
\item Hydrogen peroxide, $35 \%$, Department of Chemistry, Technische Universit\"at Kai\-sers\-lau\-tern, Germany.
\item Sulphuric acid, $95 - 97 \%$, Department of Chemistry, Technische Universit\"at Kai\-sers\-lau\-tern, Germany.
\item Sodium hypochlorite solution, approximately $13 \%$ of active chlorine, Department of Chemistry, Technische Universit\"at Kai\-sers\-lau\-tern, Germany.
\item Dehydrated toluene, work group of Professor Thiel, Department of Chemistry, Technische Universit\"at Kai\-sers\-lau\-tern, Germany.
\end{itemize}

\subsection{Sample preparation}

\subsubsection{Dental material}
The dental implant materials and the samples of bovine enamel are provided by the work group of Professor Matthias Hannig, Universit\"atsklinikum Homburg, in pieces of roughly five millimetres  by five millimetres size. 
The dental implant materials are polymer matrices containing apatite particles. The two examined types are called FAT and FAW for fluoroapatite with a more transparent or more whitish appearance. They are made of the following substances:
\begin{itemize}
\item silanized (for FAW) or unsilanized (for FAT) fluoroapatite ($\mathsf{Ca_5(PO_4)_3 F}$) particles
\item bis-phenol-A-glycidyl-di(methacrylate)
\item tri(ethylene glycol)-di(methacrylate)
\item poly(methacryl)oligo(maleic acid)
\item camphor quinone
\item strontium
\end{itemize}
The surfaces are polished. 
Before being used, the substrates are ultrasonically cleaned for five minutes in a solution of one percent of sodium hypochlorite. To remove a possible deposit of sodium hypochlorite, they are ultrasonically cleaned for another five minutes in bidistilled water and rinsed three times in bidistilled water. 

\subsubsection{Protein films on silanised substrates}
\begin{figure}[tbp]
	\centering
		\includegraphics{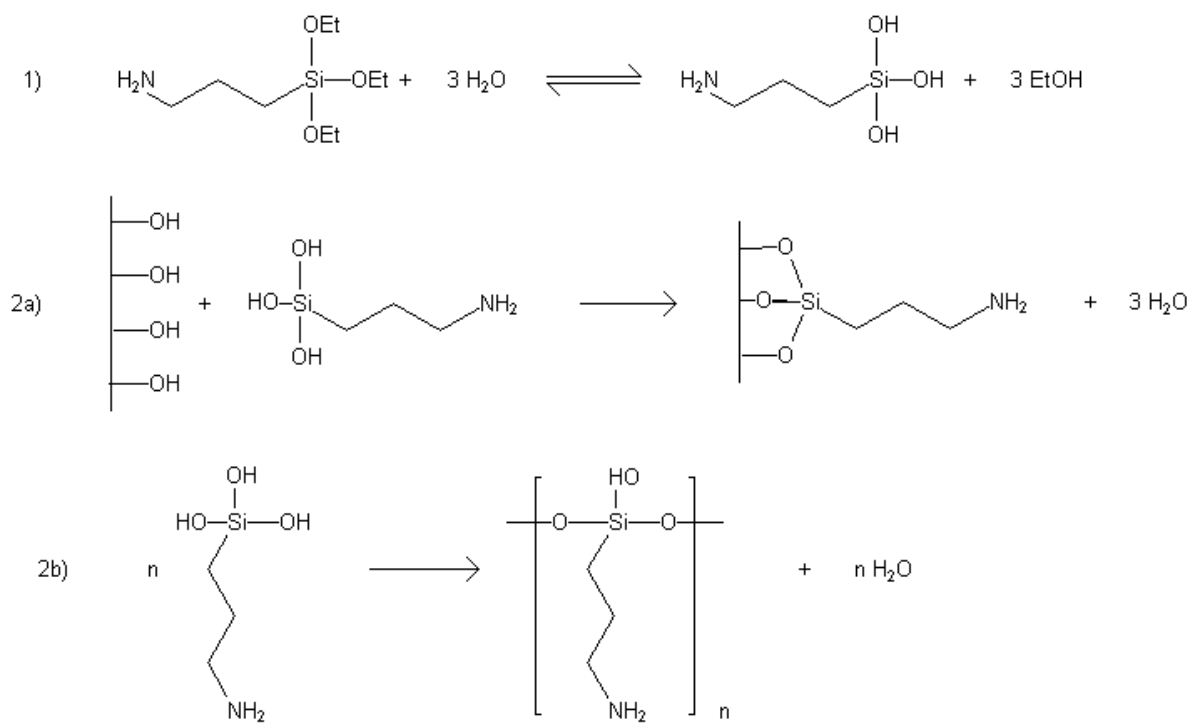}
	\caption[Hydrolysis, binding and polymerisation of APTES]{Hydrolysis (1), binding (2a) and polymerisation (2b) of APTES}
	\label{fig:APTESBinding}
\end{figure}

\begin{figure}
	\centering
		\includegraphics{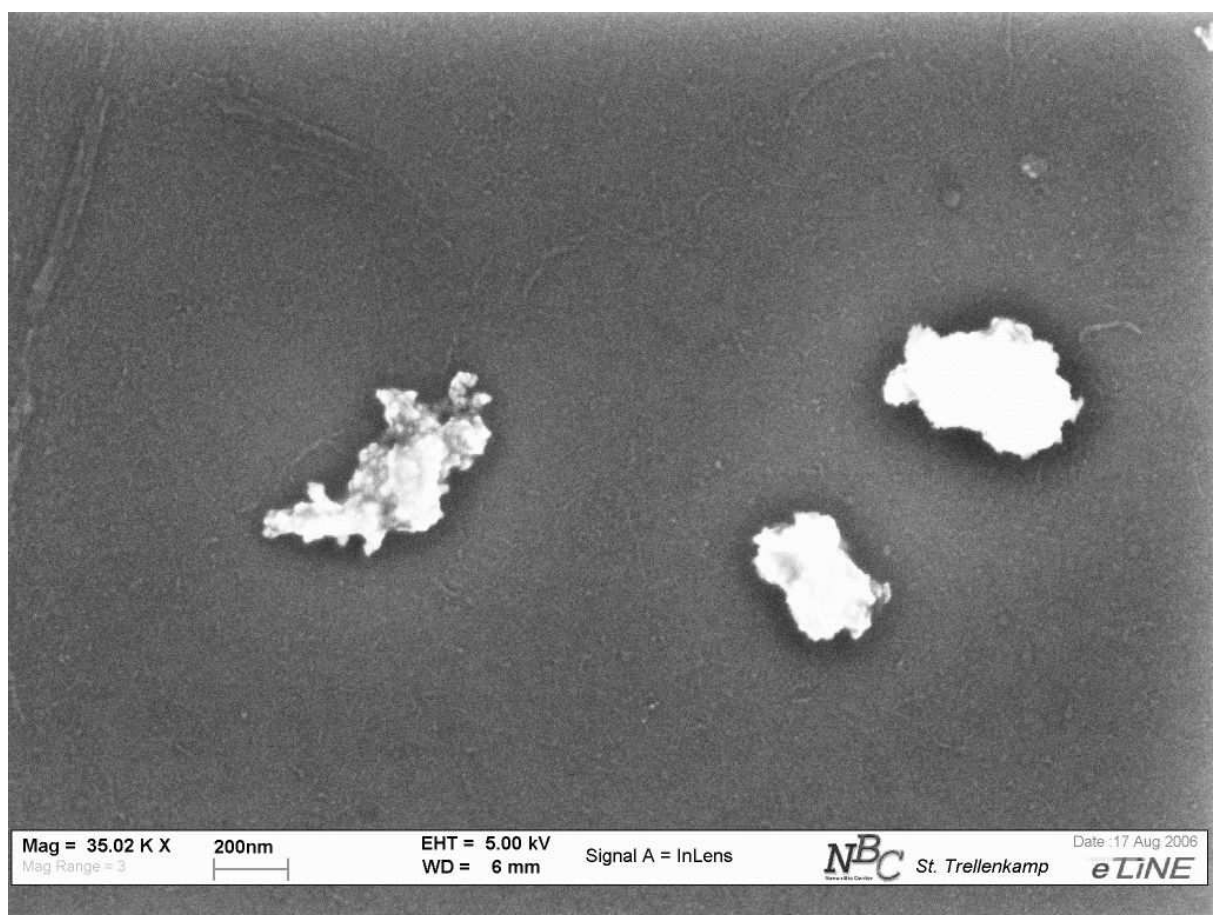}
	\caption{SEM image of a silanized substrate with aggregates of polymerised silane}
	\label{fig:polymerisiert}
\end{figure}

Approximately ten millimetres by five millimetres large silicon wafer pieces serve as substrate in this case. First they are cleaned for fifteen minutes in a solution consisting to two thirds of concentrated sulphuric acid and to one third of a thirty-five percent solution of hydrogen peroxide. This cleaning solution is called \emph{piranha solution} for its ability to remove most organic matter. Afterwards the substrates are rinsed with bidistilled water and dried with nitrogen. Then they are placed for one hour under protective gas in a mixture of 20 ml of dry toluene and 0.5 ml of 3-aminopropyl-tri(ethoxy)silane (APTES). First the APTES hydrolises  with the residual water in the solution to form a silanol. The latter can covalently bind to the oxide atoms of the silicon oxide to form a monolayer as shown in figure \ref{fig:APTESBinding}. At the same time the silane can also polymerise. To remove polymerised silane from the surface, the samples are given into an ultrasonic bath for fifteen minutes in ethanol.Then they are rinsed with bidistilled water.
Silanised substrates are imaged with a scanning electron microscope (SEM) by Dr. Stefan Trellenkamp, Nano+BioCenter Kai\-sers\-lau\-tern. When the samples are not sonicated in ethanol, many aggregates of polymerised silane can be found on the surface (see figure \ref{fig:polymerisiert}). Their number is strongly reduced after sonication. 
 
\begin{figure}[tbp]
	\centering
		\includegraphics{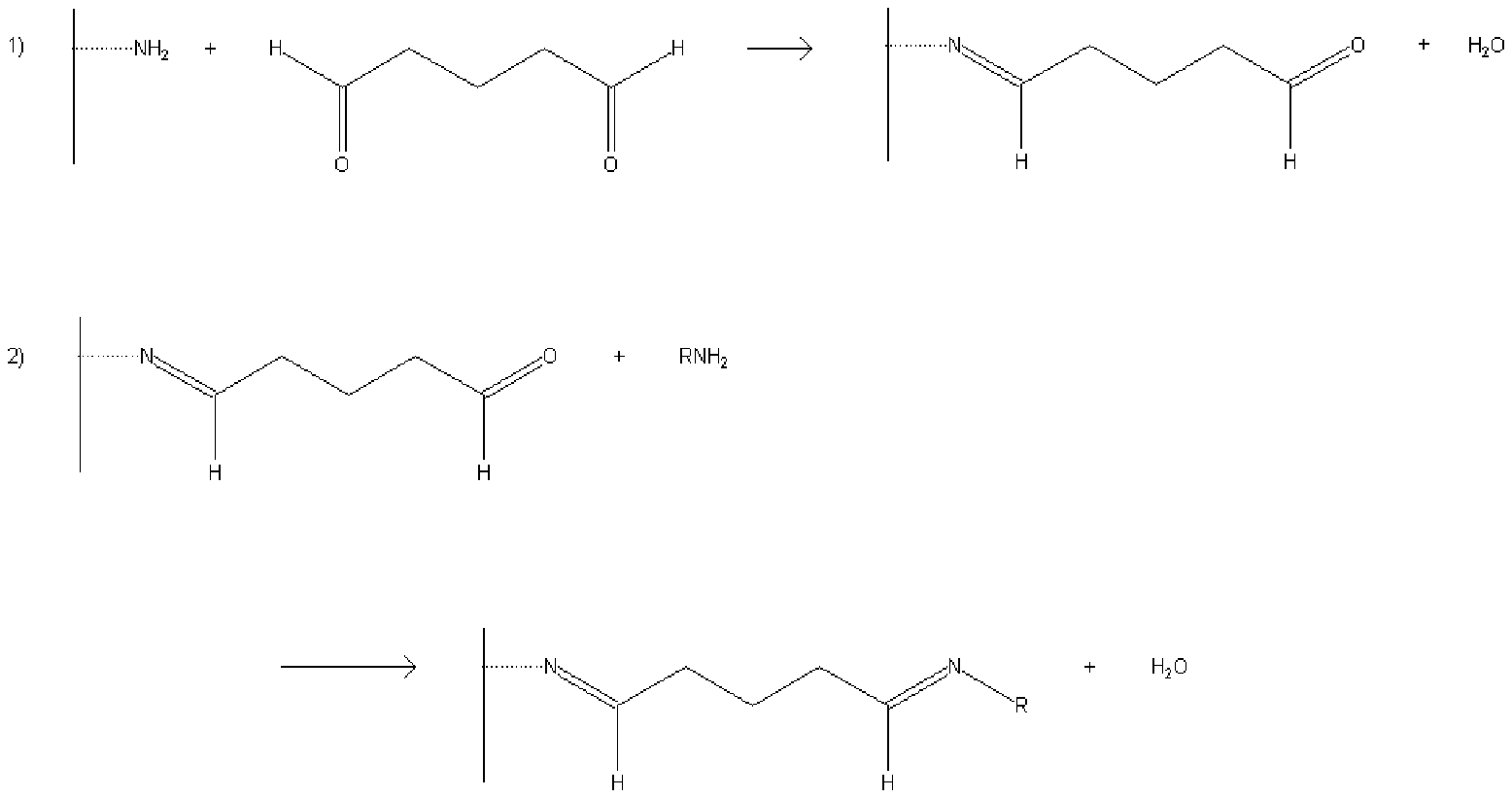}
	\caption[Binding of glutardialdehyde to the silane and of a protein to glutardialdehyde]{Binding of glutardialdehyde to the silane (1) and of a protein R to glutardialdehyde (2)}
	\label{fig:GluBinding}
\end{figure}

Glutardialdehyde serves as link between the silane and the protein. It shall bind to the silane as shown in figure \ref{fig:GluBinding}.
Therefore the samples are placed for one hour in a five percent (volume / volume) solution of glutardialdehyde in pH 7 buffer solution. The pH-value of the buffer solution is adjusted by mixing a 0.1 molar solution of sodium di(hydrogen)phosphate ($ \mathsf{NaH_2PO_4} $) and a 0.1 molar solution of disodium hydrogenphosphate  ($ \mathsf{Na_2HPO_4} $). It is controlled with a pH electrode ``Testo 252'' by ``Testo'', Germany. After the adsorption of glutardialdehyde the samples are rinsed with buffer solution. 
The actual protein adsorption also takes place in pH 7 buffer solution for one hour with protein concentrations of two grammes per litre. All adsorption steps are effectuated at room temperature. To remove loosely bound protein, the samples are rinsed with fresh buffer solution. In a last step they are rinsed another three times with bidistilled water to remove buffer salts and dried with argon.

\subsubsection{Protein films on silicon substrates}
To prevent the possible influence of polymerised silane on the protein mass spectra, protein films are prepared directly on silicon. As in the previous section ten by five millimetre silicon wafer pieces are used as substrates. These are cleaned for fifteen minutes in piranha solution. Afterwards the substrates are rinsed with bidistilled water and given into protein solutions with a concentration of one to five grammes per litre. The solvent for the proteins is either bidistilled water or a pH 7 phosphate buffer as described above. After two hours of adsorption the samples are rinsed three times with bidistilled water to remove loosely bound protein and buffer salts and dried with argon.

\subsubsection{Protein films on dental implant materials}
Substrates of the dental implant materials FAT and FAW are ultrasonically cleaned for ten minutes in a two percent solution of sodium hypochlorite ($ \mathsf{NaOCl} $). Afterwards the substrates are given for another five minutes in an ultrasonic bath in bidistilled water and rinsed with bidistilled water to remove remainders of sodium hypochlorite. The proteins are solved with molarities of about $10^{-4}$ mols of protein per litre pH 7.4 buffer solution, which is prepared from 0.01 molar solutions of disodium hydrogenphosphate and sodium di(hydrogen)phosphate as described above. For protein adsorption the cleaned substrates are given into protein solutions for two hours at room temperature. Then the samples are rinsed three times with bidistilled water to remove buffer salts and loosely bound protein and dried with argon.    

\subsubsection{Samples for fluorescence microscopy}
\begin{figure}
	\centering
		\includegraphics{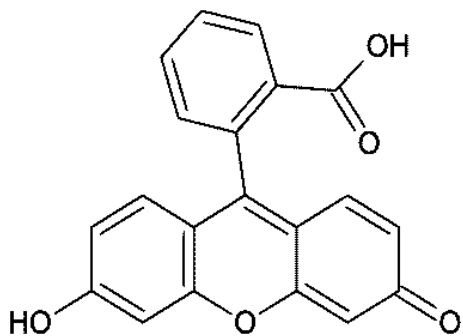}
	\caption{Chemical structure of fluorescein}
	\label{fig:Fluorescein}
\end{figure}

Protein layers are prepared on silicon substrates for fluorescence microscopy. The substrate pieces of ten by five millimetres size are cleaned for twenty minutes in piranha solution and rinsed with bidistilled water. Afterwards proteins are adsorbed from a phosphate buffer solution with pH 7.4. Either only BSA conjugated with the fluorophore fluorescein ($\mathsf{C_{20} H_{12} O_5}$, see figure \ref{fig:Fluorescein} for chemical structure) or a 1:1 (weight:weight) mixture of BSA fluorescein conjugate and non-marked lysozyme are adsorbed for two hours at room temperature. According to the manufacturer Molecular Probes (USA) the absorption and emission maxima of the BSA fluorescein conjugate are at wavelengths of 494 nm and 520 nm. After protein adsorption the samples are either rinsed with bidistilled water and air-dried or air-dried without prior rinsing.   

On one sample the protein is adsorbed onto a silane layer. Therefore the cleaned substrate is given into a solution of 0.5 ml of 3-aminopropyl-tri(ethoxy)silane (APTES) in 20 ml of dry toluene at 80 °C to 90 °C for two hours under protective gas. Then the sample is baked out at 150 °C for another two hours to favour the formation of an APTES monolayer. Next the sample is given into a $1 \%$ glutardialdehyde solution in pH 7.4 phosphate buffer for one hour. This way a monolayer of glutardialdehyde should build on the silane layer as described above. Before protein adsorption the sample is rinsed with bidistilled water. 

\subsection{Time-of-flight secondary ion mass spectrometry}
All measurements are made with a ``TRIFT II'' (TRIple Focusing Time-of-flight) apparatus by ``Physical Electronics'', USA. The use of this instrument was kindly allowed by its owner, the ``Institut f\"ur Oberfl\"achen- und Schichtanalytik, Kai\-sers\-lau\-tern (IfOS). The ion source is a liquid metal gallium gun that produces primary ions with an energy of 25 kilo-electron volts. An electron flood gun compensates charging effects with electrons of an energy of 20 electron volts.
The initial energy spread of the secondary ions is compensated by a system of three electrostatic 90° sector fields.

Since the gallium stock of the source ran out during the measurements, the source had to be replaced.
Before the replacement the mass resolution was determined on a silicon wafer as $m/\mbox{FWHM}\approx 300$ at 
$m=28$ amu/z. The mass resolution is defined as ratio between the mass $m$ and the full width at half maximum (FWHM) of the corresponding peak in the mass spectrum. With the new source the mass resolution is $m/\mbox{FWHM}\approx 420$ at $m=28$ amu/z.
 
Of all the samples mass spectra of cations and anions in the mass range of 1 amu/z to 400 amu/z are acquired at several sample sites. The scanning size is $A=(120 \ \mathrm{\mu m})^2$. The extractor current of the ion source is maintained at $1.5 \ \mathrm{\mu A}$ corresponding to a primary ion current of below one nanoampère in unpulsed mode. The acquisition time is usually five minutes.   

The primary ion dose $D$ can be calculated with the primary ion current $I$, the acquisition time $t$, the scanning size $A$, the  
repetition frequency $f$, the pulse length $\Delta t$ and the elementary charge $e=1.6 \cdot 10^{-19}$ C by
\begin{equation}
D=\frac{I t f \Delta t}{e A}.
\end{equation}
Here it is $I \approx 10^{-9}$ A, $t = 300$ s, $f \approx 10^4$ Hz and $\Delta t = 12 \cdot 10^{-9}$ s. Hence the primary ion dose
\begin{equation}
D \approx 2 \cdot 10^{12} \ \mbox{cm}^{-2}
\end{equation}
lies well in the regime of static analysis (see equation (\ref{eq:StaticLimit})).

To see whether the surfaces of the dental implant materials are contaminated, they are also examined after sputtering with the unpulsed primary ion beam. For sputtering the scanning size is augmented to $A=(240\mathrm{\mu m})^2$ to exclude the influence of gating effects on the measurements. When sputtering a surface one does not obtain a crater with vertical walls. Instead, there is a transition region at the edges from the crater ground to the unsputtered surface. To obtain a measurement only from the crater ground, one has to chose the scanning size for analysis smaller than the one for sputtering. 

In the protein films, depth profiles are acquired. Therefore the following two steps are alternated:
\begin{enumerate}
	\item Acquisition of a mass spectrum as described above.
	\item Unpulsed sputtering of the same sample site for five seconds with 1.44 square millimetres scanning size. This is the largest possible scanning size. It was chosen to obtain a slow removal of the protein layer.
\end{enumerate}
Here only mass spectra of cations are acquired, because according to Wagner et al. \cite{wagner:2001} 
the spectra of anions do not allow a discrimination between different proteins, since they contain only signals of the unspecific protein backbone (mainly $ \mathsf{CN^-} $ and $ \mathsf{CNO^-} $).

\subsection{Fluorescence microscopy}
The samples are imaged using an epifluorescence microscope ``Axioskop 2 mot'' (Carl Zeiss, Germany) equipped with a CCD camera ``AxioCam HRc'' (Carl Zeiss, Germany). The samples are illuminated by a mercury lamp with a filter transmitting light with wavelengths of 450 nm to 490 nm. Fluorescence emission of the sample is detected for wavelengths longer than 515 nm. This setup suits well the fluorescence properties of the BSA fluorescein conjugate, because its maximum exciting and emission wavelengths are 494 nm and 520 nm (according to the manufacturer Molecular Probes, USA). 

\subsection{Scanning force microscopy}
Measurements are carried out on a ``Molecular Force Probe MFP-3D'' scanning force microscope (Asylum Research, USA). The samples are imaged in air in contact mode with constant force acting on the cantilever tip. Scanning areas are varied between fifteen micrometres by fifteen micrometres and two micrometres by two micrometres.

\subsection{Determination of amylase activity}
The activity of amylase immobilized on silicon substrates is determined by the group of Dr. Christian Hannig, Universit\"atsklinikum Freiburg, Germany, with a method described in \cite{hannig:2004}. For these experiments silicon substrates with 25 square millimetres surface area are used. They are cleaned with piranha solution for fifteen minutes and rinsed with bidistilled water. Proteins are solved in ten millimolar phosphate buffer solutions with pH 7.4 at molarities of about $5 \cdot 10^{-5}$ mol per litre. On one sample only amylase is adsorbed for two hours. A second sample is given into a lysozyme solution for two hours, rinsed with buffer solution and given into an amylase solution afterwards for another two hours. The samples are rinsed three times with bidistilled water and dried with argon after the last adsorption step. Then they are sent to Freiburg in micro centrifuge tubes for activity measurements. 

The determination of amylase activity is based upon its ability to directly hydrolyse the synthetic trisaccharide 2-chloro-4-nitrophenyl-4-O-$\beta$-D-galactopyranosylmaltotrioside (GalG2CNP) without any auxiliary enzymes. One product of this reaction is aglycone 2-chloro-4-nitrophenolate (CNP). According to \cite{hannig:2004}, the formation of CNP is stoichiometric with respect to incubation time and occurs at a constant rate. CNP is detected photometrically by its absorption at a wavelength of 405 nanometres. 
The samples are incubated for 10 minutes in 300 microlitres test solution at a temperature of 25 °C. The test solution contains five millimol GalG2CNP per litre and is buffered at pH 6.0. Immediately after removal of the sample, the absorption is read against reagent blank at 405 nanometres. One unit of amylase activity is defined as hydrolytic production of 1 micromol CNP per minute. The absorbance $A_0$ of one micromol CNP in the given experimental setup can be determined independently. Thus the activity $a$ can be calculated from the measured change in absorbance within ten minutes $\Delta A$:
\begin{equation}
a=\frac{\Delta A}{10 A_0}.
\end{equation}
With the known surface $S$ (25 mm$^2$) of the samples, the immobilized activity per square centimetre of sample surface can be calculated as
\begin{equation}
\frac{a}{S}=\frac{\Delta A}{2.5 \, \mathrm{cm^2}A_0}
\end{equation}

\subsection{Data analysis}
The mass spectra are acquired and calibrated with the software CADENCE (version 2.0, Physical Electronics, USA). A newer version of the same software (WinCadence 3.7.1) is used to export the data to unit mass mass spectra in ASCII format. These are analysed in a \textsc{Matlab} 6.5 environment (The MathWorks, USA) with functions written for this purpose, which can be found in appendix \ref{sec:sourcecode}. First the mass spectra are extracted from the WinCadence export files with the functions ``einlesen'', ``voll'' and ``matrix''. A background spectrum can be subtracted (``silanweg'') and peaks related to amino acid fragments can be selected (``auswahl'') from the mass spectra. Principal component analysis with an algorithm based upon the PCA tutorial by Shlens \cite{shlens:2005} or discriminant principal component analysis with an algorithm based upon Yendle's article \cite{yendle:1989} are performed using the functions ``pca'' or ``dpca''. The results are plotted by the function ``plotpc2d''. In addition to the data points, ``plotdpc2dmitellipse2'' also plots probability ellipses around groups in the scores plot. Therefore the function ``ellipse'' is used. To calculate the size of the probability ellipses, it needs the critical value of an F distribution provided by the function ``Fdistribution'', which contains critical values taken from the ``User's guide to principal components'' \cite{jackson:1991}. The quality of the DPCA models is evaluated by leave-one-out tests. The left out measurements are assigned to a group either by their projection's position with respect to the probability ellipses of the groups (function ``loo'') or by the euclidean distance of the projections to the groups' centres of gravity (function ``loo3''). New spectra can be projected into existing scores plots using the functions ``pcaprojektion'' and ``dpcaprojektion''. They can be assigned to existing groups with the function ``zuordnen''.
\pagebreak
\section{Results}
\subsection{ToF-SIMS of dental implant materials}
Comparison of the mass spectra of the dental implant materials FAT and FAW before and after sputtering shows a significant decrease in intensity of the peaks assigned to organic compounds. Hence spectra acquired without prior sputtering are distorted by organic surface contaminants. These can be compounds not removed by the cleaning process or molecules adsorbed between cleaning and the analysis by ToF-SIMS. In the following, only spectra acquired with prior sputtering are used for analysis.

To compare the surface composition of the dental implant materials and bovine tooth enamel, mass spectra of cations and anions are acquired at three sites on samples of the implant materials FAT and FAW as well as bovine tooth enamel. The peaks showing the strongest signals in the mass spectra are selected and scaled by the total intensity of the selected peaks to compensate for a possible shift in the total secondary ion yield. Using their masses and isotope patterns, the peaks are assigned to the ions that caused them.  

\begin{figure}
	\centering
		\includegraphics{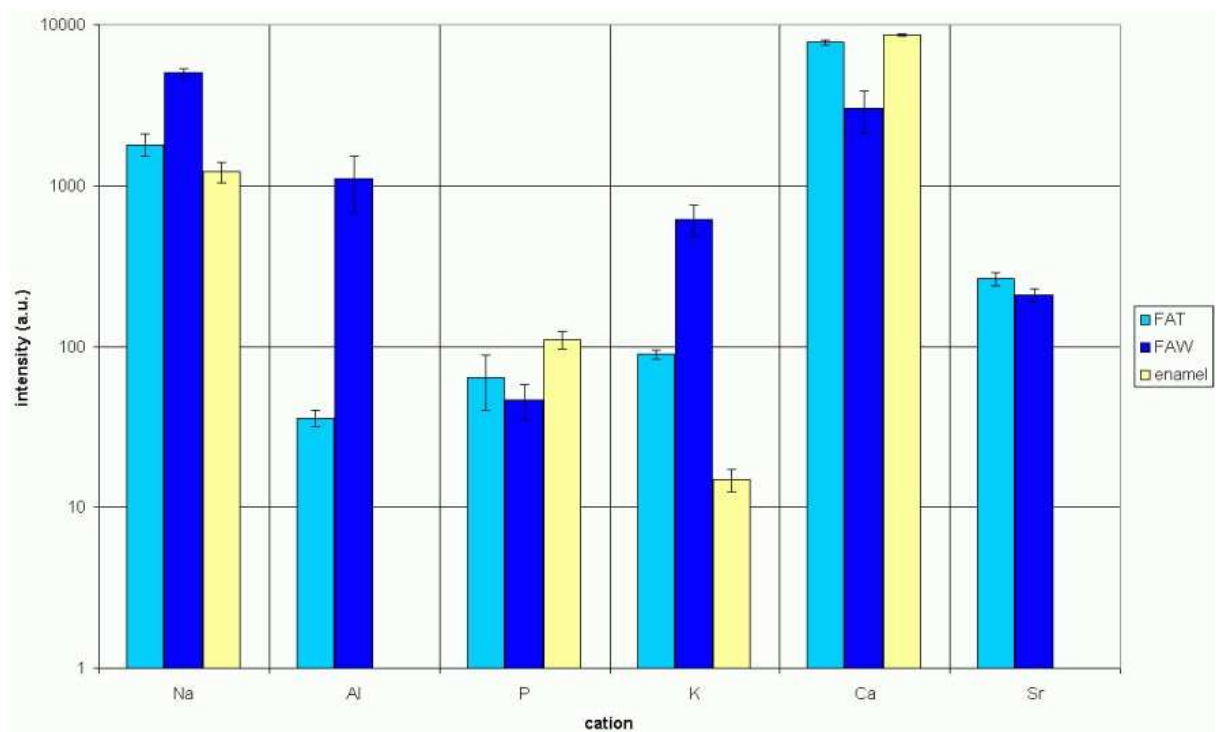}
	\caption{Intensities of cations in mass spectra acquired on dental materials}
	\label{fig:FatFawCations}
\end{figure}

In the mass spectra of positively charged ions, six peaks show a noteworthy intensity. These are shown in figure \ref{fig:FatFawCations}.
Comparing the two implant materials, FAW shows much higher intensities for the aluminium and potassium signals. The other four cations show comparable intensities for the two materials with a slightly stronger sodium signal and a little weaker phosphorus, calcium and strontium signals for FAW than for FAT. In contrast to the implant materials the enamel sample does not show any detectable signal for aluminium or strontium and a much weaker one for potassium. The signals of calcium and phosphorus are a little stronger and the one of sodium is slightly weaker than for the implant materials. 

\begin{figure}
	\centering
		\includegraphics{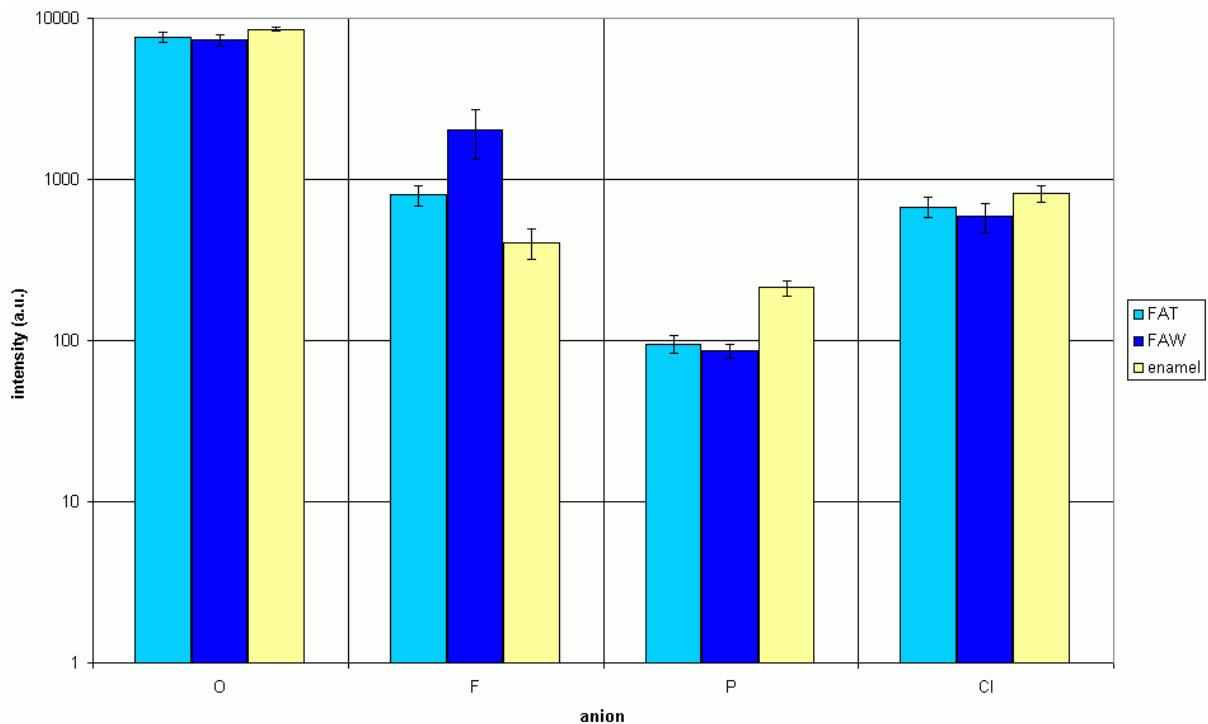}
	\caption{Intensities of anions in mass spectra acquired on dental materials}
	\label{fig:FatFawAnions}
\end{figure}

In the mass spectra of negatively charged ions, the peaks of four components of the samples can be identified. Their intensities are shown in figure \ref{fig:FatFawAnions}.
Here the two implant materials show quite similar intensities. The fluorine signal is a little stronger and the phosphorus (as in the spectra of cations) and chlorine signals are a little weaker for FAW than for FAT. The bovine enamel sample does not show any strongly differing behaviour either. Its fluorine signal is slightly weaker while the phosphorus and chlorine signals are stronger than for the implant materials. All the samples show similar intensities for the oxygen signal.  

Due to the different matrix effects and ionisation probabilities, no exact conclusions to the surface composition of the samples can be drawn from these data. Anyway it can be stated that all three materials resemble in their main components. They contain principally sodium, phosphorus, potassium, calcium, oxygen, fluorine and chlorine. In the implant materials aluminium and strontium are also abundantly present. Strontium is incorporated in the implant materials to make them visible in X-ray imaging. The aluminium originates probably from the polishing of the samples.  

\subsection{ToF-SIMS of protein films on silanised substrates}
\begin{figure}
	\centering
		\includegraphics{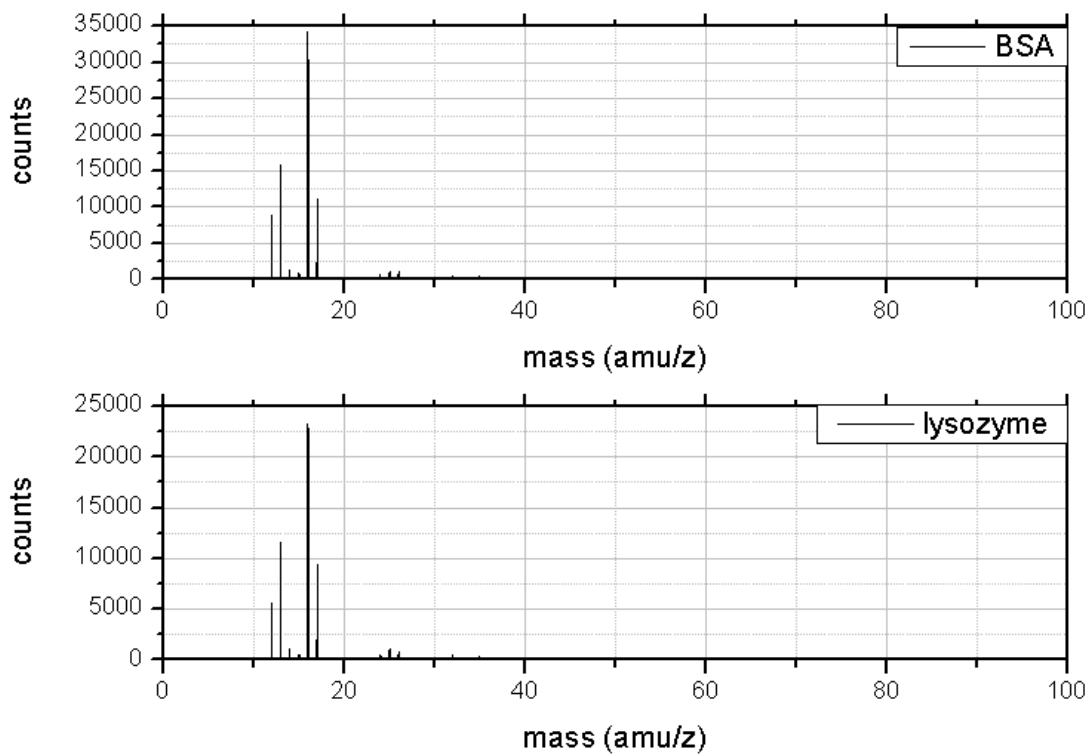}
	\caption{Exemplary mass spectra of anions of protein films adsorbed to silane}
	\label{fig:spectralysbsa5n}
\end{figure}

Regarding the mass spectra of different protein films, there are no clearly visible differences between the different films. In all the spectra of negative ions the oxygen (16 amu/z), hydroxide (17 amu/z), carbon (12 amu/z) and  hydrocarbon anions (13 amu/z) as well as the cyanide anion ($\mathsf{CN^-}$ : 26 amu/z) and $ \mathsf{C_2H^-} $ (25 amu/z) with much weaker intensities are the only clearly visible peaks (see figure \ref{fig:spectralysbsa5n}).

\begin{figure}
	\centering
		\includegraphics{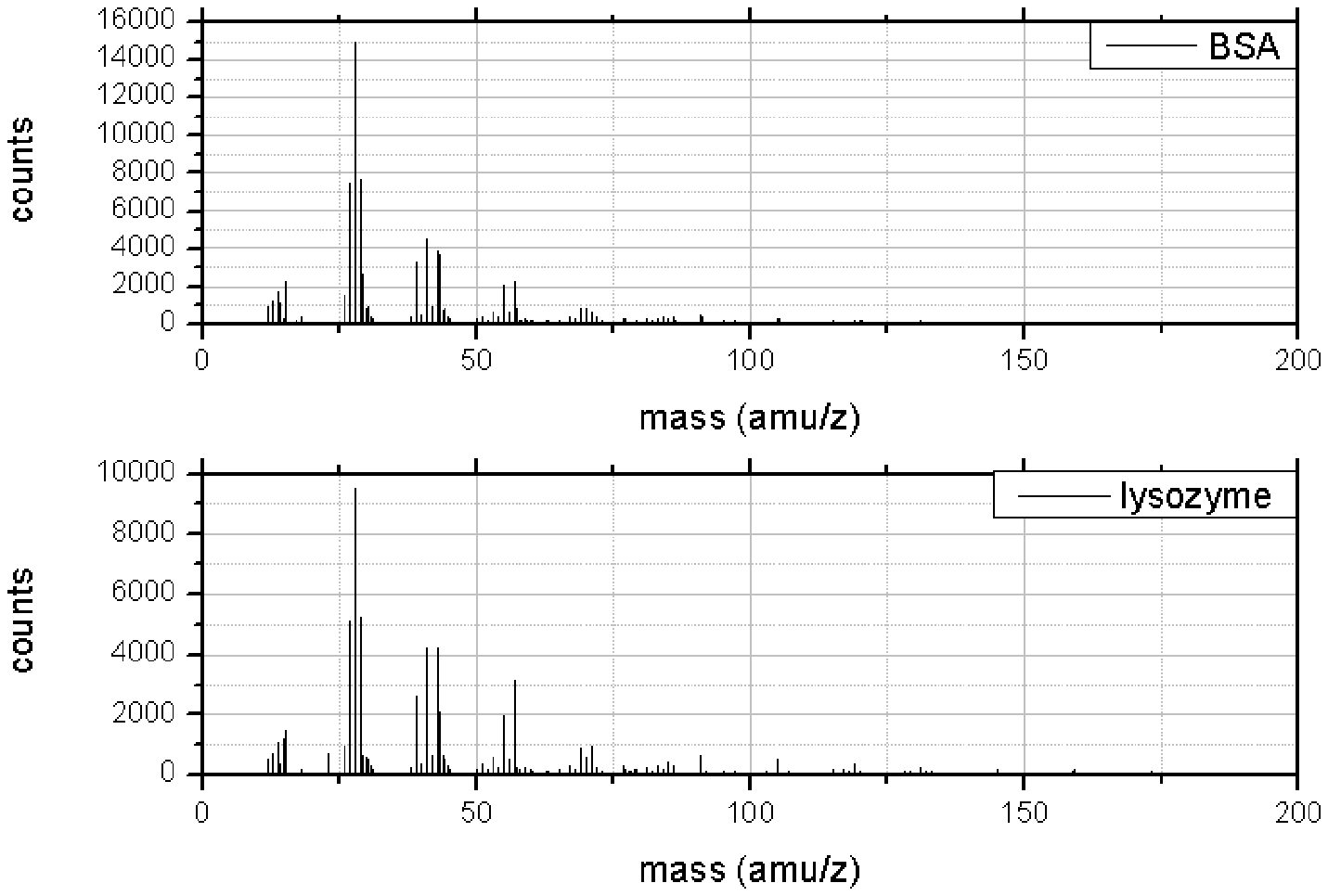}
	\caption{Exemplary mass spectra of cations of protein films adsorbed to silane}
	\label{fig:spectralysbsa5p}
\end{figure}

\begin{figure}
	\centering
		\includegraphics{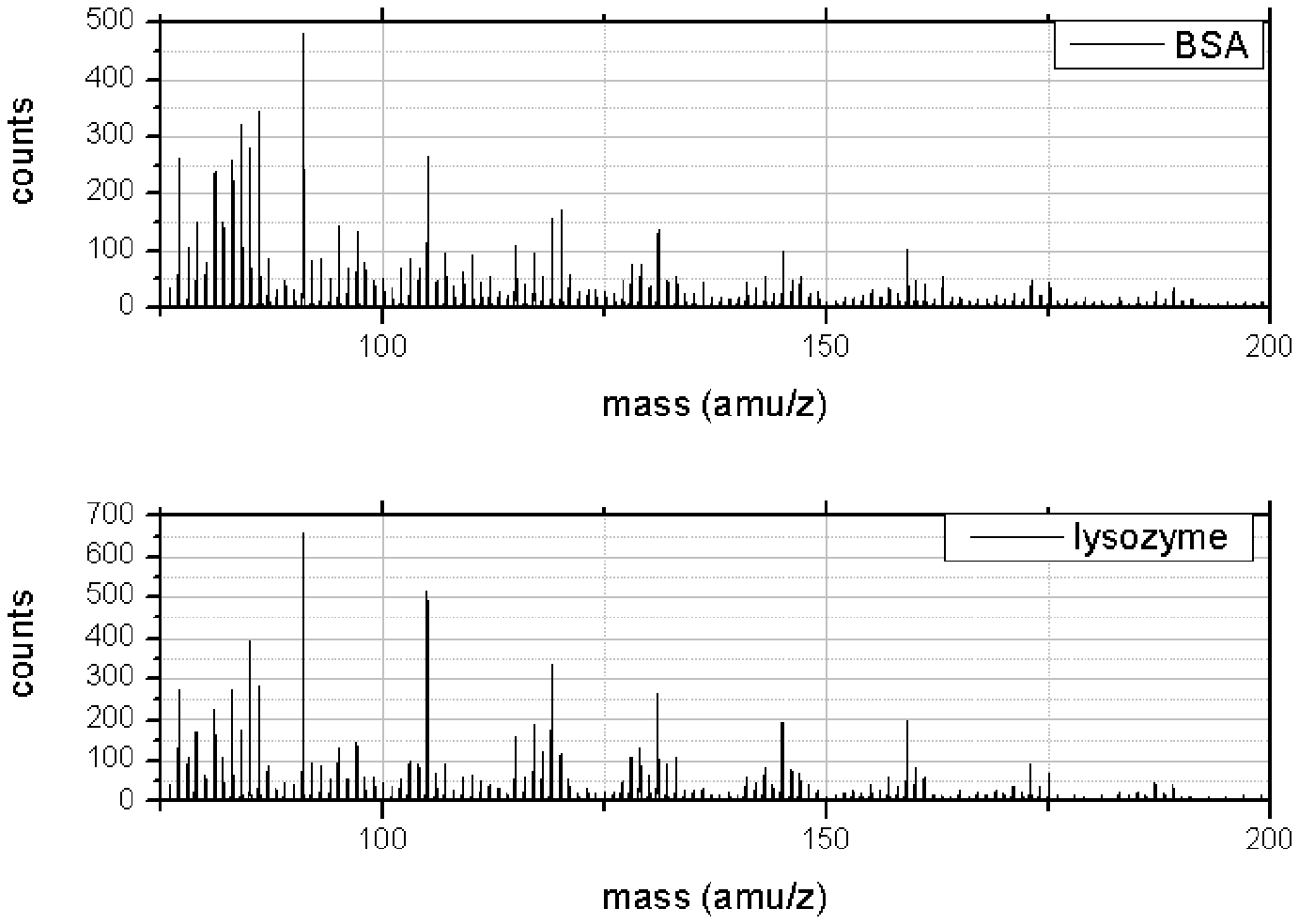}
	\caption{Exemplary mass spectra of cations of protein films adsorbed to silane}
	\label{fig:spectralysbsa5highp}
\end{figure}

In the spectra of positive ions the peak of the silicon cation (28 amu/z) is the most intense one. This is probably due to polymerised silane that was not removed from the surface by the cleaning step in ethanol. In the mass range from 27 amu/z to 70 amu/z nearly every mass shows a peak of noteworthy intensity. Furthermore up to 200 amu/z the density of peaks stays large (see figures \ref{fig:spectralysbsa5p} and \ref{fig:spectralysbsa5highp}). 

Some sample sites exhibit a strong peak at 23 amu/z due to sodium from the buffer solution. At these sites no spectra for analysis are acquired because according to Wagner et al. \cite{wagner:2001} sodium causes a strong matrix effect that might distort the measurements.

For principal component analysis only the mass range between 1 amu/z and 200 amu/z is used. The spectra are scaled by their total intensity to compensate for changes in the total number of ions detected per spectrum.

\begin{figure}
\centering
\includegraphics{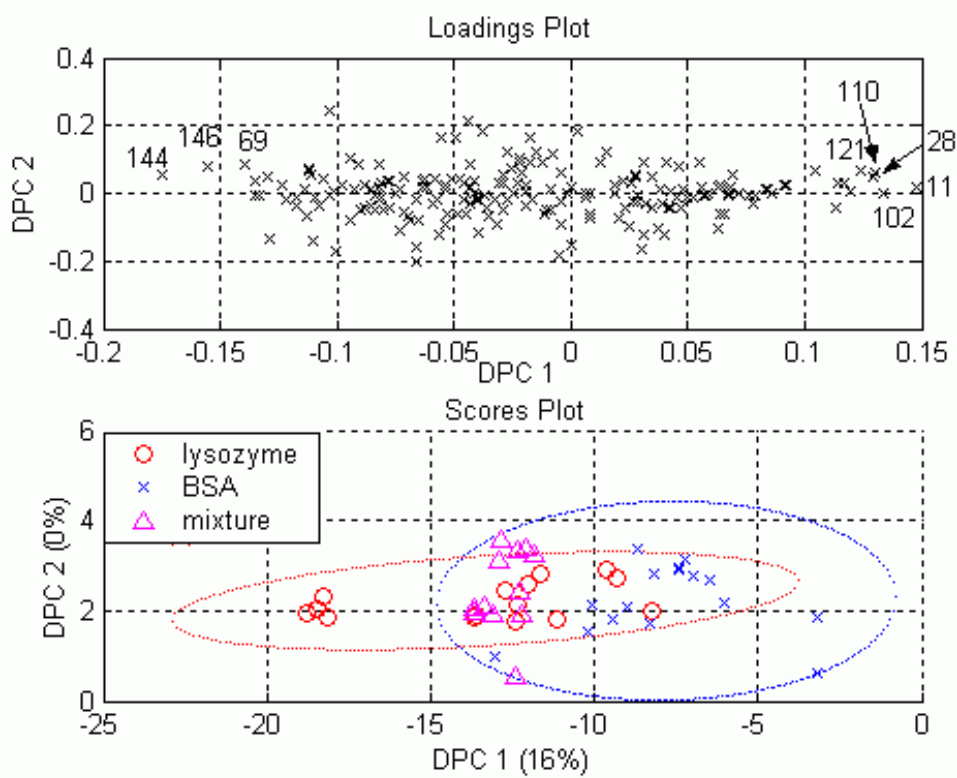}
\caption{DPCA results of two proteins adsorbed to silane from buffer solution}
\label{fig:LysBSA345}
\end{figure}

By doing DPCA on the spectra of cations a discrimination of samples coated with lysozyme or BSA by their scores is hardly possible. Figure \ref{fig:LysBSA345} shows the loadings and scores plots of three samples each coated with lysozyme or BSA on a silane and glutardialdehyde layer. On each sample mass spectra of cations were acquired at four or six sites. In addition to these data, spectra of cations acquired on three samples treated with a 1:1 (weight:weight) mixture solution  of the two proteins are projected into the scores plot. Around the data points of lysozyme and BSA spectra, $95\%$-probability ellipses are drawn. As the ellipses show a large overlap, these results cannot be used to discriminate between the two proteins. In the loadings plot the masses showing strongest influence on the first discriminant principal component are labeled with their mass to charge ratio in atomic mass units per charge number (amu/z). One of these is 28 amu/z, the mass to charge ratio of the silicon cation. To exclude the influence of silicon from polymerised silane or from the substrate on the analysis, a mean mass spectrum of cations acquired on a sample coated only with silane is subtracted from the protein mass spectra.

\begin{figure}
\centering
\includegraphics{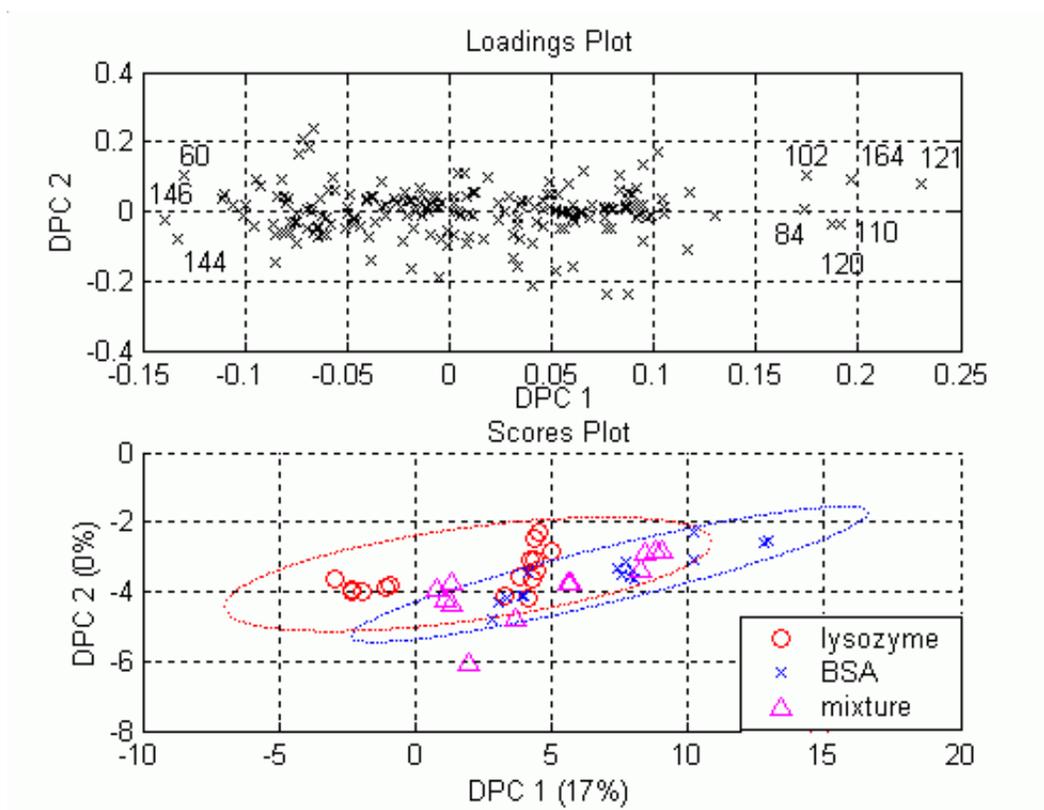}
\caption[Results of DPCA after subtraction of a silane spectrum]{Results of DPCA after subtraction of a silane spectrum of two proteins adsorbed to silane from buffer solution}
\label{fig:LysBSA345o}
\end{figure}

The new results after subtraction of the silane mass spectrum are shown in figure \ref{fig:LysBSA345o}. There still exists a large overlap of the probability ellipses.

To improve the results, in the next step only masses corresponding to intense peaks of amino acid mass spectra are taken into account. The selection was taken from the work of Lhoest et al. \cite{lhoest:2001} and is listed in table \ref{tab:PeakSelection}. At the masses 30 amu/z and 45 amu/z there are also peaks visible in the mass spectra caused by the substrate, namely $\mathsf{^{30}Si^+}$ and $ \mathsf{^{28}SiOH^+} $ that overlap with the peaks of amino acid fragments. Hence these masses are not used for analysis of protein films on silicon substrates. 
\begin{table}
	\centering
		\begin{tabular}{|c|c|c|} \hline
		mass in amu/z & fragment & corresponding amino acid \\ \hline
		$30^*$ & $ \mathsf{CH_4N} $ & glycine \\
		43  & $ \mathsf{CH_3N_2} $ & arginine \\
		$44^\dagger$ & $ \mathsf{C_2H_6N} $ & alanine \\
		$45^*$ & $ \mathsf{CHS} $ & cysteine \\
		60 & $ \mathsf{C_2H_6N0} $ & serine \\
		61 & $ \mathsf{C_2H_5S} $ & methionine \\
		68 & $ \mathsf{C_4H_6N} $ & proline \\
		$69^\dagger$ & $ \mathsf{C_4H_5O} $ & threonine \\
		70 & $ \mathsf{C_3H_4NO} $ & asparagine \\
		   & $ \mathsf{C_4H_8N} $ & proline \\
		71 & $ \mathsf{C_3H_3O_2} $ & serine \\
		72 & $ \mathsf{C_4H_{10}N} $ & valine \\
		73 & $ \mathsf{C_2H_7N_3} $ & arginine \\
		74 & $ \mathsf{C_3H_8NO} $ & threonine \\
		81 & $ \mathsf{C_4H_5N_2} $ & histidine \\
		82 & $ \mathsf{C_4H_6N_2} $ & histidine\\
		83 & $ \mathsf{C_5H_7O} $ & valine \\
		84 & $ \mathsf{C_4H_6NO} $ & glutamine, glutaminic acid\\
		   & $ \mathsf{C_5H_{10}N} $ & lysine \\
		$86^\dagger$ & $ \mathsf{C_5H_12N} $ & leucine, isoleucine \\
		$87^\dagger$ & $ \mathsf{C_3H_7N_2O} $ & asparagine \\
		$88^\dagger$ & $ \mathsf{C_3H_6NO_2} $ & asparagine,aspartic acid\\
		98 & $ \mathsf{C_4H_4NO_2} $ & asparagine \\
		100 & $ \mathsf{C_4H_{10}N_3} $ & arginine \\
		101 & $ \mathsf{C_4H_{11}N_3} $ & arginine \\
		102 & $ \mathsf{C_4H_8NO_2} $ & glutaminic acid \\
		107 & $ \mathsf{C_7H_7O} $ & tyrosine \\
		110 & $ \mathsf{C_5H_8N_3} $ & histidine \\
		112 & $ \mathsf{C_5H_8N_3} $ & arginine \\
		120 & $ \mathsf{C_8H_{10}N} $ & phenylalanine \\
		127 & $ \mathsf{C_5H_11N_4} $ & arginine \\
		130 & $ \mathsf{C_9H_8N} $ & tryptophane \\
		131 & $ \mathsf{C_9H_7O} $ & phenylalanine \\
		136 & $ \mathsf{C_8H_{10}NO} $ & tyrosine \\
		159 & $ \mathsf{C_{10}H_{11}N} $ & tryptophane \\
		170 & $ \mathsf{C_{11}H_8NO} $ & tryptophane \\	\hline
		\end{tabular}
	\caption{Selected amino acid fragment peaks. The peaks marked with $^*$ are omitted for samples prepared on silicon and the ones marked with $^\dagger$ for samples prepared on FAT or FAW because of their overlap with substrate caused peaks.}
	\label{tab:PeakSelection}
\end{table}  

\begin{figure}
\centering
\includegraphics{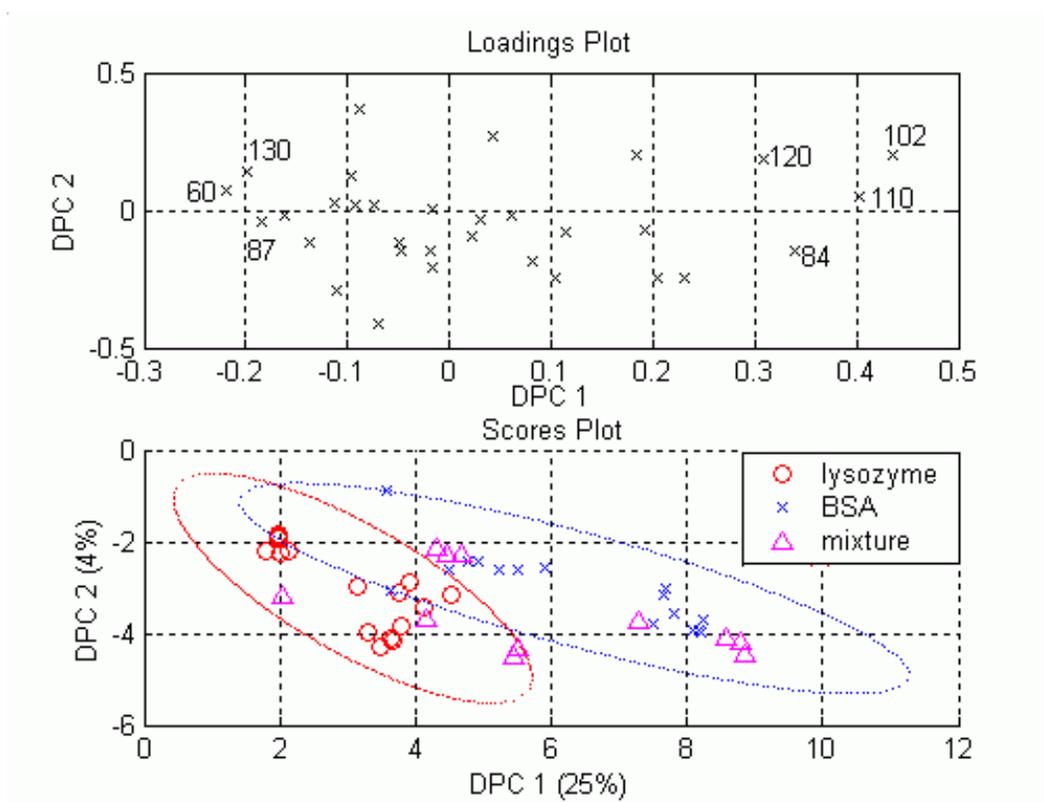}
\caption[Results of DPCA after peak selection]{Results of DPCA after peak selection of two proteins adsorbed to silane from buffer solution}
\label{fig:LysBSA345a}
\end{figure}

In the new scores plot the overlap of the probability ellipses is a lot smaller than before (see figure \ref{fig:LysBSA345a}) but there still is a large spread within the groups. Furthermore, the projections of the mass spectra of samples treated with a mixture of both proteins are spread all over the probability ellipses of the two proteins. This suggests that the composition of these samples is very variable.
In the loadings plot the masses 102 amu/z and 110 amu/z show the strongest positive influence on the first discriminant principal component, while the masses 60 amu/z, 130 amu/z and 87 amu/z show the strongest negative influence. The former are assigned to the amino acids glutamic acid and histidine and the latter are assigned to serine, tryptophane and asparagine (see table \ref{tab:PeakSelection}). 

\begin{figure}
	\centering
		\includegraphics{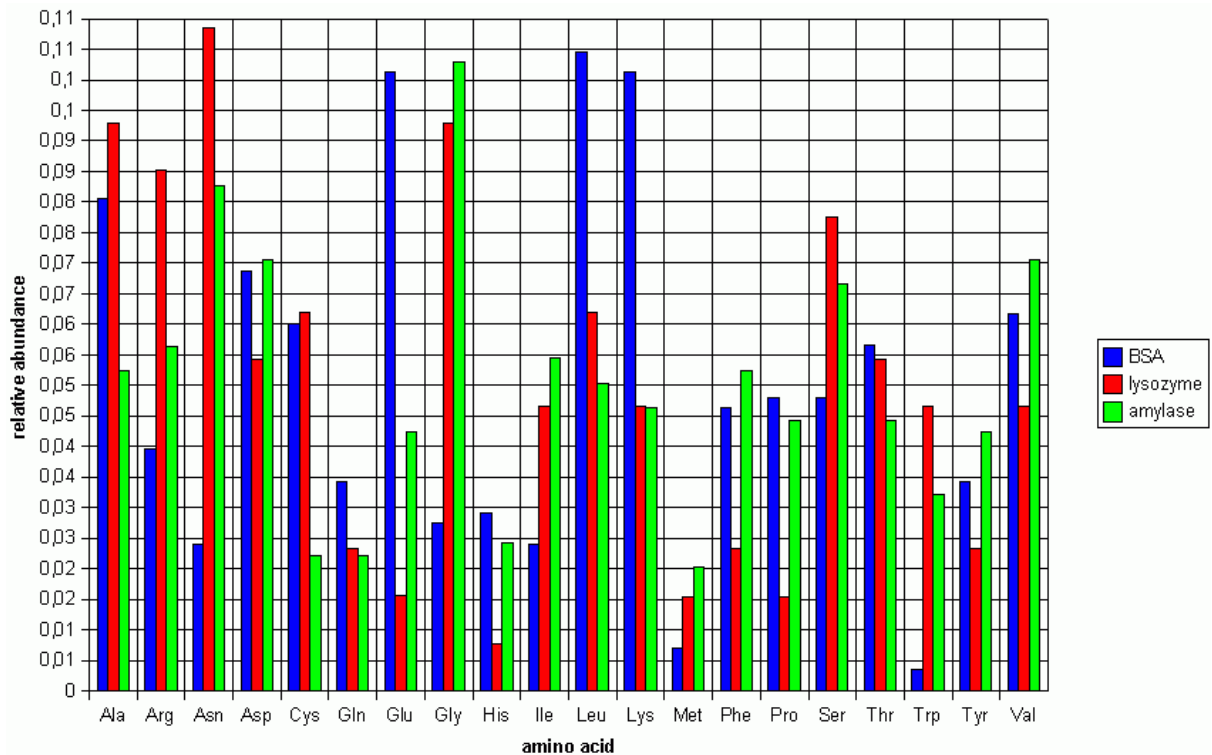}
	\caption{Relative abundances of amino acids in lysozyme, BSA and amylase}
	\label{fig:LysBSADiagram}
\end{figure}

The relative abundances of the different amino acids in lysozyme, BSA and amylase are plotted in figure \ref{fig:LysBSADiagram}. The values were taken from the work of Lhoest et al. \cite{lhoest:2001} for lysozyme and BSA and from the Protein Data Bank \cite{proteindatabank} for amylase.

It can be seen that glutamic acid and histidine are more abundant in BSA than in lysozyme while the opposite is true for serine, tryptophane and asparagine. The amino acids associated to BSA show the strongest positive loadings on the first DPC, whereas the amino acids highly abundant in lysozyme show the strongest negative loadings on the first DPC. Therefore the loadings are consistent with the fact that BSA can be found on the right side of the scores plot and lysozyme on the left side. 

To quantify the quality of the DPCA model of the spectral data, leave-one-out-tests are performed. When the samples are assigned to one of the two proteins by their euclidean distance on the first DPC to the centres of gravity of the two groups, 26 out of 32 spectra ($81 \%$) are correctly assigned. Using the probability ellipses, only 18 spectra ($56 \%$)
are correctly assigned because many spectra are projected into the overlapping region of the two ellipses and thus cannot be assigned to one group. Hence a way for better discrimination of the sample types by reducing the spread between the mass spectra of one protein, has to be found. 

Probably the spread within the sample types is due to different amounts of polymerised silane on the sample surfaces. To check this, protein films shall be adsorbed to unsilanised silicon wafers.

\subsection{ToF-SIMS of protein films on silicon substrates}
\begin{figure}
	\centering
		\includegraphics{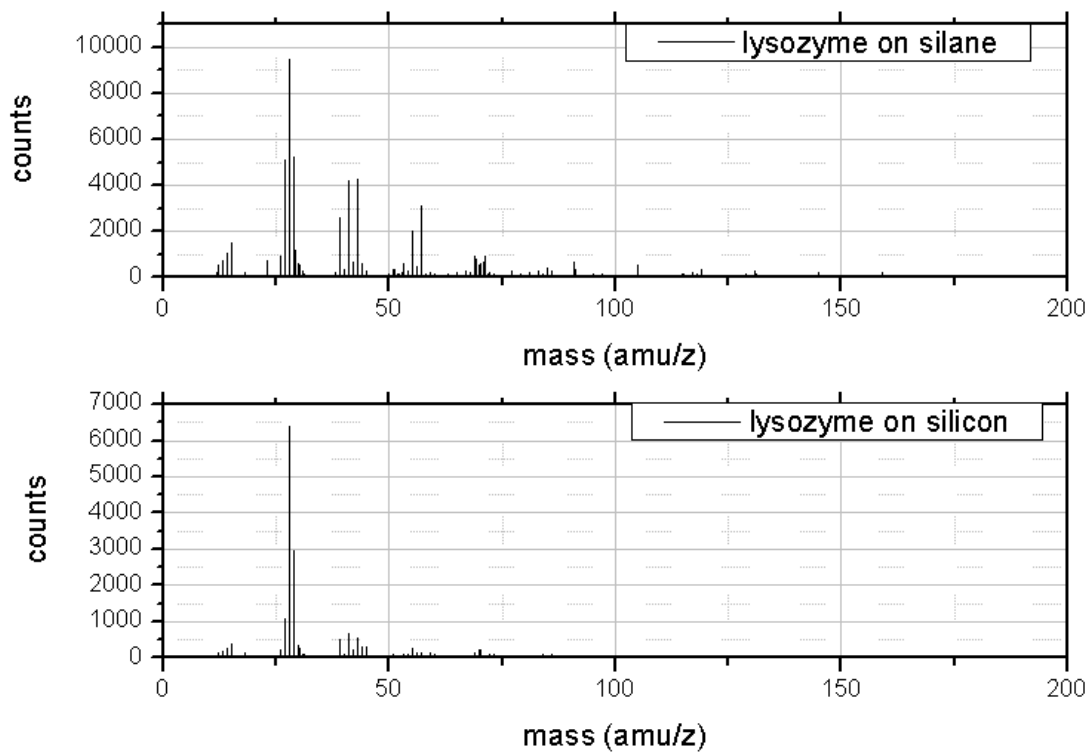}
	\caption{Comparison of mass spectra of cations from lysozyme films on silane or silicon}
	\label{fig:spectralys521p}
\end{figure}

\begin{figure}
	\centering
		\includegraphics{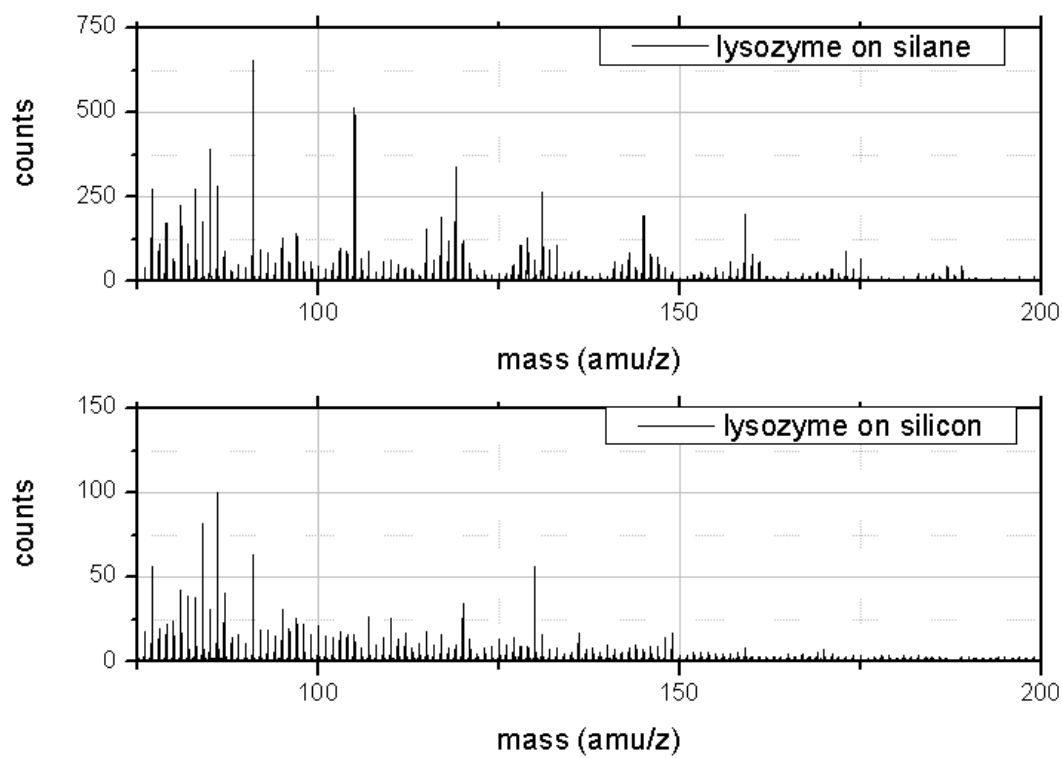}
	\caption{Comparison of mass spectra of cations from lysozyme films on silane or silicon}
	\label{fig:spectralys521high}
\end{figure}

To the naked eye the mass spectra acquired from protein films on silicon do not significantly differ from those acquired from protein films on silane. The intensity of the silicon peak at 28 amu/z is even stronger than in the spectra of protein films on silane (see figure \ref{fig:spectralys521p}). Now it can only be caused by the silicon from the substrate. Since the sampling depth of ToF-SIMS is only a few \AA ngstr\"oms, no silicon should be detected through a compact protein layer of several nanometres thickness. Probably the protein film is porous due to removal of part of the protein in the rinsing step. 
In the higher mass region different peaks can be found in the spectra obtained from protein adsorbed to silicon, but still there are no characteristic peaks for the different proteins (see figure \ref{fig:spectralys521high}). 

\subsubsection{Proteins solved in water}
First the spectra of proteins solved in water are treated. On four samples coated with lysozyme, BSA or amylase, mass spectra of cations are acquired at four sites. The spectra are unit-mass binned and amino acid related peaks (see table \ref{tab:PeakSelection}) are selected before doing discriminant principal component analysis. Additional silicon substrates are treated with solutions of 1:1 (weight:weight) mixtures of two of the three proteins or with a solution of a 1:1:1 (weight:weight:weight) mixture of all three proteins solved in water. On two samples of each mixture type, cation mass spectra are acquired at four sites. These undergo the same pretreatment as the spectra of pure protein films.

\begin{figure}
\centering
\includegraphics{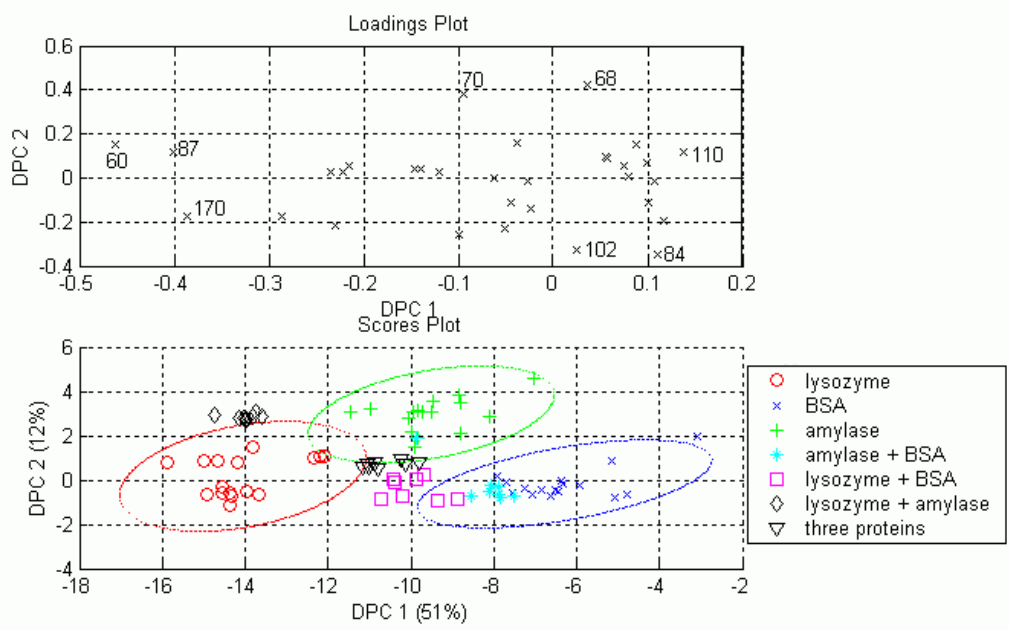}
\caption{Results of DPCA on spectra of three proteins adsorbed to silicon from water}
\label{fig:LysBSAAmy611}
\end{figure}
The results of DPCA are shown in figure \ref{fig:LysBSAAmy611}. In the scores plot there is only little overlap of the $95 \%$ probability ellipses of lysozyme and amylase, while BSA is completely separated from the two other sample types. The mass spectra acquired on mixture treated samples are projected into the scores plot. The position of the ternary mixture and the mixture of lysozyme and BSA suggest that all proteins have adsorbed in these cases because the projections are found between the regions of the pure protein spectra. On the other hand the projections of the mixture of lysozyme and amylase are very close to the lysozyme region and the projections of the mixture of amylase and BSA lie (with exception of one spectrum) completely in the probability ellipse of BSA. It can thus be presumed that in these two cases there is a preferential adsorption of lysozyme and BSA compared to amylase.

In the loadings plot the masses 60 amu/z, 87 amu/z and 170 amu/z show the strongest negative values on the first DPC. These masses are assigned to the amino acids serine, asparagine and tryptophane (see table \ref{tab:PeakSelection}). All of them are most abundant in lysozyme (see figure \ref{fig:LysBSADiagram}) causing lysozyme spectra to be projected to the left side of the scores plot. The mass 110 amu/z (histidine) has the highest positive loading on the first DPC. Since histidine is most abundant in BSA, this protein is projected to the right side of the scores plot. The positive side of the second discriminant principal component is dominated by proline (masses 68 amu/z and 70 amu/z) which is present in BSA and amylase at nearly equal proportions. Since the masses with strongest negative influence on the second DPC 84 amu/z (glutamine, glutaminic acid, lysine) and 102 amu/z (glutaminic acid) belong to amino acids most abundant in BSA, this protein is found in the lower part of the scores plot and amylase is found in the upper part.

Leave-one-out tests of the DPCA model show the following results: If spectra are assigned by their euclidean distances on the first two discriminant principal components, 47 out of 48 spectra ($98\%$) are associated with the correct group. Using the probability ellipses for assignment, still 43 spectra ($90\%$) are correctly recognised. This shows that the data are well represented by the DPCA model.

\subsubsection{Proteins solved in 100 millimolar buffer solution}
When the proteins are solved in 100 millimolar buffer solution, many sample sites exhibit a strong signal at 23 amu/z caused by sodium adsorbed from the buffer solution. Because of the matrix effect caused by sodium, spectra for analysis are acquired only at sites with small sodium concentrations but these are difficult to find. Four samples each are coated with lysozyme or BSA and four spectra of cations are acquired on each sample. The spectra are unit-mass binned and amino acid related masses are selected before doing DPCA.

\begin{figure}
	\centering
		\includegraphics{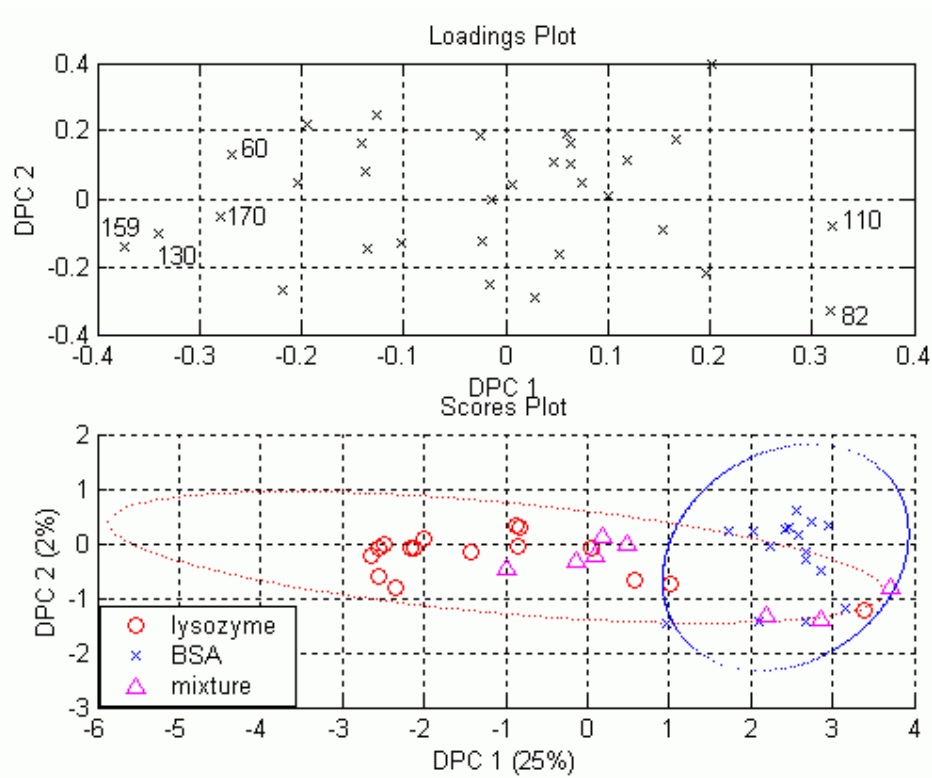}
	\caption{Results of DPCA of two proteins adsorbed on silicon from 100 mM buffer solution}
	\label{fig:lysbsa1215}
\end{figure}

In the scores plot (see figure \ref{fig:lysbsa1215}) the two sample types are separated on the first discriminant principal component but the $95 \%$-probability ellipses of BSA and lysozyme coated samples overlap to a large extent. The negative side of the first principal component is most strongly influenced by the masses 159 amu/z, 130 amu/z, 170 amu/z and 60 amu/z. These are linked to fragments of the amino acids tryptophane and serine (see table \ref{tab:PeakSelection}), which are more abundant in lysozyme than in BSA (see figure \ref{fig:LysBSADiagram}). On the other hand the masses 110 amu/z and 82 amu/z show the strongest positive loading. They are linked to histidine which is more abundant in BSA than in lysozyme. This explains why lysozyme can be found mostly in the negative range and BSA in the positive range of the first DPC in the scores plot.

In addition to the pure protein films, mass spectra acquired on two samples treated with a solution of lysozyme and BSA of equal mass concentrations in 100 millimolar buffer solution are projected into the scores plot. They are spread widely within the probability ellipses of lysozyme and BSA.

The leave-one-out-test shows that in this case only 19 out of 32 spectra ($59 \%$) are assigned to the right protein using the probability ellipses. Due to the large within group spread, many of the spectra are projected into the overlap of the two ellipses preventing an unambiguous assignment. Using the euclidean distance on the first DPC, 28 spectra ($88 \%$) are correctly assigned. So the data are not very well represented by the two-dimensional DPCA model. Probably the high spread within the groups is caused by differing amounts of buffer salts remaining on the sample surfaces.

\subsubsection{Proteins solved in 10 millimolar buffer solution}
\begin{figure}
\centering
\includegraphics{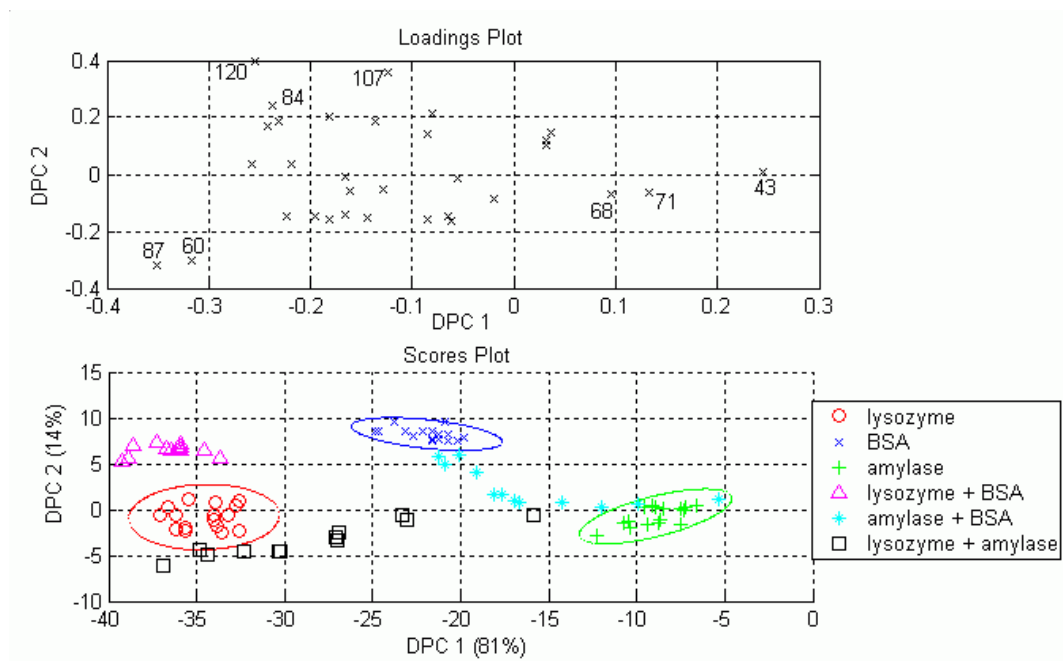}
\caption{Results of DPCA of three proteins adsorbed on silicon from 10 mM buffer solution}
\label{fig:LysBSAAmy2023}
\end{figure}

In the following experiments instead of a 100 millimolar buffer solution of sodium di\-(hydrogen)\-phosphate ($\mathsf{NaH_2PO_4}$) and disodium hydrogenphosphate ($\mathsf{Na_2HPO_4}$), a ten millimolar buffer solution with pH 7.4 is used. Onto twelve silicon substrates lysozyme, amylase or BSA are adsorbed from this ten millimolar buffer solution. Protein concentrations are about $10^{-4}$ mols per litre. On each sample mass spectra of positively charged ions are acquired at four sites. The spectra are unit-mass binned and peaks resulting from amino acid fragments (see table \ref{tab:PeakSelection}) are selected before doing discriminant principal component analysis. The resulting scores and loadings plots are shown in figure \ref{fig:LysBSAAmy2023}. 
In the scores plot a very good separation of the different sample types is visible. This supports the assumption that in the preceding experiments with proteins adsorbed from buffer solution the mass spectra were strongly influenced by buffer salts. Now due to the smaller buffer concentration, a better removal of the buffer salts is possible. Hence the spread within the groups in the scores plot is largely reduced.
Spectra acquired on nine samples treated with mixture solutions of two of the three proteins are projected into the scores plot. In the protein solutions used in this case equal molarities (approximately $10^{-5}$ mol/l) of two proteins are present. The projections of the mixtures of lysozyme and BSA can be found in a small region closer to the lysozyme probability ellipse than to the one of BSA indicating that more lysozyme than BSA adsorbed from the solution. With exception of one outlier the projections of mixtures of lysozyme and amylase are situated closer to lysozyme than to amylase. Thus it is assumed that mostly lysozyme adsorbed in this case. The projections of the three samples treated with mixtures of amylase and BSA are clearly differing from one another. Depending on the sample, the projections lie either close to the BSA region or close to the amylase region or between them. Hence the concentration ratio of the two proteins at the sample surface differs significantly between the three samples. Thus the co-adsorption process of amylase and BSA seems to be more sensitive to external influences like contaminations or temperature than the co-adsorption processes of the other mixtures.

The masses 87 amu/z and 60 amu/z show the strongest negative loadings on the first and second discriminant principal component. They are associated to the amino acids asparagine and serine (see table \ref{tab:PeakSelection}). These have highest abundance in lysozyme (see figure \ref{fig:LysBSADiagram}) which is thus projected in the lower left part of the scores plot. The positive side of the first DPC is most strongly influenced by the masses 43 amu/z (arginine), 71 amu/z (serine) and 68 amu/z (proline). The former two are most abundant in lysozyme and the latter one in BSA. But all of them are also highly abundant in amylase explaining its position on the right side of the scores plot. The amino acids phenylalanine (120 amu/z), tyrosine (107 amu/z) and glutamine, glutamic acid and lysine (84 amu/z) are all highly abundant in BSA. With their strongly positive loadings on the second DPC they project the BSA spectra in the upper part of the scores plot. 

The leave-one-out test confirms the good fitness of the DPCA model to represent the data. Using the euclidean distance on the first two DPC for assignment, all 48 spectra are correctly recognised. With the $95 \%$ probability ellipses, 46 spectra ($96 \%$) are assigned to the correct protein.

Similarly prepared samples of the three proteins on silicon substrates are sputtered with the unpulsed primary ion beam. Mass spectra of cations are acquired after 0, 5 ,10, 15 and 20 seconds of sputtering.
Using the primary ion current ($I \approx 10^{-9}$ A), the maximum sputter time ($t=20$ s), the elementary charge ($e=1.6 \cdot 10^{-19}$ C) and the scanning size ($A=1.44 \cdot 10^{-2} \mbox{cm}^2$), the maximum primary ion dose $D$ can be estimated as
\begin{equation}
D=\frac{I t}{e A} \approx 9 \cdot 10^{12} \mbox{cm}^{-2}.
\label{eq:sputterdose}
\end{equation}  

\begin{figure}
	\centering
	\includegraphics{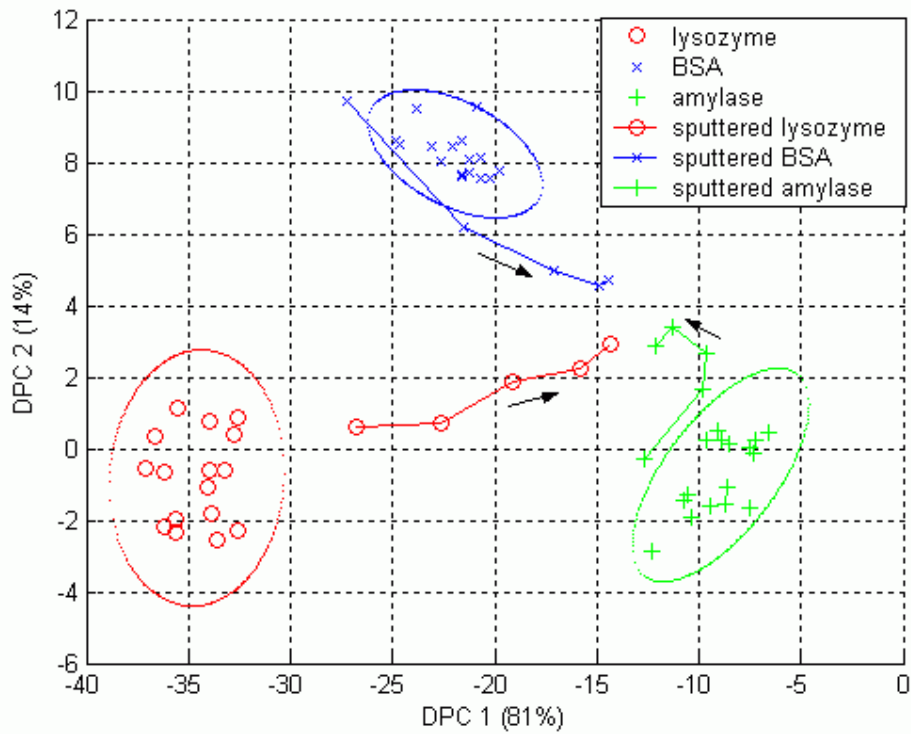}
	\caption[Projections of mass spectra of proteins adsorbed to silicon after sputtering]{Projections of mass spectra of proteins adsorbed to silicon after sputtering into the scores plot. The arrows indicate the direction of increased sputter time.}
	\label{fig:SputterSi1}
\end{figure}

These spectra are unit-mass binned and peak selection is performed before projecting them into the scores plot built with the preceding DPCA model. The projections are shown in figure \ref{fig:SputterSi1}. The arrows indicate the direction of increasing sputter time. The projections of samples of all three proteins quickly approach the same region in the scores plot. Thus even after short sputtering the protein layers are destroyed to a degree that makes it impossible to recognize the protein by DPCA. In the corresponding mass spectra, the intensities of all peaks with exception of the ones caused by the substrate are strongly reduced by sputtering. 

\begin{figure}
\centering
\includegraphics{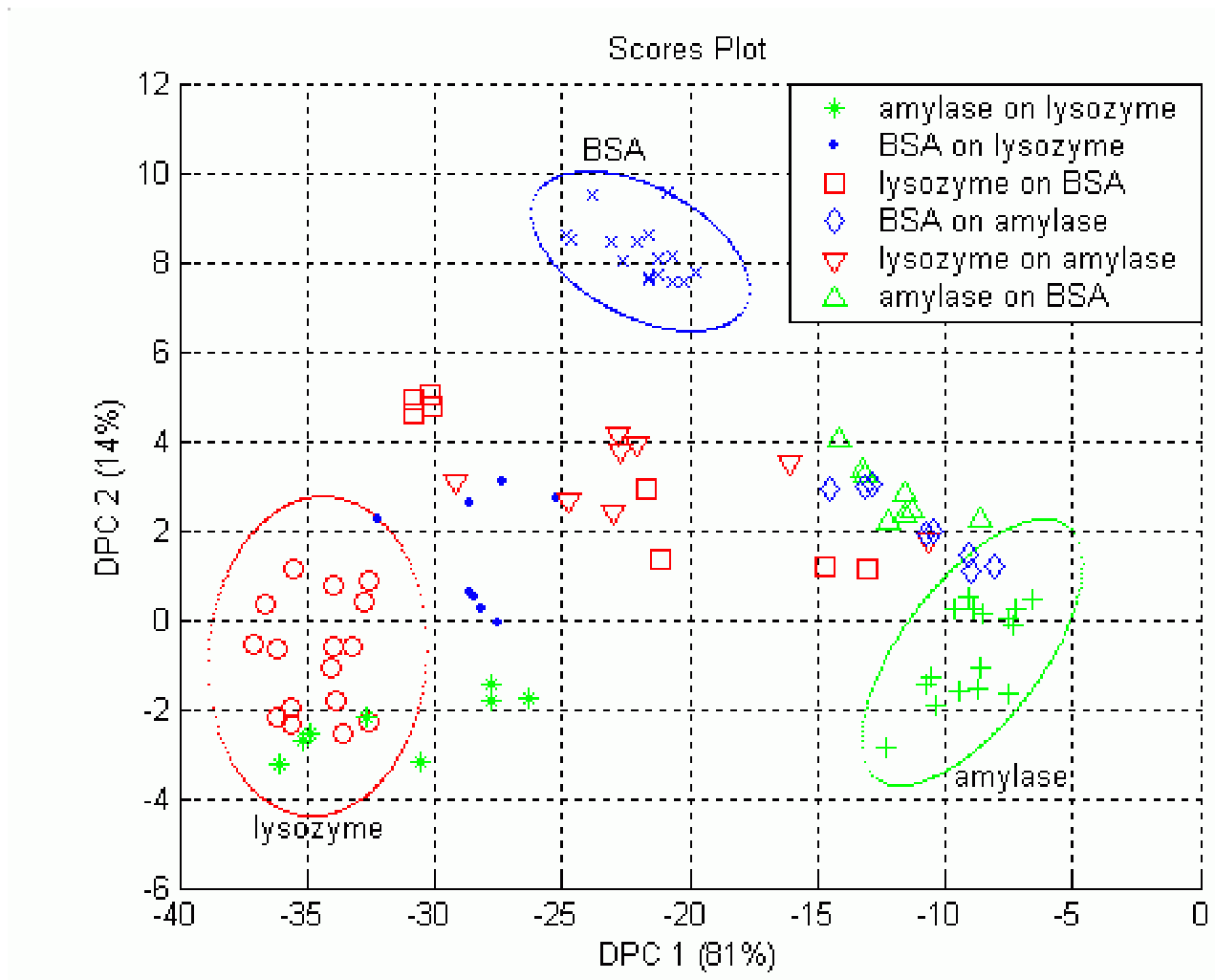}
\caption[Projections of mass spectra of protein double layers adsorbed to silicon]{Projections of mass spectra of protein double layers adsorbed to silicon into the scores plot}
\label{fig:DoppelschichtenSi}
\end{figure}

Next it is tried to prepare double layers of two different proteins on silicon substrates. Therefore, after adsorption of the first protein, the samples are rinsed and given into a solution of the second protein for two hours. Protein molarities are approximately $10^{-4}$ mols per litre buffer solution. As before, a ten millimolar phosphate buffer solution is used. Two samples are prepared for each of the six possible combinations of the three proteins and on each sample mass spectra of cations are acquired at four different sites. The mass spectra are unit-mass binned and amino acid related peaks are selected. Then they are projected into the scores plot of the DPCA model created with the mass spectra of single component protein layers on silicon. The projections and the positions of the single component spectra on the first two discriminant principal components are shown in figure \ref{fig:DoppelschichtenSi}.

The projections can be interpreted as follows:
\begin{description}
\item[Amylase on lysozyme:] These spectra are projected very closely to or into the lysozyme probability ellipse. Hence very little or no amylase adsorbs to the lysozyme layer or it is only loosely bound and removed in the rinsing step.
\item[BSA on lysozyme:] Again the projections are situated close to the lysozyme region suggesting that BSA adsorbs only in little amounts on the lysozyme layer.
\item[Lysozyme on BSA:] These projections show a large spread but they are all at approximately equal distances from the probability ellipses of lysozyme and BSA. Thus the two proteins are detected at equal amounts. This can be caused by mixing of the two proteins in the second adsorption step or by a not tightly packed lysozyme layer. The latter would make the detection of BSA via holes possible. A non homogeneous lysozyme layer can also explain the large spread of the data points.
\item[BSA on amylase:] Since these spectra are projected closely to the amylase region, it is assumed that only little amounts of BSA adsorb onto the amylase film.
\item[Lysozyme on amylase:] Within the projections of these spectra the largest spread is found. Their positions range nearly from the lysozyme probability ellipse to the amylase ellipse. Thus these samples present a very inhomogeneous surface composition and there are large local differences in the amount of lysozyme adsorbed.
\item[Amylase on BSA:] These projections are situated close to the amylase region. Hence amylase adsorbs to the BSA layer and forms a relatively compact layer masking BSA from detection by ToF-SIMS.    
\end{description}

The results for the adsorption of amylase on lysozyme are confirmed by amylase activity measurements. The group of Dr. Christian Hannig, Universit\"atsklinikum Freiburg, determined the immobilized activity of amylase adsorbed directly to silicon as 0.028 units per square centimetre. The activity measured for amylase adsorbed to lysozyme coated silicon is only ten percent as high (0.0028 U/cm$^2$). This is compared to the amount of amylase estimated by the DPCA projections of the mass spectra of samples treated first with lysozyme and then with amylase solution. Figure \ref{fig:DoppelschichtenSi} shows that lysozyme and amylase coated samples are separated by the first discriminant principal component. Thus the amylase fraction of a sample's surface composition $f_{amy}$ is estimated by the distance of its projection on the first axis to the centres of the amylase ($d_{amy}$) and lysozyme ($d_{lys}$) probability ellipses:
\begin{equation}
f_{amy}=\frac{d_{lys}}{d_{lys}+d_{amy}}.
\end{equation}
Hence it is assumed that a mass spectrum projected into the centre of the lysozyme ellipse ($d_{lys}=0$) does not contain any amylase while a spectrum projected into the centre of the amylase ellipse ($d_{amy}=0$) is caused by a pure amylase film. Calculating the surface fraction of amylase for all eight spectra of amylase adsorbed onto lysozyme a mean value of $f_{amy}=(16 \pm 5) \%$ is obtained. The error is calculated by the standard deviation of the values for different spectra. Bearing in mind the difficulties in determining the activity of immobilized enzymes caused by possible denaturation or obstructed substrate diffusion and the problems of quantification for secondary ion mass spectra due to differing ionization probabilities, the amylase activity fraction of ten percent lies remarkably close to the surface fraction estimated by ToF-SIMS and DPCA. 

Unfortunately it was not possible to measure the activity of surface immobilized lysozyme. Probably it is deactivated by the adsorption or the subsequent drying. To determine the amount of BSA adsorbed to the samples, fluorescence microscopy is used. The obtained results are described in section \ref{sec:FluorescenceResults}.  

The results of samples coated consecutively with two proteins can be correlated with the ones obtained from binary protein mixtures. From a mixture of lysozyme and amylase, little amounts of amylase adsorb to the silicon surface. Equally, only very small quantities of amylase adsorb to a surface coated with lysozyme. Thus in both cases lysozyme hinders the adsorption of amylase. A similar but weaker effect can be observed for BSA and lysozyme. More lysozyme than BSA is adsorbed from the mixture, and BSA does not adsorb very well on a lysozyme coated surface. Hence lysozyme hinders the adsorption of BSA in both experiments. Due to the high spread in the data of amylase BSA mixtures, these cannot be compared to the data from the double layer experiments.     

\begin{figure}
\centering
\includegraphics{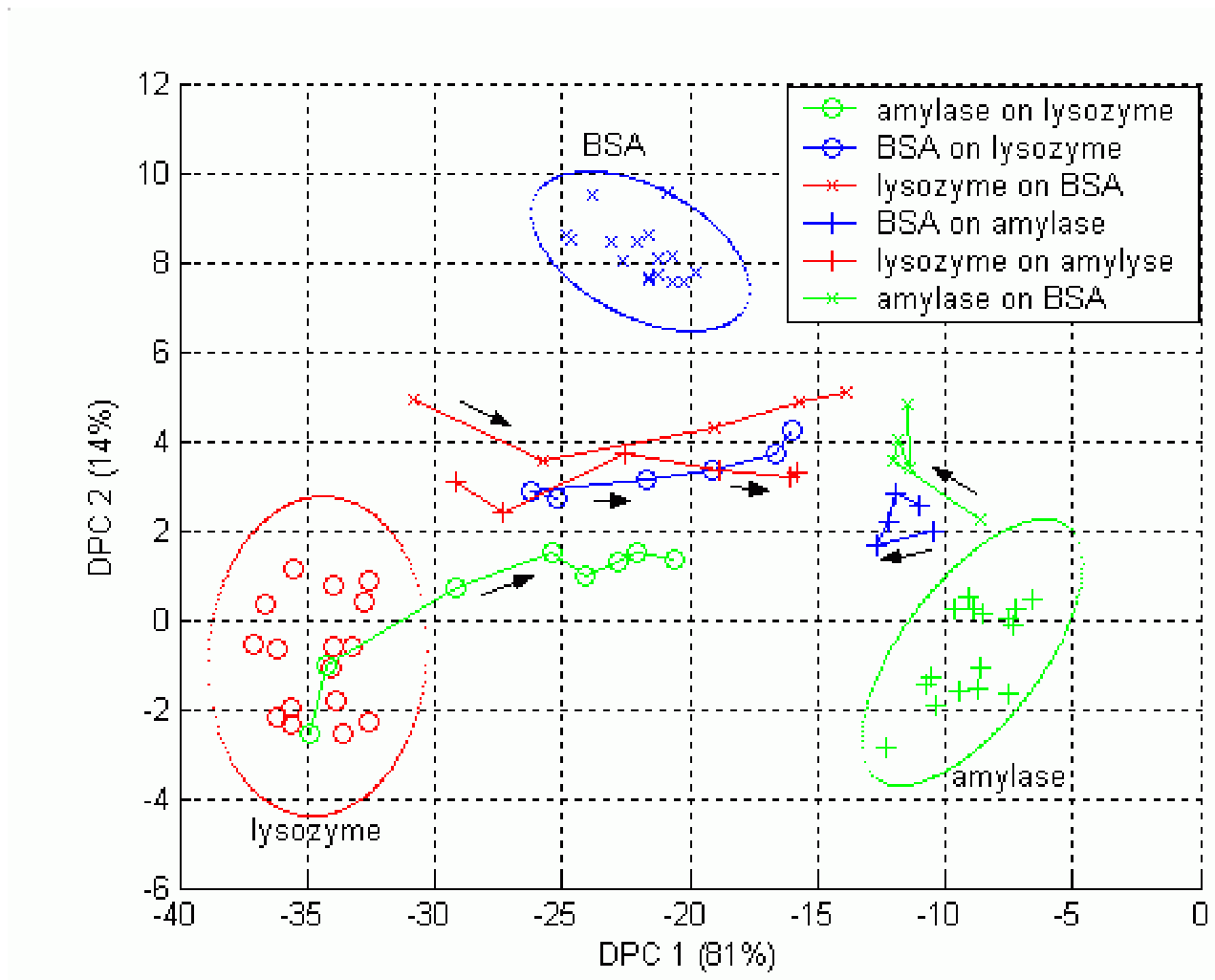}
\caption[Projections of mass spectra of sputtered protein double layers adsorbed to silicon]{Projections of mass spectra of sputtered protein double layers adsorbed to silicon into the scores plot. The arrows indicate the direction of increased sputter time.}
\label{fig:SputterSi2}
\end{figure}

On the samples coated consecutively with two proteins, mass spectra of cations are also acquired after sputter times of zero to 35 seconds. The mass spectra are unit-mass binned and amino acid related peaks are selected. Then they are projected into the scores plot of the DPCA model created with the mass spectra of single component protein layers on silicon (see figure \ref{fig:SputterSi2}). The arrows indicate the direction of increasing sputter time. 

For all samples the projections develop directly to the same region under sputtering. This is the same region, the projections of single protein layers developed to in figure \ref{fig:SputterSi1}. Thus sputtering destroys rapidly the whole protein layer making a discrimination of proteins after sputtering impossible. In particular the assumed double layer structure of the protein films cannot be detected. The experiments were performed twice for each sample type giving essentially the same results. For clarity figure \ref{fig:SputterSi2} shows only the results of one series of experiments. 

\subsection{Fluorescence microscopy of protein films on silicon}
\label{sec:FluorescenceResults}
\begin{figure}
	\centering
		\includegraphics{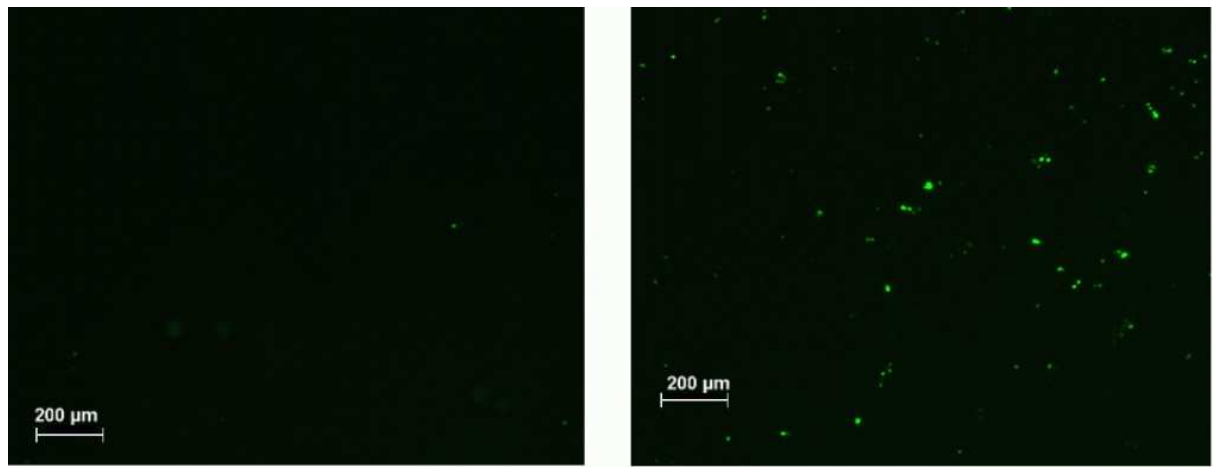}
	\caption[Fluorescence images of BSA fluorescein conjugate and lysozyme adsorbed on silicon]{Fluorescence microscopy images of pure BSA fluorescein conjugate (left) and a mixture of BSA and lysozyme (right) adsorbed on silicon}
	\label{fig:FluoreszenzBsa16Lb15}
\end{figure}
\begin{figure}
	\centering
		\includegraphics{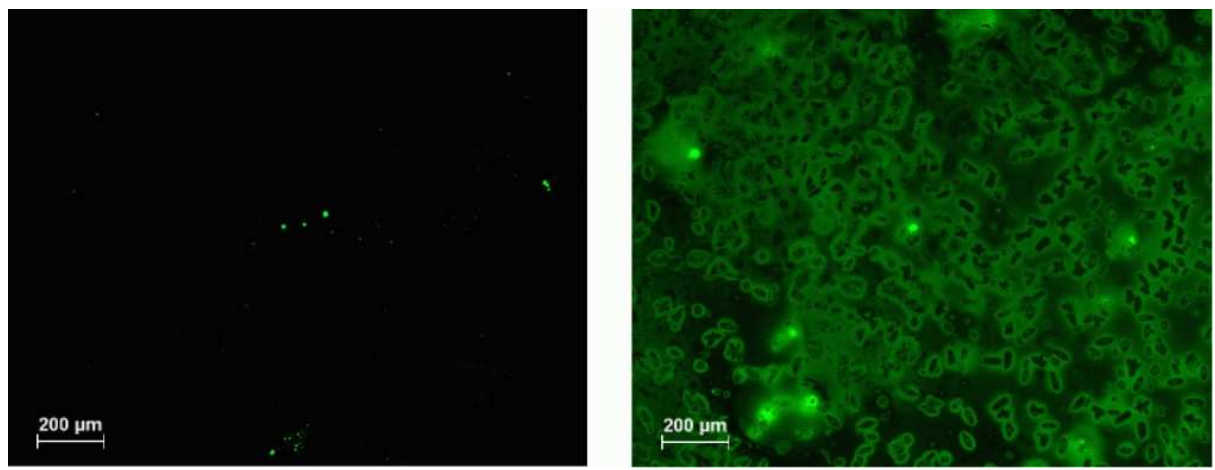}
	\caption[Fluorescence images of BSA fluorescein conjugate adsorbed to silane and to silicon]{Fluorescence microscopy images of BSA fluorescein conjugate adsorbed to silane (left) and to silicon without rinsing of the sample  after adsorption (right)}
	\label{fig:FluoreszenzBsa19Bsa18}
\end{figure}
Figures \ref{fig:FluoreszenzBsa16Lb15} and \ref{fig:FluoreszenzBsa19Bsa18} show typical fluorescence microscopy images obtained from different samples. If the protein is adsorbed to silicon and the samples are rinsed after adsorption only few fluorescence spots are visible (see figure \ref{fig:FluoreszenzBsa16Lb15}). The sample treated with a mixture of BSA fluorescein conjugate and lysozyme shows more fluorescence spots than the sample treated only with BSA fluorescein conjugate. Possibly most of the protein is removed in the rinsing step. This would also explain the strong substrate signal in the mass spectra acquired on similarly prepared samples. The images confirm the theory that BSA adsorbs to the substrate from a protein mixture as well as from a single protein solution. There might be different explanations for the stronger fluorescence shown by the mixture treated samples. It could be caused by possibly non-identical rinsing times removing more or less of the protein layer. The mixture of the two proteins might form a denser layer more difficult to remove by rinsing. Or the presence of lysozyme might enhance the fluorescence yield of fluorescein. 

On the sample where the BSA fluorescein conjugate was adsorbed to silane, the fluorescence is only visible at a few spots on the sample surface, too. As the previous ones this sample was rinsed after protein adsorption. But since the proteins should bind covalently to the glutardialdehyde (see figure \ref{fig:GluBinding}), it should not be possible to remove them by rinsing. There are the following possible explanations for the weak fluorescence: The fluorophore could loose its fluorescence properties by conformational changes of BSA upon adsorption. Or the protein does not bind covalently in the first place and can thus be removed by rinsing. 

If the samples are not rinsed after protein adsorption to silicon, a strong fluorescence is visible. This supports the idea that most of the protein is removed from the silicon substrate in the rinsing step. The structures visible in the fluorescence image are probably crystallised buffer salts.

\begin{figure}
	\centering
		\includegraphics{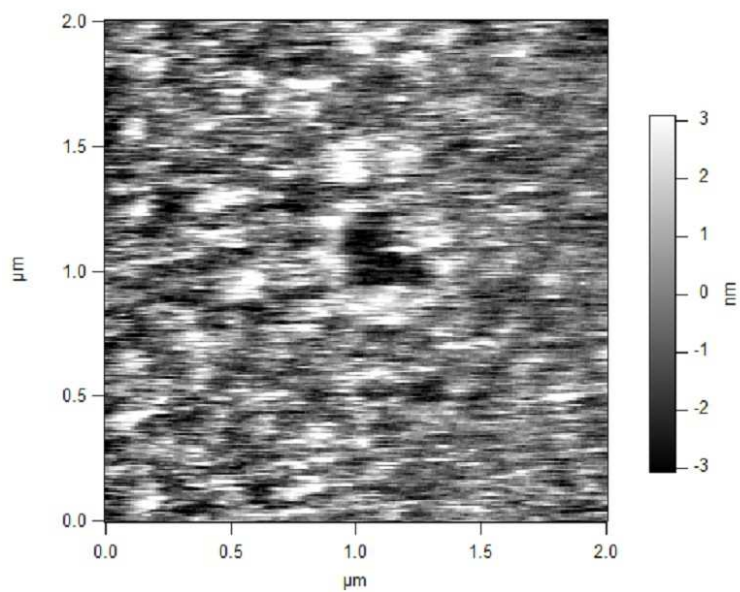}
	\caption[Scanning force microscope image of BSA fluorescein conjugate and lysozyme on silicon]{Scanning force microscope image of a silicon sample treated with a solution of BSA fluorescein conjugate and lysozyme rinsed after protein adsorption. The lower region in the centre of the image was scratched with the cantilever tip}
	\label{fig:AfmLb14}
\end{figure}

\begin{figure}
	\centering
		\includegraphics{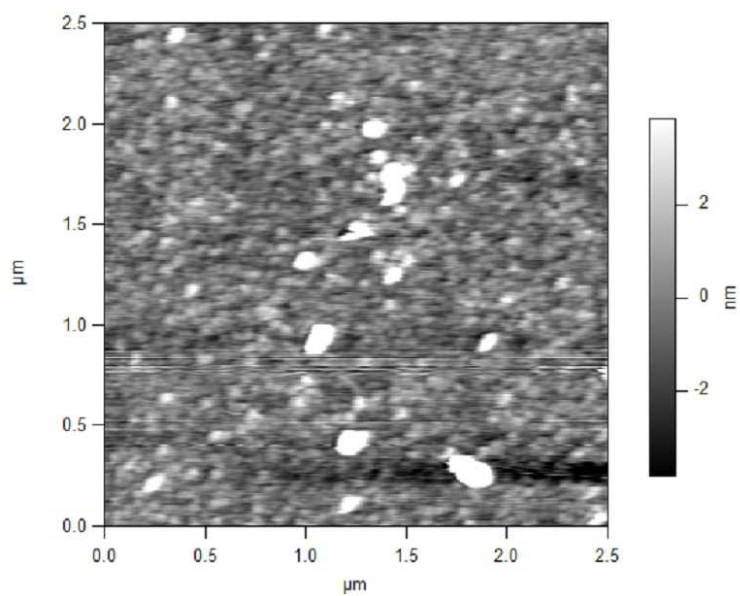}
	\caption[Scanning force microscope image of BSA fluorescein conjugate and lysozyme on silicon]{Scanning force microscope image of a silicon sample treated with a solution of BSA fluorescein conjugate and lysozyme rinsed after protein adsorption}
	\label{fig:AfmLb15}
\end{figure}

To find out whether they are covered with protein or not, the sample surfaces are investigated with scanning force microscopy in contact mode. On all of the rinsed samples globular structures with height differences of approximately four nanometres are visible covering most of the surface. The structures vary in their lateral size. In addition, larger structures with some tens of nanometres height and up to some hundreds of nanometres in diameter are found on the surface. Probably the small structure is a protein film, while the larger ones are protein agglomerates. As examples, images obtained on the two samples treated with a solution of BSA fluorescein conjugate and lysozyme in buffer solution are shown in figures \ref{fig:AfmLb14} and \ref{fig:AfmLb15}. In the centre of figure \ref{fig:AfmLb14} a smaller region was scanned for ten minutes with higher frequency and higher force than usual. Afterwards the surface level lies about two nanometres deeper in this scratched area because the protein film has been removed.  

\begin{figure}
	\centering
		\includegraphics{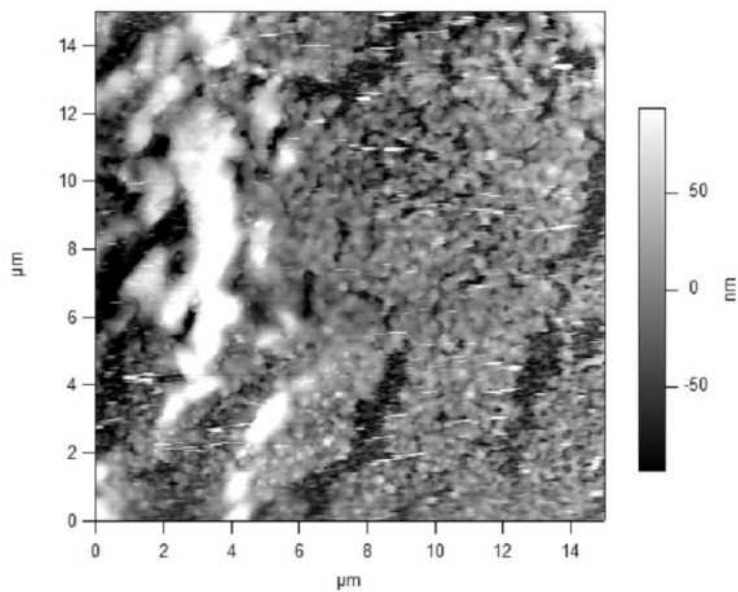}
	\caption[Scanning force microscope image of BSA fluorescein conjugate on silicon]{Scanning force microscope image of a silicon sample treated with a solution of BSA fluorescein conjugate not rinsed after protein adsorption}
	\label{fig:AfmBSA18}
\end{figure}

On the sample that was not rinsed after protein adsorption larger structures are visible (see figure \ref{fig:AfmBSA18}). Their lateral size is some micrometres and their height is some hundreds of nanometres. These are probably crystallized buffer salts. The protein layer should be found underneath the buffer crystals and is not detectable by scanning force measurements.
 
From the scanning force microscopic images it can be concluded that most of the samples' surfaces is covered with a protein layer even after rinsing. Thus the fluorescence patterns on the samples do not reflect the protein distribution. Presumably the fluorescence yield of fluorescein is reduced due to interactions which can occur when the protein adsorbs to the surface and changes its conformation. 

\subsection{ToF-SIMS of protein films on dental implant materials}
\begin{figure}
\centering
\includegraphics{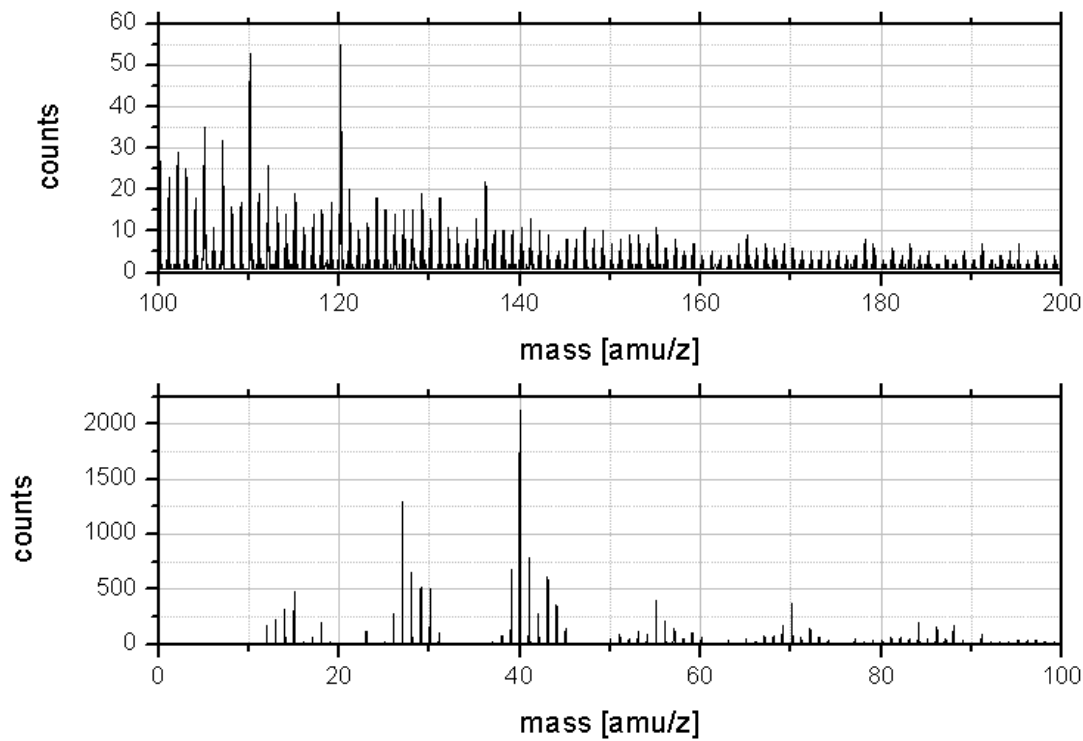}
\caption{Mass spectrum of cations acquired on a FAT substrate coated with BSA}
\label{fig:spectrabsa25p}
\end{figure}

Single protein films are prepared on substrates of FAT and FAW. Of each protein three samples per substrate type are made. Mass spectra of cations are acquired by ToF-SIMS at four sites on each sample. Regardless of the substrate or protein used the spectra have a similar appearance. As an example a mass spectrum acquired on BSA coated FAT is shown in figure \ref{fig:spectrabsa25p}. A strong signal at 40 amu/z (calcium) and weaker ones at 23 amu/z (sodium) and 88 amu/z (strontium) are caused by the substrate. Additionally at nearly every mass up to 200 amu/z peaks caused by protein fragments or organic contaminants are visible. The visibility of substrate caused peaks shows that once again the proteins are not closely packed on the surface. 
The mass spectra are unit-mass binned and amino acid related peaks are selected. Of the peaks listed in table \ref{tab:PeakSelection}, the ones at 44 amu/z, 86 amu/z, 87 amu/z, 88 amu/z and 69 amu/z overlap with peaks caused by the substrate ($\mathsf{^{44}Ca}$, $\mathsf{^{86}Sr}$, $\mathsf{^{87}Sr}$ and $\mathsf{^{88}Sr}$) or by the primary ions ($\mathsf{^{69}Ga}$). Hence these masses are not taken into account for multivariate analysis.  

\begin{figure}
\centering
\includegraphics{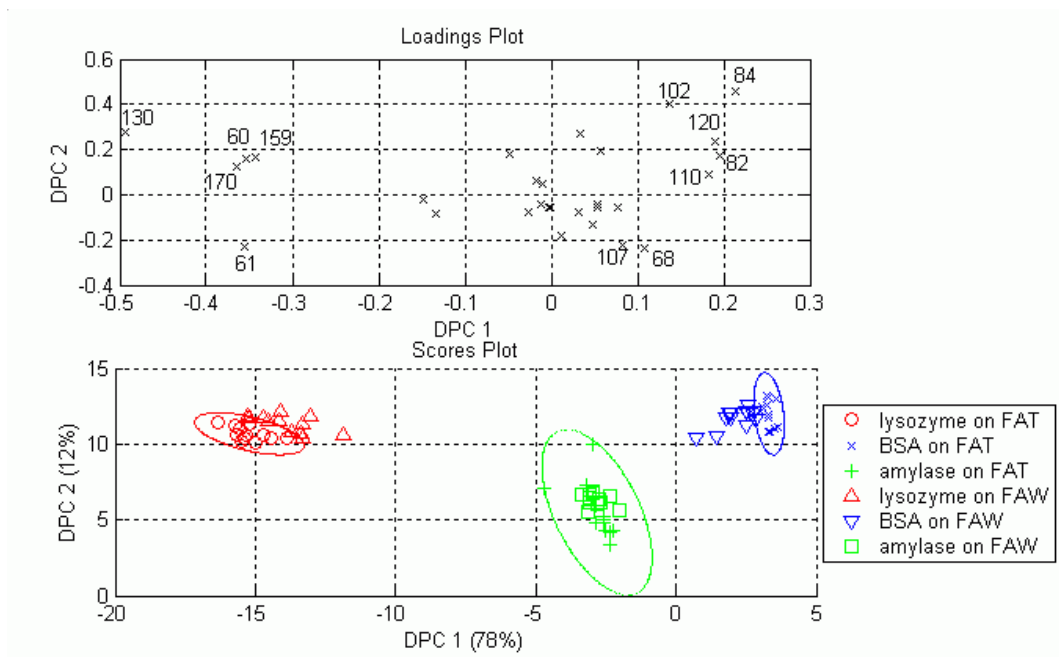}
\caption{DPCA results of proteins adsorbed to FAT from 10 mM buffer solution}
\label{fig:LysBSAAmy2536}
\end{figure}

Performing DPCA on the data acquired on FAT substrates leads to the results shown in figure \ref{fig:LysBSAAmy2536}. In the scores plot the $95 \%$-probability ellipses of the three proteins are well seperated. In addition to the data acquired on FAT, spectra acquired on FAW substrates are projected into the scores plot using the loadings calculated with the FAT data. The projections lie very close to or within the probability ellipses of the corresponding protein. This shows that the intensity patterns of the selected peaks are only weakly influenced by the substrate. 

In the loadings plot the masses 130 amu/z, 170 amu/z, 159 amu/z, 60 amu/z and 61 amu/z have the strongest negative values on the first discriminant principal component. They are caused by fragments of tryptophane, serine and methionine (see table \ref{tab:PeakSelection}). The former two are most abundant in lysozyme while methionine is most abundant in amylase but can be found to nearly equal amounts in lysozyme as well (see figure \ref{fig:LysBSADiagram}). Together these masses cause the lysozyme spectra to be projected to the left of the scores plot. The highest positive loadings on the first DPC are achieved by the masses 84 amu/z (glutamine, glutamic acid and lysine), 82 amu/z (histidine), 120 amu/z (phenylalanine) and 110 amu/z (histidine). All the corresponding amino acids but phenylalanine are most abundant in BSA projecting its spectra to the right of the scores plot. Finally, fragments of proline (68 amu/z), methionine (61 amu/z) and tyrosine (107 amu/z) have the strongest negative loadings on the second DPC. Since they are highly abundant in amylase, spectra of this protein are projected in the lower part of the scores plot. 

A leave-one-out test confirms the quality of the DPCA model. Using the euclidean distance on the first two DPC to the centres of gravity of the protein groups for assignment, all 36 spectra are correctly assigned. With the probability ellipses, still $94\%$ of the spectra are assigned to the right protein. 

\begin{figure}
\centering
\includegraphics{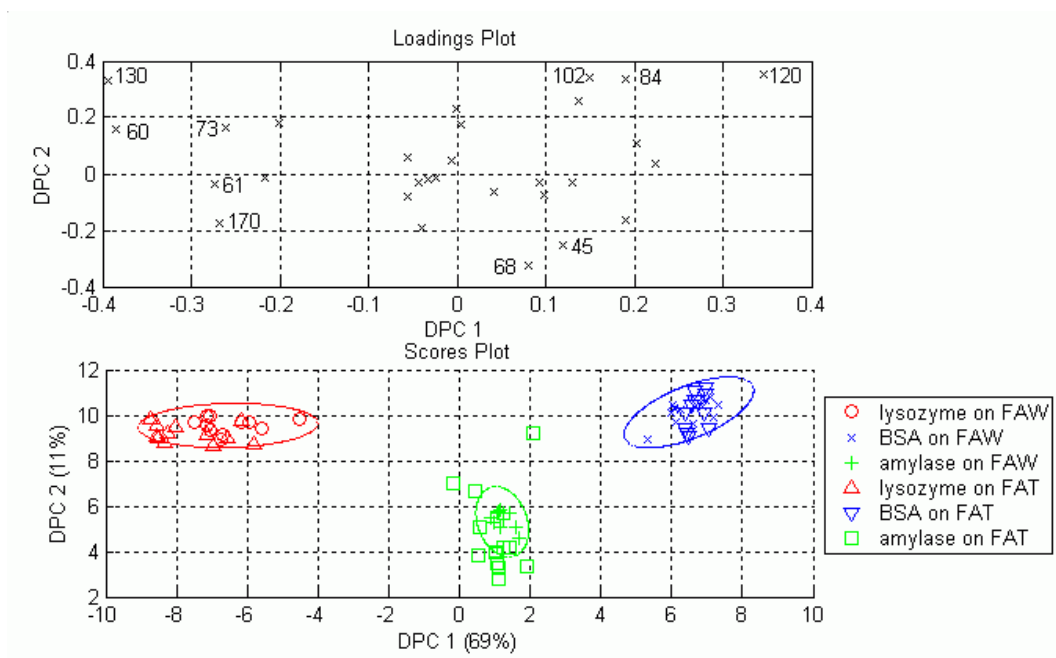}
\caption{DPCA results of proteins adsorbed to FAW from 10 mM buffer solution}
\label{fig:LysBSAAmy2839}
\end{figure}

By performing DPCA on the spectra of proteins adsorbed to FAW similar results are obtained (see figure \ref{fig:LysBSAAmy2839}). Again the probability ellipses of the three proteins are well separated in the scores plot and the projections of spectra acquired on protein films on FAT lie mostly within the probability ellipse of the right protein.

The loadings plot shows that the positioning of the proteins in the scores plot is mainly determined by the same amino acid fragments as before. Only two new fragments of significant loadings occur. On the negative side of the first discriminant principal component arginine (73 amu/z), which is most abundant in lysozyme, supports its positioning on the left side of the scores plot. Cysteine (45 amu/z) contributes with its negative loading on the second DPC in the positioning of amylase because it is highly abundant in this protein.  

In the leave-one-out test the DPCA model for FAW substrates performs slightly worse than the one for FAT substrates. Using the euclidean distance, still all 36 spectra are correctly assigned but with the probability ellipses the percentage of spectra assigned to the right protein is reduced to $89\%$.

The similarities between the DPCA results of protein layers on the two substrate types show that very similar single component protein layers are formed on FAT and FAW substrates.

\subsubsection{Adsorption from binary protein mixtures}
\begin{figure}
\centering
\includegraphics{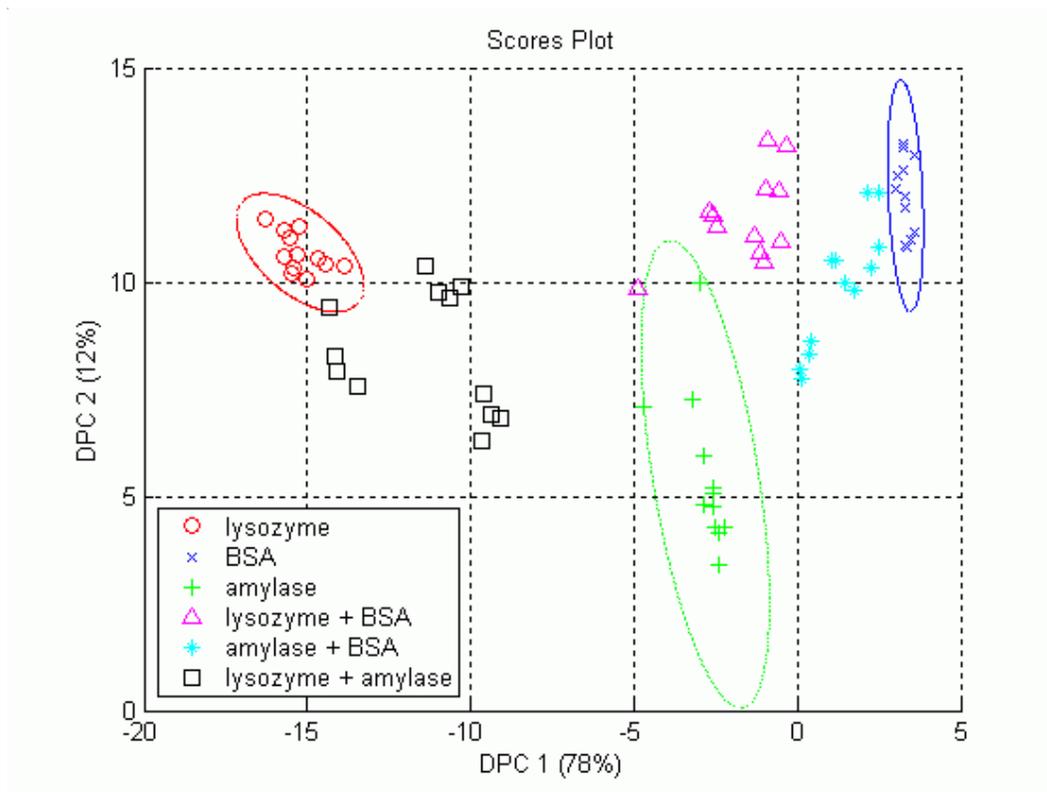}
\caption{Projections of mass spectra of binary protein mixtures on FAT}
\label{fig:mixfat}
\end{figure}

\begin{figure}
\centering
\includegraphics{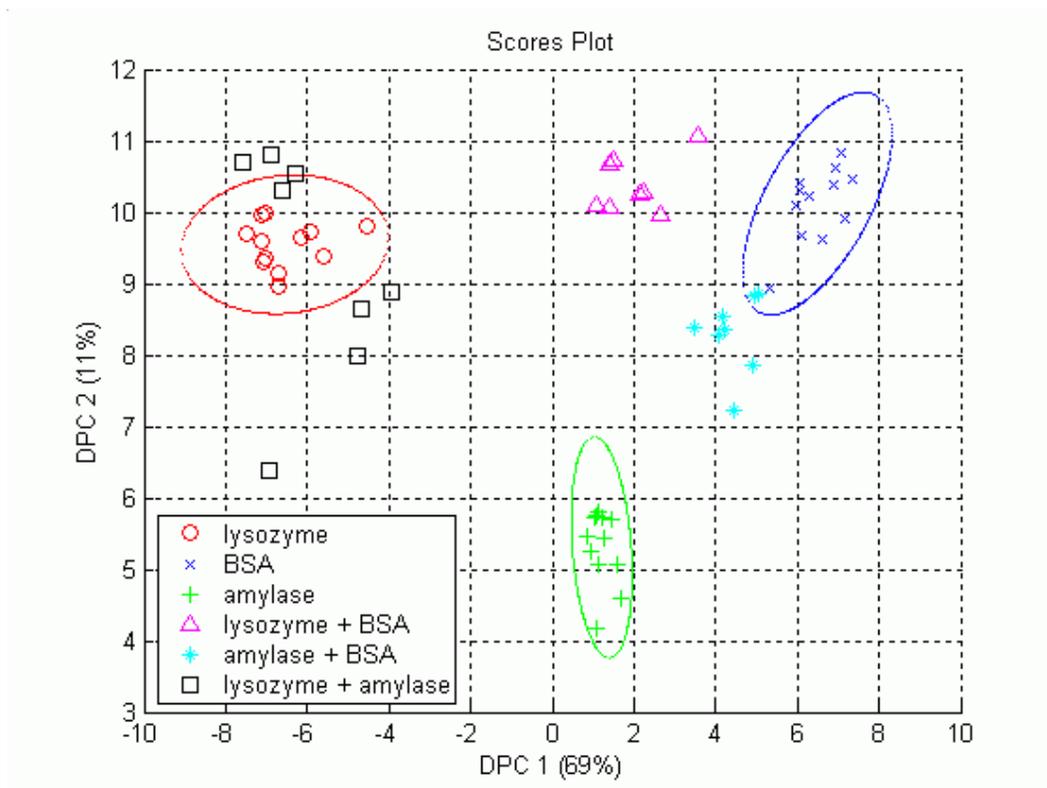}
\caption{Projections of mass spectra of binary protein mixtures on FAW}
\label{fig:mixfaw}
\end{figure}

Substrates of FAT and FAW are also treated with solutions of binary mixtures of two of the three proteins. Equal molarities of about $5\cdot 10^{-5}$ mol/l of the proteins are solved in 10 mM buffer solution at pH 7.4. After protein adsorption time-of-flight mass spectra of cations are acquired at four sites on each sample. Of every possible binary mixture three samples are prepared on FAT and two samples on FAW. The spectra are unit-mass binned and amino acid related peaks are selected. They are projected into the scores plots of the DPCA models built with single protein spectra on the respective substrates. The resulting projections can be seen in figure \ref{fig:mixfat} for FAT substrates and in figure \ref{fig:mixfaw} for FAW substrates. 

The spectra of protein films on FAT substrates show a relatively large spread between different samples treated with the same type of mixture. The mass spectra of a mixture of lysozyme and amylase are projected in a region from the lysozyme probability ellipse to the middle between the probability ellipses of lysozyme and amylase. Thus it is assumed that mostly lysozyme adsorbs from this mixture but the proportions of lysozyme and amylase are different from experiment to experiment. Similarly the projections of the mass spectra of the amylase BSA mixture lie in a region from the BSA ellipse to the middle between the BSA and amylase ellipses. This indicates that BSA forms the major component of the adsorbed films but their exact composition differs between the experiments. In contrast the mass spectra of a mixture of lysozyme and BSA show less spread. They are found closer to BSA than to lysozyme. Hence in this case BSA seems to be the major component of the adsorbed film. 

When the protein films are adsorbed to FAW substrates there are smaller differences between the projections of mass spectra of different samples treated with the same type of mixture. Now, with one exception, the projections of a mixture of lysozyme and amylase are situated very close to the lysozyme probability ellipse. Thus probably only lysozyme has adsorbed to the substrate. The mass spectra of BSA amylase mixtures are projected close to the BSA ellipse indicating that this protein is the major component of the adsorbed film. The projection of mixtures of lysozyme and BSA are closer to the BSA ellipse than to the one of lysozyme. Thus more BSA than lysozyme is detected on the samples. So the composition deduced from the DPCA projections of the protein films created with binary mixtures is similar on the two dental implant materials but it differs from the one on silicon substrates. This is not surprising, as an effect of the substrate on protein adsorption should be expected. These results are summarized in table \ref{tab:mixtures}. They suggest that BSA has the highest affinity to adsorb to the dental implant materials followed by lysozyme while amylase has the lowest affinity. On silicon substrates lysozyme shows the highest affinity while the relation between the ones of BSA and amylase is varying.  
\begin{table}
	\centering
	\begin{tabular}{|c||c|c|c|}\hline
	 proteins &  \multicolumn{3}{c|}{substrate} \\
	 in solution & silicon & FAT & FAW \\ \hline \hline
	 lysozyme & more lysozyme & more BSA & more BSA \\
	 and BSA & & & \\ \hline
	 amylase & spread from & spread from equality to & spread from more to\\
	 and BSA & amylase to BSA & much more BSA & much more BSA\\ \hline
	 lysozyme & spread from more to & spread from equality to & much more lysozyme \\
	 and amylase & much more lysozyme  & more lysozyme & \\ \hline 
	 \end{tabular}
	\caption{DPCA projection deduced compositions of protein films adsorbed from binary mixtures}
	\label{tab:mixtures}
\end{table}

\subsubsection{Sputtering of single protein layers}
\begin{figure}
\centering
\includegraphics{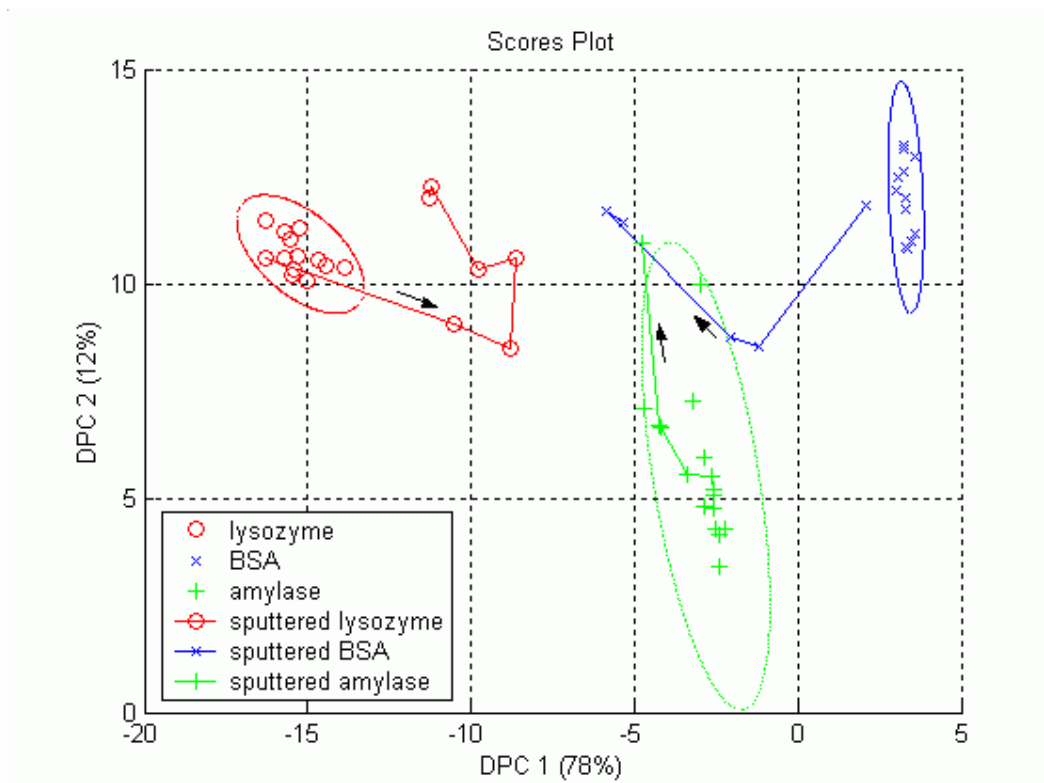}
\caption{Projections of sputtered protein films on FAT}
\label{fig:sputterFAT2}
\end{figure}

\begin{figure}
\centering
\includegraphics{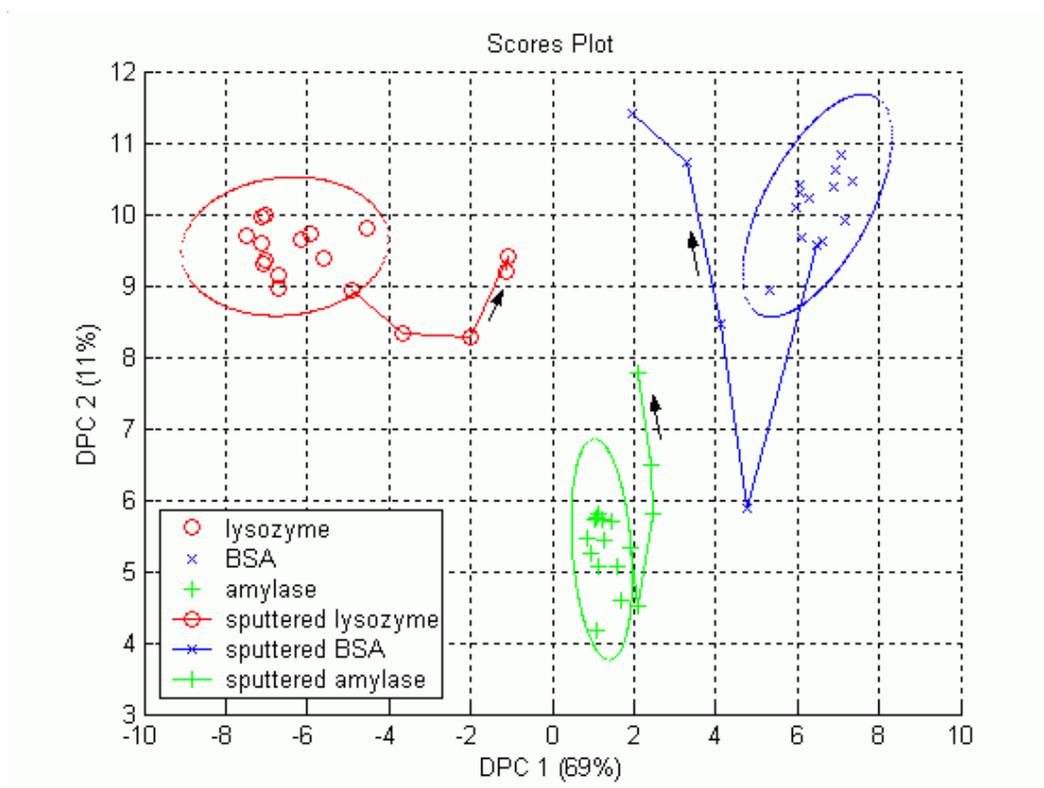}
\caption{Projections of sputtered protein films on FAW}
\label{fig:sputterFAW3}
\end{figure}

On two other sets of samples coated with single protein films, mass spectra are acquired after sputtering the surface with the unpulsed primary ion beam. The samples are prepared like the ones used to build the DPCA model. On one sample of lysozyme, BSA and amylase adsorbed to FAT and FAW substrates, mass spectra are acquired after sputtering for 0, 5, 10, 15, 20, 25 and 30 seconds. 
This corresponds to maximum primary ion doses of about $10^{13} \mbox{cm}^{-2}$ (see equation \ref{eq:sputterdose}).
The spectra are unit-mass binned, amino acid related peaks are selected, and the spectra are projected in the scores plots using the DPCA models created before. The projections are shown in figure \ref{fig:sputterFAT2} for FAT substrates and in figure \ref{fig:sputterFAW3} for FAW substrates. Under sputtering the projections move from the probability ellipses of the respective protein into the region between the three ellipses. After short sputtering times a recognition of the different proteins by their DPCA projections is not possible any more. At this stage it can be seen in the mass spectra that most of the organic material has been removed from the sample surface. The first one or two spectra corresponding to sputtering times of five to ten seconds of samples coated with lysozyme or BSA show a development heading to the amylase probability ellipse. Probably the different amino acids are not equally destroyed by sputtering. Thus amino acids related to lysozyme and BSA can be destroyed before the ones related to amylase. This makes the spectra more similar to amylase spectra and explains the development of sputtered lysozyme and BSA in the scores plot. 

\subsubsection{Consecutive adsorption of two proteins}
As on the silicon substrates, the consecutive adsorption of two proteins is studied on the dental implant materials, too. Therefore the substrates are rinsed with buffer solution after the first adsorption step. Then they are put into a solution of the second protein. Proteins are solved at molarities of about $5 \cdot 10^{-5}$ mols per litre buffer solution. The latter is a ten millimolar phosphate buffer as before. Two samples are prepared for each of the six possible combinations of two proteins on the two substrate types. Mass spectra of cations are acquired at four sites on each sample. At one site on each sample mass spectra are also acquired after sputtering the surface with the unpulsed primary ion beam for five to thirty seconds. This corresponds to maximum primary ion doses of about $10^{13} \mbox{cm}^{-2}$ (see equation \ref{eq:sputterdose}). The mass spectra are unit-mass binned and amino acid related peaks are selected. Then they are projected into the scores plots of the DPCA models built with mass spectra of single protein films.

The projections of the mass spectra of proteins adsorbed to FAT substrates are shown in figure \ref{fig:doppelschichtenfat} and reveal the following: 
\begin{figure}
\centering
\includegraphics{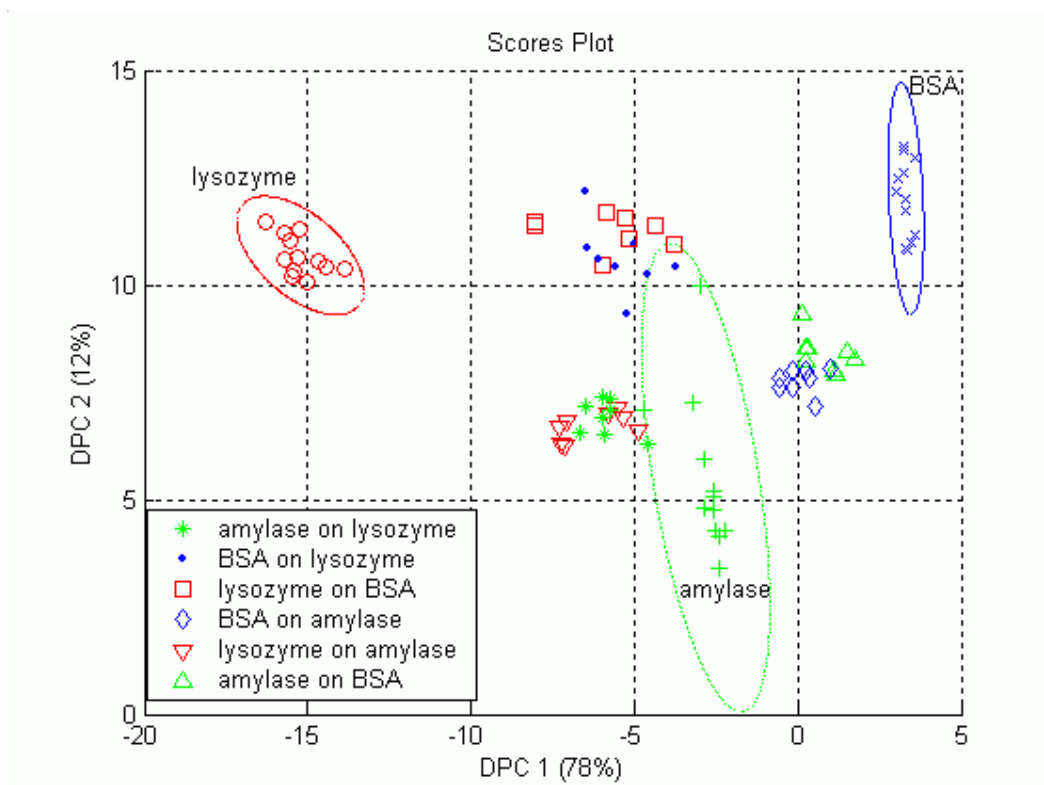}
\caption{Projections of protein double layers adsorbed to FAT substrates}
\label{fig:doppelschichtenfat}
\end{figure}
\begin{description}
\item[Amylase on lysozyme:] The projections lie close to the amylase probability ellipse. Hence amylase is detected in high amounts indicating that it adsorbs well on the lysozyme layer.
\item[Lysozyme on amylase:] These projections are also close to the amylase ellipse. Thus only little amounts of lysozyme are detected indicating that it does not adsorb well on the amylase coated substrate.
\item[BSA and lysozyme:] The spectra of both types of samples are projected in the middle between the BSA and lysozyme ellipses. As both proteins are detected, either the secondly adsorbed protein forms a porous layer making detection of the lower layer possible, or the two proteins form one mixed layer in the second adsorption step. 
\item[BSA and amylase:] Here the same observations and conclusions as for BSA and ly\-so\-zyme are made.  
\end{description}

Comparing these conclusions to the ones drawn from experiments of simultaneous adsorption of two proteins, there seems to be a contradiction in the interaction of lysozyme and amylase. On the one hand amylase adsorbs well on lysozyme and lysozyme does not adsorb well on amylase. On the other hand more lysozyme than amylase adsorbs from a mixture of the two proteins. This raises the question why lysozyme is not covered by amylase as in the two step experiment. A possible answer can be found in the rinsing step between the two adsorption steps. It might induce conformational changes in the adsorbed lysozyme that favour the adsorption of amylase on it.  

\begin{figure}
\centering
\includegraphics{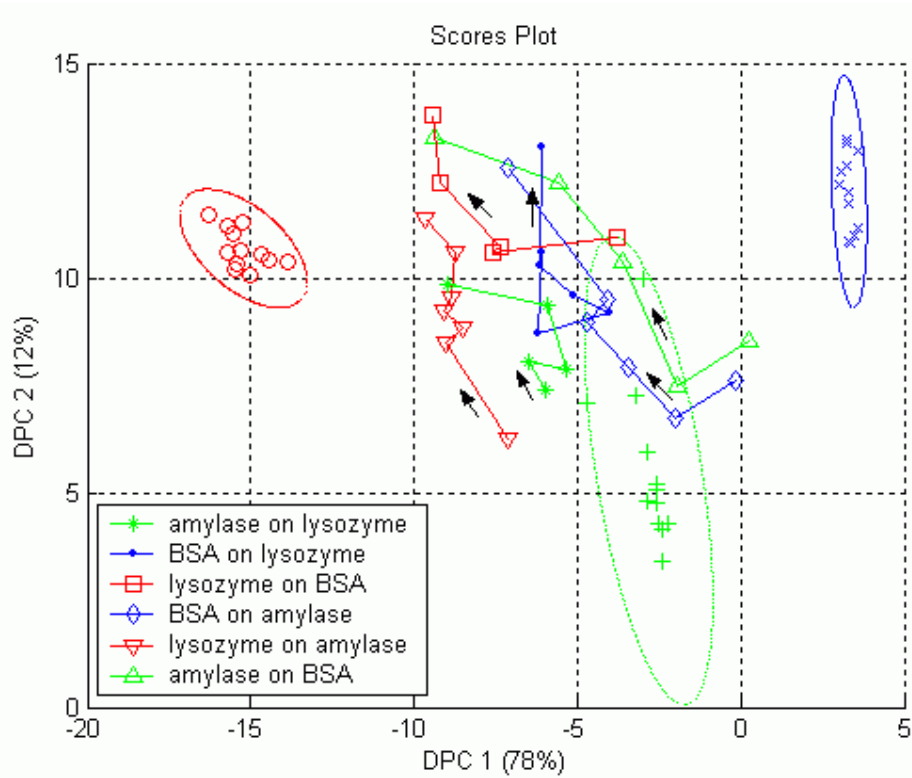}
\caption{Projections of protein double layers adsorbed to FAT after sputtering. The arrows indicate the direction of increasing sputter time.}
\label{fig:sputterfatdoppel3}
\end{figure}

Figure \ref{fig:sputterfatdoppel3} shows the effect of sputtering on the projections. The arrows indicate the direction of increasing sputter time in steps of five seconds between two consecutive spectra. For clarity only the projections of one series of mass spectra are shown. The projections of the other series develop mainly in the same way.
As expected, the projections develop into a region between the three probability ellipses of single protein spectra. The development is not represented by straight lines because of unequal destruction of different amino acids by sputtering and because of noise, which gets more important with increasing sputter times because of the vanishing of the amino acid related peaks. The projections of spectra of samples coated with BSA and amylase behave in a very similar way. This indicates that, regardless of the order of the adsorptions, one mixed layer of the two proteins develops. If there were two layers one would expect a differing behaviour depending on the order of adsorptions as it can be seen for lysozyme and BSA. The effect that the projections of layers of amylase and BSA move closer to the amylase ellipse after the first five seconds of sputtering is also seen for pure BSA layers (see figure \ref{fig:sputterFAT2}). Thus it does not show the detection of an increasing amount of amylase. 

In figure \ref{fig:doppelschichtenfaw} it can be seen that many of the projections of mass spectra of protein double layers adsorbed to FAW substrates are situated differently than the ones of FAT substrates (see figure \ref{fig:doppelschichtenfat}). This shows an influence of the substrate onto the interactions between different proteins upon adsorption. In detail the following observations are made for FAW substrates:
\begin{figure}
\centering
\includegraphics{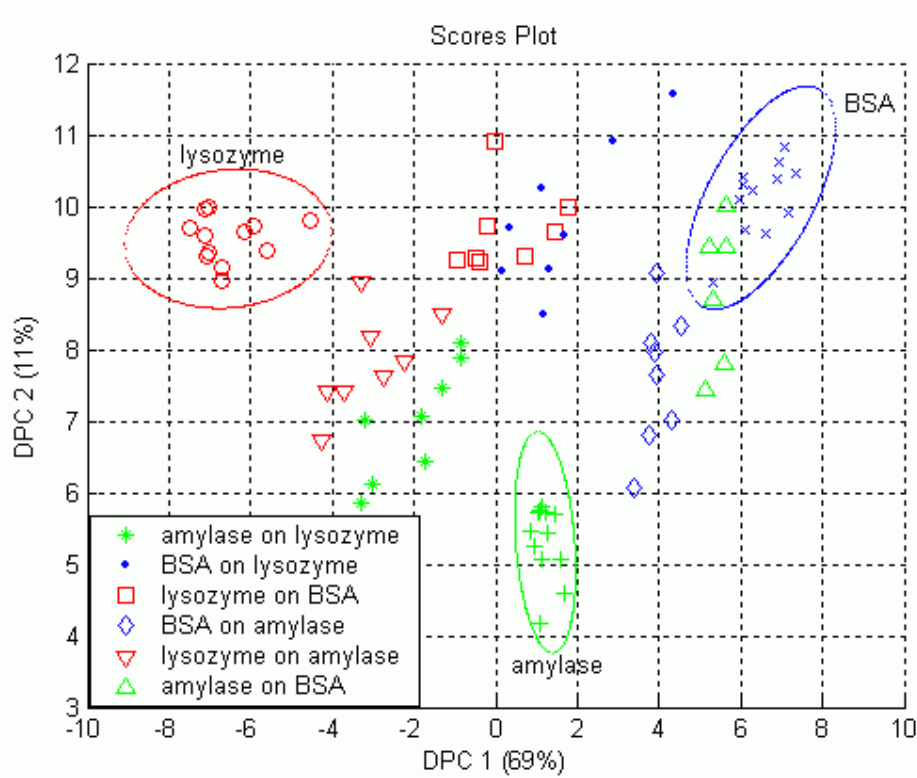}
\caption{Projections of protein double layers adsorbed to FAW substrates}
\label{fig:doppelschichtenfaw}
\end{figure}
\begin{description}
\item[Amylase and lysozyme:] Contrary to the spectra on FAT substrates, the projections are situated more or less in the middle between the probability ellipses of amylase and lysozyme. Hence both proteins are detected at equal amounts. This can be explained either by the formation of a porous layer of the secondly adsorbed protein or by mixing of the two proteins in the second adsorption step.
\item[BSA and lysozyme:] Like the spectra on FAT substrates, the spectra are mostly projected into the middle between the ellipses of BSA and lysozyme. This leads to the same conclusions as before.
\item[Amylase on BSA:] These projections lie close to or within the BSA ellipse. Thus the amount of amylase detected is very small. It can be concluded that in contrast to the case of FAT substrates, amylase does not adsorb well to BSA coated FAW substrates. 
\item[BSA on amylase:] Compared to the spectra on FAT substrates, these are projected in a larger region closer to the BSA ellipse than to the amylase ellipse. Thus BSA is detected in larger quantities than before. Hence BSA adsorbs in higher amounts to an amylase coated FAW substrate than to a corresponding FAT substrate. 
\end{description}
The projections of the spectra of all sample types show a higher spread on FAW substrates than on FAT substrates. 

Again the results of the consecutive adsorption of two proteins can be compared to the ones of the simultaneous adsorption. In contrast to the experiments on FAT substrates, the two types of experiments on FAW substrates lead partly to the same conclusions. From a solution of BSA and amylase mainly BSA adsorbs to the substrate. Equally only little amounts of amylase adsorb on a BSA coated sample. Thus in both experiments BSA hinders the adsorption of amylase.
Since the projections of consecutive adsorptions of lysozyme and BSA or amylase and lysozyme are situated in the middle between the probability ellipses of their components, these proteins do not significantly hinder or favour the adsorption of one another. Hence their are no conclusions to be compared to the one-step adsorption experiments.  

Table \ref{tab:doublelyers} summarizes the results of the consecutive adsorption experiments of two proteins for all three substrate types.
\begin{table}
	\centering
		\begin{tabular}{|c||c|c|c|} \hline
		& \multicolumn{3}{c|}{substrate} \\
		proteins & silicon & FAT & FAW \\ \hline \hline
		lysozyme on BSA & medium & medium & medium \\ \hline
		lysozyme on amylase & spread from & little & spread from \\
		& very little to high & & medium to high \\ \hline
		BSA on lysozyme &  little & medium & medium\\ \hline
		BSA on amylase & spread from & medium & spread from \\ 
		& very little to little & & medium to high \\ \hline
		amylase on lysozyme & very little & high & medium \\ \hline
		amylase on BSA & high & medium & spread from \\ 
		& & & very little to little \\ \hline	
		\end{tabular}
\caption{Results of DPCA on mass spectra of protein double layers. This table summarizes the amount detected of the secondly adsorbed protein}
\label{tab:doublelyers}	
\end{table}

\begin{figure}
\centering
\includegraphics{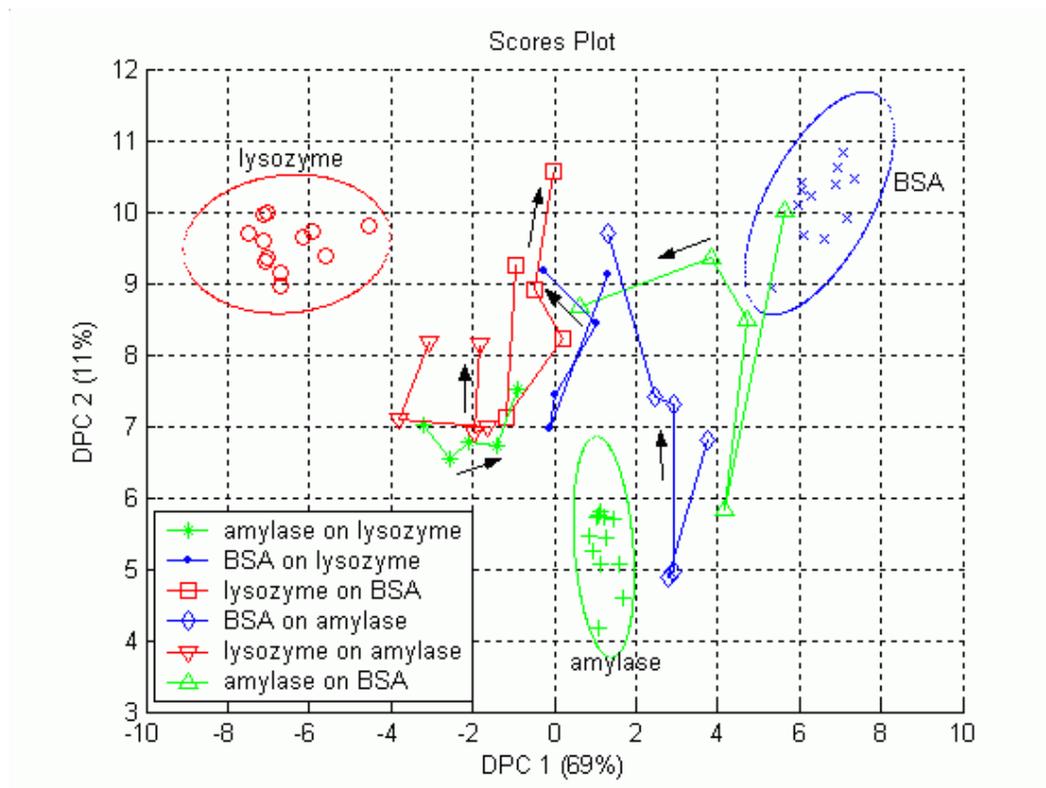}
\caption{Projections of protein double layers adsorbed to FAW after sputtering. The arrows indicate the direction of increasing sputter time.}
\label{fig:sputterfawdoppel2}
\end{figure}

Figure \ref{fig:sputterfawdoppel2} shows the development of the projections of mass spectra of protein double layers on FAW substrates under sputtering. The arrows indicate the direction of increasing sputter time in steps of five seconds between two consecutive spectra. For clarity only the projections of one series of mass spectra are shown. The projections of the other series develop mainly in the same way.

As before the projections of the spectra of all samples develop into the region between the three protein probability ellipses. Thus the protein layers are rapidly destroyed and cannot be recognized by their mass spectra any more. After the last sputtering period only the substrate generated peaks show a noteworthy intensity in the mass spectra. The projections of most samples do not develop directly to their final position but show a strong shift to smaller values on the second discriminant principal component during the first five to ten seconds of sputtering. As this behaviour is seen for most samples regardless of their composition, it is not the effect of a rising proportion of amylase, which can be found at lower scores on the second DPC. Instead it is probably caused by the not equal destruction of different amino acids by the primary ion beam. A preferential destruction of amino acids with highly positive loadings on the second DPC explains the development of the projections.    
\pagebreak
\section{Summary and Outlook}
In this work the two experimental dental implant materials FAT and FAW, made of fluoroapatite particles embedded in polymer matrices, and films of the proteins lysozyme, amylase and bovine serum albumin (BSA), adsorbed to the two dental implant materials, were investigated with time-of-flight secondary ion mass spectrometry (ToF-SIMS) and the multivariate data analysis technique discriminant principal component analysis (DPCA).

The elemental surface composition of the dental implant materials was estimated from the intensity of the peaks caused by the ions of different elements in ToF-SIMS mass spectra acquired on samples of FAT and FAW. These compositions were compared to the surface composition of bovine tooth enamel obtained in the same way. It was found that the main surface components of all three materials are sodium, phosphorus, potassium, calcium, oxygen and fluorine. Additionally the two implant materials contain high quantities of aluminium and strontium. Due to the limitations of the ToF-SIMS technique for quantitative analysis, exact surface compositions cannot be measured by this method.  

Protein adsorption experiments were at first performed on silanised silicon substrates, because silicon substrates offer a well-defined flat surface and the adsorption of proteins to silanised samples had already been studied with other methods within the work group. ToF-SIMS mass spectra of cations were acquired on samples coated with lysozyme or BSA. These spectra were dominated by a signal caused by silicon probably from polymerised silane on the sample surface. Additionally many peaks associated to amino acid fragments could be found in the mass spectra. \textsc{Matlab} functions were written to select amino acid related peaks from the mass spectra, to subtract a background spectrum and to do DPCA on the mass spectra. Nevertheless it was not possible to distinguish between different proteins by their scores resulting from DPCA because of a large spread within the data. This spread was probably caused by varying amounts of polymerised silane on the sample surfaces.  

In the next step proteins were adsorbed to unsilanised silicon substrates to prevent the influence of polymerised silane on the mass spectra. In this case, after selection of amino acid related peaks and DPCA,  discrimination of samples coated with amylase, lysozyme or BSA by their mass spectra of cations was possible. The concentration of the buffer solution was varied to obtain the best discrimination of the proteins at a concentration of ten millimols per litre of the buffer salts sodium di(hydrogen)phosphate and disodium hydrogenphosphate. The results of DPCA were evaluated with leave-one-out (LOO) tests. These tests showed that the data were very well represented by the DPCA model, because 46 out of 48 spectra could be assigned to the right protein. 

With the DPCA model developed with single component protein films, the adsorption behaviour in the cases of simultaneous or consecutive adsorption of two proteins to the same sample was studied. Therefore mass spectra obtained on the new samples were projected into the existing scores plot. Mutual influences between the different proteins upon adsorption were observed. It was concluded that the presence of lysozyme hindered the adsorption of amylase and BSA in the two studied cases. These results were partly confirmed by enzymatic activity measurements of amylase. 

Finally, amylase, lysozyme and BSA were adsorbed to the dental implant materials. Again the three proteins could be distinguished by the scores of their mass spectra of cations after selection of amino acid related peaks and DPCA. The peak selection had to be slightly modified, because some of the peaks were influenced by substrate generated peaks of the same nominal mass. A correct assignment of at least $89 \%$ of the spectra in LOO tests showed a good representation of the data by the DPCA models. Furthermore, a great similarity of the DPCA results obtained on the two different substrate types was observed. This shows the formation of similar single component protein films on the two dental implant materials. 

Samples coated with two proteins adsorbed simultaneously or consecutively were also studied. Mass spectra acquired on these samples were projected into the scores plots of the corresponding DPCA model built with spectra from single component protein films. On both substrate types the experiments of simultaneous adsorption of two proteins led to the same results: BSA shows the strongest affinity to adsorb from the mixed solutions followed by lysozyme, while amylase has the smallest affinity. The consecutive adsorption of lysozyme and BSA on both substrates led to the detection of equal amounts of both proteins regardless of the order of adsorption. Thus either one mixed layer of the two proteins or a porous layer of the second protein had developed. On FAT there seemed to be a contradiction in the behaviour of lysozyme and amylase between the two types of experiments: Lysozyme hindered the simultaneous adsorption of amylase while it favoured the consecutive adsorption of amylase. This could be explained by conformational changes of lysozyme in the rinsing step between the consecutive adsorptions. By contrast both types of experiments on FAW suggest that the presence of lysozyme hinders the adsorption of amylase. Hence in this case the differences between the two substrate types clearly influence the co-adsorption of amylase and lysozyme.
\vspace{3ex}

Lysozyme has bacteriolytical effects in the acquired enamel pellicle while amylase favours the adsorption of bacteria and thus the formation of plaque. Thus the mutual influence of these two proteins upon adsorption is of special interest for the design of dental implant materials. This influence, as well as the ones between the other protein combinations, has been shown to be dependent on the substrate used. It should be further investigated why a given substrate favours the adsorption of one protein or another. Possible reasons are the electrical charge and the hydrophilicity of the substrate and the protein.

The co-adsorption results can be verified using other observation techniques. Nicole Lawrence, a member of our work group, is examining the adsorption of proteins on differently prepared silicon substrates with dynamic contact angle measurements and ellipsometry in the context of her Ph.D. thesis.

The solutions of maximum three proteins used in this work are by far less complex than human saliva. In future works more proteins should be used to approach the in vivo situation. Furthermore, the parameters pH and temperature are not constant in the oral cavity. So their influence should also be examined.

The influence of the buffer solution merits further investigation, too. It has been seen that the results strongly depend on the concentration of the buffer salts. The use of other than phosphate buffers could lead to new results.

So far no use was made of the possibility to image the sample surface by ToF-SIMS. It could be tried to  bring different proteins on the sample at separated sites and to examine the resulting surface distribution.

It was not possible to create depth profiles of the two layer protein films because the unpulsed gallium ion beam used for sputtering the upper protein layer also destroyed the lower protein layer at the same time. This is due to the high penetration depth of the gallium ions.
A possible solution to this problem is the use of a cluster ion source instead of the gallium ion source. If for example a buckminsterfullerene ($\mathsf{C_{60}}$) is accelerated to the same total energy as a gallium ion, the energy per atom and thus the penetration depth is much smaller. Boussofiane-Baudin and others estimated that a buckminsterfullerene projectile at 20 kilo-electron volts deposits its energy within the first 30 \AA ngst\"oms of an organic surface \cite{boussofiane-baudin:1994}. Hence most of the damage is limited to the removed layer and the lower layer remains intact for analysis. An additional effect of the lower penetration depth is a strongly increased secondary ion yield. Weibel and others have shown that using a buckminsterfullerene source improves the secondary ion yield on various organic surfaces by a factor of at least 30 compared to a gallium source \cite{weibel:2003}. For high mass fragments the effect is even stronger. Thus sample damage could be minimised by reducing the primary ion dose while maintaining strong secondary ion intensities due to the high yield. Other cluster ion sources as gold clusters ($\mathsf{Au_n^+}$ with $n=2-5$) or sulfur pentafluoride ($\mathsf{SF_5^+}$) show comparable secondary ion yields as buckminsterfullerenes but the sample damage is smallest for the latter \cite{weibel:2003}.
Another approach to make depth profiling possible can be the use of a cryo stage. By cooling down the samples the protein layers should be more stable against in depth destruction by the sputter beam. 

The protein layers are certainly denatured due to dehydration upon exposure to the ultra high vacuum conditions in the ToF-SIMS apparatus. To preserve the protein layers in their physiological state for analysis, one should try to prevent denaturation for example by use of a cryo stage or by fixation with glutardialdehyde. 

All these possibilities of further investigations show that this work has by no means exhaustively dealt with the investigation of protein films on dental implant materials by ToF-SIMS. Instead it has established a method of investigation, which can be used for future studies, that can finally lead to many new insights into the subject. 
\pagebreak
\begin{appendix}
\section{Matlab functions}
\label{sec:sourcecode}

This section contains the source code of the \textsc{Matlab} functions written for data analysis.
\small
\subsection*{einlesen.m}
\def\dash{\raise2.1pt\hbox{\rule{5pt}{0.3pt}}
\hspace{1pt}}\begin{tabbing}
{\textbf{function}}\hspace{6pt}out=einlesen(prefix,start,stop,suffix,m)\\
{\it{\%out=einlesen(prefix,start,stop,suffix,m))}}\\
{\it{\%reads\hspace{6pt}the\hspace{6pt}files\hspace{6pt}{'}c:$\backslash$tof\dash{}sims$\backslash$ascii$\backslash$prefixstartsuffix.asc{'}}}\\
{\it{\%to\hspace{6pt}{'}c:$\backslash$tof\dash{}sims$\backslash$ascii$\backslash$prefixstopsuffix.asc{'}\hspace{6pt}containing}}\\
{\it{\%unit\dash{}mass\hspace{6pt}mass\hspace{6pt}spectra\hspace{6pt}exported\hspace{6pt}from\hspace{6pt}WinCadence\hspace{6pt}and}}\\
{\it{\%forms\hspace{6pt}a\hspace{6pt}(mx(stop\dash{}start))\dash{}data\hspace{6pt}matrix\hspace{6pt}with\hspace{6pt}the\hspace{6pt}}}\\
{\it{\%first\hspace{6pt}m\hspace{6pt}masses\hspace{6pt}in\hspace{6pt}(stop\dash{}start)\hspace{6pt}spectra}}\\
\\
longprefix={[}{\texttt{{'}c:$\backslash$tof\dash{}sims$\backslash$ascii$\backslash${'}}},prefix{]};\\
\\
k=start;\\
filename={[}longprefix,int2str(k),suffix,{\texttt{{'}.asc{'}}}{]};\\
\\
{\it{\%reading\hspace{6pt}the\hspace{6pt}first\hspace{6pt}WinCadence\hspace{6pt}export:}}\\
x=load(filename);\\
\\
{\it{\%extracting\hspace{6pt}the\hspace{6pt}counts\hspace{6pt}at\hspace{6pt}the\hspace{6pt}first\hspace{6pt}m\hspace{6pt}masses:}}\\
out=voll(x,m);\\
\\
{\it{\%loop\hspace{6pt}for\hspace{6pt}the\hspace{6pt}other\hspace{6pt}files:}}\\
{\textbf{for}}\hspace{6pt}k=start+1:stop\\
\hspace{24pt}filename={[}longprefix,int2str(k),suffix,{\texttt{{'}.asc{'}}}{]};\\
\hspace{24pt}x=load(filename);\\
\hspace{24pt}{\it{\%new\hspace{6pt}data\hspace{6pt}is\hspace{6pt}appended\hspace{6pt}to\hspace{6pt}the\hspace{6pt}existing\hspace{6pt}data\hspace{6pt}matrix}}\\
\hspace{24pt}out=matrix(out,x);\\
{\textbf{end}}
\end{tabbing}

\subsection*{voll.m}
{{\def\dash{\raise2.1pt\hbox{\rule{5pt}{0.3pt}}\hspace{1pt}}\begin{tabbing}
{\textbf{function}}\hspace{6pt}filled\hspace{6pt}=\hspace{6pt}voll(in,m);
\\
{\it{\%filled\hspace{6pt}=\hspace{6pt}voll(in,m)}}\\
{\it{\%extracts\hspace{6pt}counts\hspace{6pt}from\hspace{6pt}a\hspace{6pt}WinCadence\hspace{6pt}output\hspace{6pt}unit\hspace{6pt}mass}}\\
{\it{\%mass\hspace{6pt}spectrum}}\\
{\it{\%in:\hspace{6pt}(?x3)\dash{}matrix:\hspace{6pt}1.\hspace{6pt}column:\hspace{6pt}channel\hspace{6pt}numbers;}}\\
{\it{\%\hspace{18pt}2.\hspace{6pt}column:\hspace{6pt}atomic\hspace{6pt}mass\hspace{6pt}units;\hspace{6pt}3.\hspace{6pt}column:\hspace{6pt}counts;}}\\
{\it{\%\hspace{18pt}masses\hspace{6pt}with\hspace{6pt}zero\hspace{6pt}counts\hspace{6pt}are\hspace{6pt}not\hspace{6pt}exported}}\\
{\it{\%m:\hspace{6pt}highest\hspace{6pt}mass\hspace{6pt}extracted}}\\
{\it{\%filled:\hspace{6pt}(mx1)\dash{}vector:\hspace{6pt}counts\hspace{6pt}for\hspace{6pt}m\hspace{6pt}masses}}\\

\\
c=0;
\\
go=1;
\\
k=1;
\\
\\
{\it{\%delete\hspace{6pt}first\hspace{6pt}row\hspace{6pt}if\hspace{6pt}it\hspace{6pt}contains\hspace{6pt}mass\hspace{6pt}0:
}}\\
{\textbf{if}}\hspace{6pt}in(1,2)==0
\\
\hspace{24pt}in=in(2:m+1,:);
\\
{\textbf{end}}
\\
\\
{\it{\%if\hspace{6pt}the\hspace{6pt}mass\hspace{6pt}spectrum\hspace{6pt}is\hspace{6pt}complete,}}\\
{\it{\%no\hspace{6pt}masses\hspace{6pt}with\hspace{6pt}0\hspace{6pt}counts\hspace{6pt}must\hspace{6pt}be\hspace{6pt}added:
}}\\
{\textbf{if}}\hspace{6pt}in(m,2)==m
\\
\hspace{24pt}go=0;
\\
\hspace{24pt}filled(:,1)=in(1:m,3);
\\
{\textbf{end}}
\\

\\
{\it{\%add\hspace{6pt}masses\hspace{6pt}with\hspace{6pt}0\hspace{6pt}counts:}}\\
{\it{\%loop\hspace{6pt}for\hspace{6pt}different\hspace{6pt}masses:
}}\\
{\textbf{while}}\hspace{6pt}go==1\\
\\
\hspace{24pt}{\it{\%check\hspace{6pt}if\hspace{6pt}mass\hspace{6pt}and\hspace{6pt}row\hspace{6pt}number\hspace{6pt}match}}\\
\hspace{24pt}{\it{\%and\hspace{6pt}add\hspace{6pt}the\hspace{6pt}missing\hspace{6pt}mass\hspace{6pt}if\hspace{6pt}they\hspace{6pt}do\hspace{6pt}not:
}}\\
\hspace{24pt}{\textbf{if}}\hspace{6pt}in(k\dash{}c,2)\~{}=k
\\
\hspace{48pt}filled(k,1)=0;\\
\hspace{48pt}{\it{\%c\hspace{6pt}accounts\hspace{6pt}for\hspace{6pt}the\hspace{6pt}shift\hspace{6pt}induced\hspace{6pt}by\hspace{6pt}adding\hspace{6pt}masses:
}}\\
\hspace{48pt}c=c+1;
\\
\hspace{24pt}{\textbf{else}}\hspace{6pt}
\\
\hspace{48pt}filled(k,1)=in(k\dash{}c,3);
\\
\hspace{24pt}{\textbf{end}}\\
\\
\hspace{24pt}{\it{\%stop\hspace{6pt}when\hspace{6pt}highest\hspace{6pt}mass\hspace{6pt}is\hspace{6pt}reached:
}}\\
\hspace{24pt}{\textbf{if}}\hspace{6pt}k==m
\\
\hspace{48pt}go=0;
\\
\hspace{24pt}{\textbf{end}}
\\
\hspace{24pt}k=k+1;
\\
{\textbf{end}}
\end{tabbing}}}

\subsection*{matrix.m}
{{\def\dash{\raise2.1pt\hbox{\rule{5pt}{0.3pt}}\hspace{1pt}}\begin{tabbing}
{\textbf{function}}\hspace{6pt}X\hspace{6pt}=\hspace{6pt}matrix(X,in)
\\
{\it{\%X\hspace{6pt}=\hspace{6pt}matrix(X,in)}}\\
{\it{\%adds\hspace{6pt}a\hspace{6pt}new\hspace{6pt}measurement\hspace{6pt}to\hspace{6pt}a\hspace{6pt}data\hspace{6pt}matrix}}\\
{\it{\%X:\hspace{6pt}(mxn)\dash{}matrix:\hspace{6pt}values\hspace{6pt}of\hspace{6pt}m\hspace{6pt}variables\hspace{6pt}in\hspace{6pt}n\hspace{6pt}measurements}}\\
{\it{\%in:\hspace{6pt}WinCadence\hspace{6pt}output:\hspace{6pt}unit\hspace{6pt}mass\hspace{6pt}mass\dash{}spectrum}}\\

\\
{[}m,n{]}=size(X);\\
\\
{\it{\%extracting\hspace{6pt}the\hspace{6pt}counts\hspace{6pt}from\hspace{6pt}WinCadence\hspace{6pt}output:
}}\\
in=voll(in,m);\\
\\
{\it{\%appending\hspace{6pt}the\hspace{6pt}new\hspace{6pt}data:
}}\\
X(:,n+1)=in(:,1);
\end{tabbing}}}

\subsection*{silanweg.m}
{{\def\dash{\raise2.1pt\hbox{\rule{5pt}{0.3pt}}\hspace{1pt}}\begin{tabbing}
{\textbf{function}}\hspace{6pt}output=silanweg(input,background);\\
{\it{\%output=silanweg(input,background);}}\\
{\it{\%substracts\hspace{6pt}a\hspace{6pt}background\hspace{6pt}spectrum\hspace{6pt}from\hspace{6pt}mass\hspace{6pt}spectra;}}\\
{\it{\%the\hspace{6pt}background\hspace{6pt}is\hspace{6pt}scaled\hspace{6pt}to\hspace{6pt}have\hspace{6pt}the\hspace{6pt}same\hspace{6pt}value\hspace{6pt}as}}\\
{\it{\%the\hspace{6pt}measurement\hspace{6pt}at\hspace{6pt}the\hspace{6pt}28th\hspace{6pt}variable\hspace{6pt}}}\\
{\it{\%input:\hspace{6pt}(mxn)\dash{}matrix:\hspace{6pt}values\hspace{6pt}of\hspace{6pt}m\hspace{6pt}variables\hspace{6pt}in\hspace{6pt}n\hspace{6pt}spectra}}\\
{\it{\%background:\hspace{6pt}(mx1)\dash{}vector:\hspace{6pt}background\hspace{6pt}values\hspace{6pt}of\hspace{6pt}m\hspace{6pt}variables\hspace{6pt}}}\\
\\
{[}m,n{]}=size(input);\\
{[}mm,nn{]}=size(background);\\
\\
{\textbf{if}}\hspace{6pt}mm\~{}=m\\
\hspace{24pt}fprintf(1,{\texttt{{'}Error:\hspace{6pt}Measurements\hspace{6pt}and\hspace{6pt}background\hspace{6pt}must\hspace{6pt}have}} \\
{\texttt{the\hspace{6pt}same\hspace{6pt}number\hspace{6pt}of\hspace{6pt}rows!{'}}})\\
\hspace{24pt}{\textbf{return}}\\
{\textbf{end}}\\
\\
{\it{\%loop\hspace{6pt}to\hspace{6pt}treat\hspace{6pt}different\hspace{6pt}measurements:}}\\
{\textbf{for}}\hspace{6pt}k=1:n\\
\hspace{24pt}factor=input(28,k)/background(28,1);\\
\hspace{24pt}output(:,k)=input(:,k)\dash{}factor$\ast$background(:,1);\\
{\textbf{end}}
\end{tabbing}}}

\subsection*{auswahl.m}
{{\def\dash{\raise2.1pt\hbox{\rule{5pt}{0.3pt}}\hspace{1pt}}\begin{tabbing}
{\textbf{function}}\hspace{6pt}out=auswahl(in)\\
{\it{\%out=auswahl(in)}}\\
{\it{\%selects\hspace{6pt}amino\hspace{6pt}acid\hspace{6pt}related\hspace{6pt}peaks\hspace{6pt}from\hspace{6pt}unit\hspace{6pt}mass}}\\
{\it{\%mass\hspace{6pt}spectra
}}\\
out(1,:)=in(30,:);\hspace{6pt}{\it{\%Gly
}}\\
out(2,:)=in(43,:);\hspace{6pt}{\it{\%Arg
}}\\
out(3,:)=in(44,:);\hspace{6pt}{\it{\%Ala
}}\\
out(4,:)=in(45,:);\hspace{6pt}{\it{\%Cys
}}\\
out(5,:)=in(60,:);\hspace{6pt}{\it{\%Ser
}}\\
out(6,:)=in(61,:);\hspace{6pt}{\it{\%Met
}}\\
out(7,:)=in(68,:);\hspace{6pt}{\it{\%Pro
}}\\
out(8,:)=in(69,:);\hspace{6pt}{\it{\%Thr
}}\\
out(9,:)=in(70,:);\hspace{6pt}{\it{\%Asn,\hspace{6pt}Pro
}}\\
out(10,:)=in(71,:);\hspace{6pt}{\it{\%Ser
}}\\
out(11,:)=in(72,:);\hspace{6pt}{\it{\%Val
}}\\
out(12,:)=in(73,:);\hspace{6pt}{\it{\%Arg
}}\\
out(13,:)=in(74,:);\hspace{6pt}{\it{\%Thr
}}\\
out(14,:)=in(81,:);\hspace{6pt}{\it{\%His
}}\\
out(15,:)=in(82,:);\hspace{6pt}{\it{\%His
}}\\
out(16,:)=in(83,:);\hspace{6pt}{\it{\%Val
}}\\
out(17,:)=in(84,:);\hspace{6pt}{\it{\%Gln,\hspace{6pt}Glu,\hspace{6pt}Lys
}}\\
out(18,:)=in(86,:);\hspace{6pt}{\it{\%Ile,\hspace{6pt}Leu
}}\\
out(19,:)=in(87,:);\hspace{6pt}{\it{\%Asn
}}\\
out(20,:)=in(88,:);\hspace{6pt}{\it{\%Asn,\hspace{6pt}Asp
}}\\
out(21,:)=in(98,:);\hspace{6pt}{\it{\%Asn
}}\\
out(22,:)=in(100,:);\hspace{6pt}{\it{\%Arg
}}\\
out(23,:)=in(101,:);\hspace{6pt}{\it{\%Arg
}}\\
out(24,:)=in(102,:);\hspace{6pt}{\it{\%Glu
}}\\
out(25,:)=in(107,:);\hspace{6pt}{\it{\%Tyr
}}\\
out(26,:)=in(110,:);\hspace{6pt}{\it{\%His
}}\\
out(27,:)=in(112,:);\hspace{6pt}{\it{\%Arg
}}\\
out(28,:)=in(120,:);\hspace{6pt}{\it{\%Phe
}}\\
out(29,:)=in(127,:);\hspace{6pt}{\it{\%Arg
}}\\
out(30,:)=in(130,:);\hspace{6pt}{\it{\%Trp
}}\\
out(31,:)=in(131,:);\hspace{6pt}{\it{\%Phe
}}\\
out(32,:)=in(136,:);\hspace{6pt}{\it{\%Tyr
}}\\
out(33,:)=in(159,:);\hspace{6pt}{\it{\%Trp
}}\\
out(34,:)=in(170,:);\hspace{6pt}{\it{\%Trp}}
\end{tabbing}}}

\subsection*{pca.m}
{{\def\dash{\raise2.1pt\hbox{\rule{5pt}{0.3pt}}\hspace{1pt}}\begin{tabbing}
{\textbf{function}}\hspace{6pt}{[}PC,\hspace{6pt}SCORE,\hspace{6pt}V,\hspace{6pt}mn{]}=pca(X);\\
{\it{\%{[}PC,\hspace{6pt}SCORE,\hspace{6pt}V,\hspace{6pt}mn{]}=pca(X)}}\\
{\it{\%performs\hspace{6pt}principal\hspace{6pt}component\hspace{6pt}analysis}}\\
{\it{\%X:\hspace{6pt}(mxn)\dash{}matrix:\hspace{6pt}values\hspace{6pt}of\hspace{6pt}m\hspace{6pt}variables\hspace{6pt}for\hspace{6pt}n\hspace{6pt}measurements}}\\
{\it{\%PC:\hspace{6pt}(mxm)\dash{}matrix:\hspace{6pt}loadings\hspace{6pt}of\hspace{6pt}m\hspace{6pt}PC\hspace{6pt}as\hspace{6pt}m\hspace{6pt}columns}}\\
{\it{\%SCORE:\hspace{6pt}(mxn)\dash{}matrix:\hspace{6pt}scores\hspace{6pt}of\hspace{6pt}m\hspace{6pt}PC\hspace{6pt}for\hspace{6pt}n\hspace{6pt}measurements}}\\
{\it{\%V:\hspace{6pt}(m)\dash{}vector:\hspace{6pt}variance\hspace{6pt}captured\hspace{6pt}by\hspace{6pt}the\hspace{6pt}m\hspace{6pt}PC}}\\
{\it{\%mn:\hspace{6pt}(m)\dash{}vector:\hspace{6pt}mean\hspace{6pt}values\hspace{6pt}of\hspace{6pt}the\hspace{6pt}m\hspace{6pt}variables}}\\
{\it{\%for\hspace{6pt}details\hspace{6pt}on\hspace{6pt}the\hspace{6pt}PCA\hspace{6pt}algorithm\hspace{6pt}see\hspace{6pt}J.\hspace{6pt}Shlens,\hspace{6pt}}}\\
{\it{\%A\hspace{6pt}tutorial\hspace{6pt}on\hspace{6pt}principal\hspace{6pt}component\hspace{6pt}analysis,\hspace{6pt}10/12/2005}}\\
\\
{[}m,n{]}=size(X);\\
\\
{\it{\%scaling\hspace{6pt}by\hspace{6pt}total\hspace{6pt}intensity:}}\\
{\textbf{for}}\hspace{6pt}k=1:n\\
\hspace{24pt}X(:,k)=X(:,k)/sum(X(:,k));\\
{\textbf{end}}\\
\\
{\it{\%mean\dash{}centering:}}\\
mn=mean(X,2);\\
X=X\dash{}repmat(mn,1,n);\\
\\
{\it{\%auxiliary\hspace{6pt}matrix\hspace{6pt}for\hspace{6pt}SVD:}}\\
Y=X{'}/sqrt(n\dash{}1);\\
\\
{\it{\%singular\hspace{6pt}value\hspace{6pt}decomposition:}}\\
{[}u,S,PC{]}=svd(Y);\\
\\
{\it{\%variances\hspace{6pt}\dash{}$>$\hspace{6pt}m\dash{}vector\hspace{6pt}V:}}\\
S=diag(S);\\
V=S.$\ast$S;\\
\\
{\it{\%projection\hspace{6pt}\dash{}$>$\hspace{6pt}(mxn)\dash{}matrix\hspace{6pt}SCORE:}}\\
SCORE=PC{'}$\ast$X;\\
\\
{\it{\%saving\hspace{6pt}loadings\hspace{6pt}and\hspace{6pt}scores:}}\\
save\hspace{6pt}ploading.dat\hspace{6pt}PC\hspace{6pt}\dash{}ascii;\\
save\hspace{6pt}pscore.dat\hspace{6pt}SCORE\hspace{6pt}\dash{}ascii;
\end{tabbing}}}

\subsection*{dpca.m}
{{\def\dash{\raise2.1pt\hbox{\rule{5pt}{0.3pt}}\hspace{1pt}}\begin{tabbing}
{\textbf{function}}\hspace{6pt}{[}loading,score,variance,varianceintra{]}=dpca(x,anz,rep)\\
{\it{\%{[}loading,score,relvariance,varianceintra{]}=dpca(x,anz,rep)}}\\
{\it{\%performs\hspace{6pt}discriminant\hspace{6pt}principal\hspace{6pt}component\hspace{6pt}analysis}}\\
{\it{\%x:\hspace{6pt}(mxn)\dash{}matrix:\hspace{6pt}values\hspace{6pt}of\hspace{6pt}m\hspace{6pt}variables\hspace{6pt}for\hspace{6pt}n\hspace{6pt}measurements}}\\
{\it{\%anz:\hspace{6pt}number\hspace{6pt}of\hspace{6pt}groups}}\\
{\it{\%rep:\hspace{6pt}(1xanz)\dash{}vector:\hspace{6pt}number\hspace{6pt}of\hspace{6pt}measurements\hspace{6pt}per\hspace{6pt}group}}\\
{\it{\%loading:\hspace{6pt}(mxm)\dash{}matrix:\hspace{6pt}loadings\hspace{6pt}of\hspace{6pt}m\hspace{6pt}DPC\hspace{6pt}as\hspace{6pt}m\hspace{6pt}columns}}\\
{\it{\%score:\hspace{6pt}(mxn)\dash{}matrix:\hspace{6pt}scores\hspace{6pt}of\hspace{6pt}m\hspace{6pt}DPC\hspace{6pt}for\hspace{6pt}n\hspace{6pt}measurements}}\\
{\it{\%variance:\hspace{6pt}(1xm)\dash{}vector:\hspace{6pt}variance\hspace{6pt}captured\hspace{6pt}by\hspace{6pt}the\hspace{6pt}m\hspace{6pt}DPC}}\\
{\it{\%varianceintra:\hspace{6pt}(mx1)\dash{}vector:\hspace{6pt}pooled\hspace{6pt}intra\hspace{6pt}group\hspace{6pt}variances\hspace{6pt}of\hspace{6pt}m\hspace{6pt}variables}}\\
{\it{\%for\hspace{6pt}details\hspace{6pt}on\hspace{6pt}the\hspace{6pt}PCA\hspace{6pt}algorithm\hspace{6pt}see\hspace{6pt}J.\hspace{6pt}Shlens,\hspace{6pt}}}\\
{\it{\%A\hspace{6pt}tutorial\hspace{6pt}on\hspace{6pt}principal\hspace{6pt}component\hspace{6pt}analysis,\hspace{6pt}10/12/2005}}\\
{\it{\%for\hspace{6pt}details\hspace{6pt}on\hspace{6pt}DPCA\hspace{6pt}see\hspace{6pt}P.W.\hspace{6pt}Yendle,\hspace{6pt}H.J.M.\hspace{6pt}Macfie,}}\\
{\it{\%Discriminant\hspace{6pt}principal\hspace{6pt}components\hspace{6pt}analysis,}}\\
{\it{\%J.\hspace{6pt}of\hspace{6pt}Chemometrics\hspace{6pt}3,\hspace{6pt}589\dash{}600,\hspace{6pt}1989
}}\\

\\
{[}m,n{]}=size(x);
\\
{[}sizerepx,sizerepy{]}=size(rep);
\\
{\textbf{if}}\hspace{6pt}sizerepy==1
\\
\hspace{24pt}rep=repmat(rep,1,anz);
\\
{\textbf{end}}
\\

\\
{\it{\%scaling\hspace{6pt}by\hspace{6pt}total\hspace{6pt}intensity:
}}\\
{\textbf{for}}\hspace{6pt}k=1:n
\\
\hspace{24pt}x(:,k)=x(:,k)/sum(x(:,k));
\\
{\textbf{end}}
\\
\\
{\it{\%pooled\hspace{6pt}intra\hspace{6pt}group\hspace{6pt}variances\hspace{6pt}of\hspace{6pt}the\hspace{6pt}variables:
}}\\
{\it{\%\dash{}$>$\hspace{6pt}varianceintra:\hspace{6pt}(mx1)
}}\\
varianceintra=zeros(m,1);
\\
{\textbf{for}}\hspace{6pt}k=1:m
\\
\hspace{24pt}start=1;
\\
\hspace{24pt}stop=0;
\\
\hspace{24pt}{\textbf{for}}\hspace{6pt}o=1:anz
\\
\hspace{48pt}stop=stop+rep(o);
\\
\hspace{48pt}varianceintra(k)=varianceintra(k)+var(x(k,start:stop));
\\
\hspace{48pt}start=start+rep(o);
\\
\hspace{24pt}{\textbf{end}}
\\
{\textbf{end}}
\\
\\
{\it{\%scaling\hspace{6pt}by\hspace{6pt}intra\hspace{6pt}group\hspace{6pt}standard\hspace{6pt}deviations:}}\\
{\it{\%\dash{}$>$\hspace{6pt}scaled\hspace{6pt}data\hspace{6pt}matrix\hspace{6pt}s
}}\\
{\textbf{for}}\hspace{6pt}k=1:m
\\
\hspace{24pt}s(k,:)=x(k,:)/sqrt(varianceintra(k));
\\
{\textbf{end}}
\\
\\
{\it{\%scaled\hspace{6pt}group\hspace{6pt}means:}}\\
{\it{\%\dash{}$>$\hspace{6pt}(mxanz)\dash{}matrix\hspace{6pt}sbar
}}\\
{\textbf{for}}\hspace{6pt}k=1:m
\\
\hspace{24pt}start=1;
\\
\hspace{24pt}stop=0;
\\
\hspace{24pt}{\textbf{for}}\hspace{6pt}o=1:anz
\\
\hspace{48pt}stop=stop+rep(o);
\\
\hspace{48pt}sbar(k,o)=mean(s(k,start:stop));
\\
\hspace{48pt}start=start+rep(o);
\\
\hspace{24pt}{\textbf{end}}
\\
{\textbf{end}}
\\
\\
{\it{\%(mxn)\dash{}matrix\hspace{6pt}of\hspace{6pt}scaled\hspace{6pt}group\hspace{6pt}means:
}}\\
groupmeans=repmat(sbar(:,1),1,rep(1));
\\
{\textbf{for}}\hspace{6pt}o=2:anz
\\
\hspace{24pt}groupmeans={[}groupmeans\hspace{6pt}repmat(sbar(:,o),1,rep(o)){]};
\\
{\textbf{end}}
\\
\\
{\it{\%PCA\hspace{6pt}with\hspace{6pt}SVD\hspace{6pt}\dash{}$>$\hspace{6pt}(mxm)\dash{}matrix\hspace{6pt}of\hspace{6pt}loadings\hspace{6pt}loading:}}\\
\\
{\it{\%mean\dash{}centering:
}}\\
mn=mean(groupmeans,2);
\\
groupmeans=groupmeans\dash{}repmat(mn,1,n);
\\
helpmatrix=groupmeans{'}/sqrt(n\dash{}1);
\\
{[}u,singularvalues,loading{]}=svd(helpmatrix);
\\

\\
{\it{\%projection\hspace{6pt}\dash{}$>$\hspace{6pt}(mxn)\dash{}matrix\hspace{6pt}of\hspace{6pt}scores\hspace{6pt}score
}}\\
score=loading{'}$\ast$s;
\\

\\
{\it{\%variances\hspace{6pt}of\hspace{6pt}the\hspace{6pt}DPC:}}\\
{\it{\%\dash{}$>$\hspace{6pt}m\dash{}vector\hspace{6pt}variance
}}\\
{\textbf{for}}\hspace{6pt}k=1:m
\\
\hspace{24pt}variance(k)=var(score(k,:));
\\
{\textbf{end}}
\\

\\
{\it{\%saving\hspace{6pt}loadings\hspace{6pt}and\hspace{6pt}scores:
}}\\
save\hspace{6pt}dloading.dat\hspace{6pt}loading\hspace{6pt}\dash{}ascii;
\\
save\hspace{6pt}dscore.dat\hspace{6pt}score\hspace{6pt}\dash{}ascii;
\end{tabbing}}}

\subsection*{plotpc2d.m}
{{\def\dash{\raise2.1pt\hbox{\rule{5pt}{0.3pt}}\hspace{1pt}}\begin{tabbing}
{\textbf{function}}\hspace{6pt}plotpc2d(PC,SCORE,relvariance,anz,rep,omax)\\
{\it{\%plotpc2d(PC,SCORE,relvariance,anz,rep,omax)}}\\
{\it{\%creates\hspace{6pt}2D\hspace{6pt}scores\hspace{6pt}plots\hspace{6pt}and\hspace{6pt}loadings\hspace{6pt}plots}}\\
{\it{\%PC:\hspace{6pt}(mxm)\dash{}matrix:\hspace{6pt}loadings\hspace{6pt}of\hspace{6pt}m\hspace{6pt}PC\hspace{6pt}as\hspace{6pt}m\hspace{6pt}columns}}\\
{\it{\%SCORE:\hspace{6pt}(mxn)\dash{}matrix:\hspace{6pt}scores\hspace{6pt}of\hspace{6pt}m\hspace{6pt}PC\hspace{6pt}for\hspace{6pt}n\hspace{6pt}measurements}}\\
{\it{\%relvariance:\hspace{6pt}(m)\dash{}vector:\hspace{6pt}relative\hspace{6pt}variance\hspace{6pt}captured\hspace{6pt}by\hspace{6pt}the\hspace{6pt}m\hspace{6pt}PC}}\\
{\it{\%anz:\hspace{6pt}number\hspace{6pt}of\hspace{6pt}groups}}\\
{\it{\%rep:\hspace{6pt}(anz)\dash{}vector:\hspace{6pt}number\hspace{6pt}of\hspace{6pt}measurements\hspace{6pt}per\hspace{6pt}group}}\\
{\it{\%omax:\hspace{6pt}highest\hspace{6pt}dimension\hspace{6pt}plotted\hspace{6pt}}}\\
\\
{\textbf{if}}\hspace{6pt}omax$<$2\\
\hspace{24pt}fprintf(1,{\texttt{{'}Error:\hspace{6pt}Maximum\hspace{6pt}principal\hspace{6pt}component\hspace{6pt}must\hspace{6pt}be\hspace{6pt}grater\hspace{6pt}1!{'}}})\\
\hspace{24pt}{\textbf{return}}\\
{\textbf{end}}\\
\\
{\it{\%scaling\hspace{6pt}variances\hspace{6pt}if\hspace{6pt}necessary:}}\\
{\textbf{if}}\hspace{6pt}sum(relvariance)\~{}=100\\
\hspace{24pt}relvariance=relvariance/sum(relvariance)$\ast$100;\\
{\textbf{end}}\\
\\
{[}sizerepx,sizerepy{]}=size(rep);\\
{\textbf{if}}\hspace{6pt}sizerepy==1\\
\hspace{24pt}rep=repmat(rep,1,anz);\\
{\textbf{end}}\\
\\
{\it{\%loop\hspace{6pt}for\hspace{6pt}different\hspace{6pt}plots:}}\\
{\textbf{for}}\hspace{6pt}o=2:omax\\
\\
figure\\
\\
{\it{\%loadings\hspace{6pt}plot:}}\\
subplot(2,1,1)\\
plot(PC(:,o\dash{}1),PC(:,o),{\texttt{{'}kx{'}}});\\
grid\hspace{6pt}on\\
title({\texttt{{'}Loadings\hspace{6pt}Plot{'}}});\\
xlabel({[}{\texttt{{'}PC\hspace{6pt}{'}}}\hspace{6pt}int2str(o\dash{}1){]});\\
ylabel({[}{\texttt{{'}PC\hspace{6pt}{'}}}\hspace{6pt}int2str(o){]});\\
\\
subplot(2,1,2)\\
\\
{\it{\%scores\hspace{6pt}plots:}}\\
hold\hspace{6pt}on\\
min=1;\\
max=0;\\
\\
{\it{\%loop\hspace{6pt}for\hspace{6pt}different\hspace{6pt}groups:}}\\
{\textbf{for}}\hspace{6pt}k=1:anz\\
\hspace{24pt}max=max+rep(k);\\
\hspace{24pt}{\textbf{switch}}\hspace{6pt}k\\
\hspace{48pt}{\textbf{case}}\hspace{6pt}1\\
\hspace{72pt}plot(SCORE(o\dash{}1,min:max),SCORE(o,min:max),{\texttt{{'}ro{'}}});\\
\hspace{48pt}{\textbf{case}}\hspace{6pt}2\\
\hspace{72pt}plot(SCORE(o\dash{}1,min:max),SCORE(o,min:max),{\texttt{{'}bx{'}}});\\
\hspace{48pt}{\textbf{case}}\hspace{6pt}3\\
\hspace{72pt}plot(SCORE(o\dash{}1,min:max),SCORE(o,min:max),{\texttt{{'}g+{'}}});\\
\hspace{48pt}{\textbf{case}}\hspace{6pt}4\\
\hspace{72pt}plot(SCORE(o\dash{}1,min:max),SCORE(o,min:max),{\texttt{{'}c$\ast${'}}});\\
\hspace{48pt}{\textbf{case}}\hspace{6pt}5\\
\hspace{72pt}plot(SCORE(o\dash{}1,min:max),SCORE(o,min:max),{\texttt{{'}ms{'}}});;\\
\hspace{48pt}{\textbf{case}}\hspace{6pt}6\\
\hspace{72pt}plot(SCORE(o\dash{}1,min:max),SCORE(o,min:max),{\texttt{{'}yd{'}}});\\
\hspace{48pt}{\textbf{otherwise}}\\
\hspace{66pt}plot(SCORE(o\dash{}1,min:max),SCORE(o,min:max),{\texttt{{'}k\^{}{'}}});\\
\hspace{66pt}fprintf(1,{\texttt{{'}Cannot\hspace{6pt}plot\hspace{6pt}more\hspace{6pt}than\hspace{6pt}6\hspace{6pt}groups!{'}}});\\
\hspace{24pt}{\textbf{end}}\\
\hspace{24pt}min=min+rep(k);\\
{\textbf{end}}\\
grid\hspace{6pt}on\\
title({\texttt{{'}Scores\hspace{6pt}Plot{'}}});\\
xlabel({[}{\texttt{{'}PC\hspace{6pt}{'}}}\hspace{6pt}int2str(o\dash{}1)\hspace{6pt}{\texttt{{'}\hspace{6pt}({'}}}\hspace{6pt}int2str(relvariance(o\dash{}1))\hspace{6pt}{\texttt{{'}\%){'}}}{]});\\
ylabel({[}{\texttt{{'}PC\hspace{6pt}{'}}}\hspace{6pt}int2str(o)\hspace{6pt}{\texttt{{'}\hspace{6pt}({'}}}\hspace{6pt}int2str(relvariance(o))\hspace{6pt}{\texttt{{'}\%){'}}}{]});\\
\\
{\textbf{end}}
\end{tabbing}}}

\subsection*{plotdpc2dmitellipse2.m}
{{\def\dash{\raise2.1pt\hbox{\rule{5pt}{0.3pt}}\hspace{1pt}}\begin{tabbing}
{\textbf{function}}\hspace{6pt}plotdpc2dmitellipse2(PC,SCORE,relvariance,anz,rep,omax)\\
{\it{\%plotdpc2dmitellipse2(PC,SCORE,relvariance,anz,rep,omax)}}\\
{\it{\%creates\hspace{6pt}2D\hspace{6pt}scores\hspace{6pt}plots\hspace{6pt}with\hspace{6pt}probability\hspace{6pt}ellipses\hspace{6pt}and\hspace{6pt}loadings\hspace{6pt}plots\hspace{6pt}}}\\
{\it{\%PC:\hspace{6pt}(mxm)\dash{}matrix:\hspace{6pt}loadings\hspace{6pt}of\hspace{6pt}m\hspace{6pt}DPC\hspace{6pt}as\hspace{6pt}m\hspace{6pt}columns}}\\
{\it{\%SCORE:\hspace{6pt}(mxn)\dash{}matrix:\hspace{6pt}scores\hspace{6pt}of\hspace{6pt}m\hspace{6pt}DPC\hspace{6pt}for\hspace{6pt}n\hspace{6pt}measurements}}\\
{\it{\%relvariance:\hspace{6pt}(m)\dash{}vector:\hspace{6pt}relative\hspace{6pt}variance\hspace{6pt}captured\hspace{6pt}by\hspace{6pt}the\hspace{6pt}m\hspace{6pt}DPC}}\\
{\it{\%anz:\hspace{6pt}number\hspace{6pt}of\hspace{6pt}groups}}\\
{\it{\%rep:\hspace{6pt}(anz)\dash{}vector:\hspace{6pt}number\hspace{6pt}of\hspace{6pt}measurements\hspace{6pt}per\hspace{6pt}group}}\\
{\it{\%omax:\hspace{6pt}highest\hspace{6pt}dimension\hspace{6pt}plotted\hspace{6pt}}}\\
\\
{\textbf{if}}\hspace{6pt}omax$<$2\\
\hspace{24pt}fprintf(1,{\texttt{{'}Error:\hspace{6pt}Maximum\hspace{6pt}principal\hspace{6pt}component\hspace{6pt}must\hspace{6pt}be\hspace{6pt}greater\hspace{6pt}one!{'}}})\\
\hspace{24pt}{\textbf{return}}\\
{\textbf{end}}\\
\\
{\it{\%scaling\hspace{6pt}variances:}}\\
{\textbf{if}}\hspace{6pt}sum(relvariance)\~{}=100\\
\hspace{24pt}relvariance=relvariance/sum(relvariance)$\ast$100;\\
{\textbf{end}}\\
\\
{[}sizerepx,sizerepy{]}=size(rep);\\
{\textbf{if}}\hspace{6pt}sizerepy==1\\
\hspace{24pt}rep=repmat(rep,1,anz);\\
{\textbf{end}}\\
\\
{\it{\%loop\hspace{6pt}to\hspace{6pt}create\hspace{6pt}more\hspace{6pt}than\hspace{6pt}one\hspace{6pt}plot:}}\\
{\textbf{for}}\hspace{6pt}o=2:omax\\
\\
figure\\
\\
{\it{\%loadings\hspace{6pt}plot:}}\\
subplot(2,1,1)\\
plot(PC(:,o\dash{}1),PC(:,o),{\texttt{{'}kx{'}}});\\
grid\hspace{6pt}on\\
title({\texttt{{'}Loadings\hspace{6pt}Plot{'}}});\\
xlabel({[}{\texttt{{'}DPC\hspace{6pt}{'}}}\hspace{6pt}int2str(o\dash{}1){]});\\
ylabel({[}{\texttt{{'}DPC\hspace{6pt}{'}}}\hspace{6pt}int2str(o){]});\\
\\
{\it{\%scores\hspace{6pt}plot\hspace{6pt}with\hspace{6pt}different\hspace{6pt}styles\hspace{6pt}for\hspace{6pt}different\hspace{6pt}groups:}}\\
subplot(2,1,2)\\
hold\hspace{6pt}on\\
min=1;\\
max=0;\\
\\
{\it{\%loop\hspace{6pt}for\hspace{6pt}different\hspace{6pt}groups:}}\\
{\textbf{for}}\hspace{6pt}k=1:anz\\
\\
\hspace{24pt}{\it{\%critical\hspace{6pt}value\hspace{6pt}of\hspace{6pt}F\dash{}distribution\hspace{6pt}for\hspace{6pt}2\hspace{6pt}variables,}}\\
\hspace{24pt}{\it{\%rep(k)\hspace{6pt}measurements\hspace{6pt}and\hspace{6pt}95\%\hspace{6pt}confidence\hspace{6pt}limit:}}\\
\hspace{24pt}F=Fdistribution(rep(k));\\
\\
\hspace{24pt}{\it{\%calculate\hspace{6pt}critical\hspace{6pt}TSquare\hspace{6pt}value\hspace{6pt}for\hspace{6pt}the\hspace{6pt}ellipses:}}\\
\hspace{24pt}Tsquare=2$\ast$(rep(k)\dash{}1)/(rep(k)\dash{}2)$\ast$F;\\
\\
\hspace{24pt}max=max+rep(k);\\
\hspace{24pt}{\textbf{switch}}\hspace{6pt}k\\
\hspace{48pt}{\textbf{case}}\hspace{6pt}1\\
\hspace{72pt}plot(SCORE(o\dash{}1,min:max),SCORE(o,min:max),{\texttt{{'}ro{'}}});\\
\hspace{72pt}ellipse(SCORE(o\dash{}1:o,min:max),Tsquare,{\texttt{{'}r{'}}});\\
\hspace{48pt}{\textbf{case}}\hspace{6pt}2\\
\hspace{72pt}plot(SCORE(o\dash{}1,min:max),SCORE(o,min:max),{\texttt{{'}bx{'}}});\\
\hspace{72pt}ellipse(SCORE(o\dash{}1:o,min:max),Tsquare,{\texttt{{'}b{'}}});\\
\hspace{48pt}{\textbf{case}}\hspace{6pt}3\\
\hspace{72pt}plot(SCORE(o\dash{}1,min:max),SCORE(o,min:max),{\texttt{{'}g+{'}}});\\
\hspace{72pt}ellipse(SCORE(o\dash{}1:o,min:max),Tsquare,{\texttt{{'}g{'}}});\\
\hspace{48pt}{\textbf{case}}\hspace{6pt}4\\
\hspace{72pt}plot(SCORE(o\dash{}1,min:max),SCORE(o,min:max),{\texttt{{'}c$\ast${'}}});\\
\hspace{72pt}ellipse(SCORE(o\dash{}1:o,min:max),Tsquare,{\texttt{{'}c{'}}});\\
\hspace{48pt}{\textbf{case}}\hspace{6pt}5\\
\hspace{72pt}plot(SCORE(o\dash{}1,min:max),SCORE(o,min:max),{\texttt{{'}ms{'}}});\\
\hspace{72pt}ellipse(SCORE(o\dash{}1:o,min:max),Tsquare,{\texttt{{'}m{'}}});\\
\hspace{48pt}{\textbf{case}}\hspace{6pt}6\\
\hspace{72pt}plot(SCORE(o\dash{}1,min:max),SCORE(o,min:max),{\texttt{{'}yd{'}}});\\
\hspace{72pt}ellipse(SCORE(o\dash{}1:o,min:max),Tsquare,{\texttt{{'}y{'}}});\\
\hspace{48pt}{\textbf{otherwise}}\\
\hspace{72pt}plot(SCORE(o\dash{}1,min:max),SCORE(o,min:max),{\texttt{{'}k\^{}{'}}});\\
\hspace{72pt}fprintf(1,{\texttt{{'}Cannot\hspace{6pt}plot\hspace{6pt}more\hspace{6pt}than\hspace{6pt}six\hspace{6pt}groups!{'}}});\\
\hspace{24pt}{\textbf{end}}\\
\hspace{24pt}min=min+rep(k);\\
{\textbf{end}}\\
grid\hspace{6pt}on\\
title({\texttt{{'}Scores\hspace{6pt}Plot{'}}});\\
xlabel({[}{\texttt{{'}DPC\hspace{6pt}{'}}}\hspace{6pt}int2str(o\dash{}1)\hspace{6pt}{\texttt{{'}\hspace{6pt}({'}}}\hspace{6pt}int2str(relvariance(o\dash{}1))\hspace{6pt}{\texttt{{'}\%){'}}}{]});\\
ylabel({[}{\texttt{{'}DPC\hspace{6pt}{'}}}\hspace{6pt}int2str(o)\hspace{6pt}{\texttt{{'}\hspace{6pt}({'}}}\hspace{6pt}int2str(relvariance(o))\hspace{6pt}{\texttt{{'}\%){'}}}{]});\\
\\
{\textbf{end}}
\end{tabbing}}}

\subsection*{ellipse.m}
{{\def\dash{\raise2.1pt\hbox{\rule{5pt}{0.3pt}}\hspace{1pt}}\begin{tabbing}
{\textbf{function}}\hspace{6pt}ellipse(X,Tsquare,color)\\
{\it{\%ellipse(X,Tsquare,color)}}\\
{\it{\%plots\hspace{6pt}a\hspace{6pt}probability\hspace{6pt}ellipse\hspace{6pt}around\hspace{6pt}data\hspace{6pt}points}}\\
{\it{\%X:\hspace{6pt}(2xn)\dash{}matrix\hspace{6pt}containing\hspace{6pt}n\hspace{6pt}measurements\hspace{6pt}of\hspace{6pt}2\hspace{6pt}variables}}\\
{\it{\%Tsquare:\hspace{6pt}critical\hspace{6pt}value\hspace{6pt}of\hspace{6pt}the\hspace{6pt}TSquare\hspace{6pt}distribution}}\\
{\it{\%color:\hspace{6pt}colour\hspace{6pt}(r,b,g,c,m,y\hspace{6pt}or\hspace{6pt}k)}}\\
{\it{\%for\hspace{6pt}details\hspace{6pt}on\hspace{6pt}the\hspace{6pt}algorithm\hspace{6pt}see\hspace{6pt}J.E.\hspace{6pt}Jackson,\hspace{6pt}A\hspace{6pt}user{'}s\hspace{6pt}guide\hspace{6pt}}}\\
{\it{\%to\hspace{6pt}principal\hspace{6pt}components,\hspace{6pt}John\hspace{6pt}Wiley\hspace{6pt}and\hspace{6pt}sons\hspace{6pt}inc.,\hspace{6pt}New\hspace{6pt}York,\hspace{6pt}1991
}}\\

\\
{[}m,n{]}=size(X);
\\
{\textbf{if}}\hspace{6pt}m\~{}=2\hspace{6pt}
\\
\hspace{24pt}fprintf(1,{\texttt{{'}\hspace{6pt}$\backslash$n\hspace{6pt}Error:\hspace{6pt}Data\hspace{6pt}matrix\hspace{6pt}must\hspace{6pt}be\hspace{6pt}2\hspace{6pt}x\hspace{6pt}n!\hspace{6pt}$\backslash$n{'}}});
\\
\hspace{24pt}{\textbf{return}}
\\
{\textbf{end}}
\\

\\
{\it{\%Centre\hspace{6pt}of\hspace{6pt}the\hspace{6pt}ellipse:
}}\\
mitte=mean(X,2);
\\

\\
{\it{\%Mean\dash{}centered\hspace{6pt}auxiliary\hspace{6pt}matrix:
}}\\
Y=X\dash{}repmat(mitte,1,n);
\\
Y=Y{'}/sqrt(n\dash{}1);
\\
\\
{\it{\%singular\hspace{6pt}value\hspace{6pt}decomposition\hspace{6pt}to\hspace{6pt}calculate\hspace{6pt}}}\\
{\it{\%principal\hspace{6pt}components\hspace{6pt}and\hspace{6pt}standard\hspace{6pt}deviations:
}}\\
{[}u,S,PC{]}=svd(Y);
\\

\\
{\it{\%standard\hspace{6pt}deviations:
}}\\
S=diag(S);
\\

\\
{\it{\%rescaling\hspace{6pt}of\hspace{6pt}principal\hspace{6pt}components
}}\\
PC(:,1)=S(1)$\ast$PC(:,1);
\\
PC(:,2)=S(2)$\ast$PC(:,2);
\\

\\
kmax=sqrt(Tsquare);
\\
hold\hspace{6pt}on\\

\\
{\it{\%calculation\hspace{6pt}and\hspace{6pt}plotting\hspace{6pt}of\hspace{6pt}the\hspace{6pt}ellipse;}}\\
{\it{\%number\hspace{6pt}of\hspace{6pt}points\hspace{6pt}can\hspace{6pt}be\hspace{6pt}controlled}}\\
{\it{\%by\hspace{6pt}the\hspace{6pt}stepwidth\hspace{6pt}of\hspace{6pt}the\hspace{6pt}loop:
}}\\
{\textbf{for}}\hspace{6pt}k=0:0.04:kmax
\\
\hspace{24pt}g=k;
\\
\hspace{24pt}h=sqrt(Tsquare\dash{}k\^{}2);
\\
\hspace{24pt}x=mitte+g$\ast$PC(:,1)+h$\ast$PC(:,2);
\\
\hspace{24pt}plot(x(1),x(2),color)
\\
\hspace{24pt}x=mitte+g$\ast$PC(:,1)\dash{}h$\ast$PC(:,2);
\\
\hspace{24pt}plot(x(1),x(2),color)
\\
\hspace{24pt}x=mitte\dash{}g$\ast$PC(:,1)+h$\ast$PC(:,2);
\\
\hspace{24pt}plot(x(1),x(2),color)
\\
\hspace{24pt}x=mitte\dash{}g$\ast$PC(:,1)\dash{}h$\ast$PC(:,2);
\\
\hspace{24pt}plot(x(1),x(2),color)\hspace{36pt}
\\
{\textbf{end}}
\\

\\
g=sqrt(Tsquare);
\\
h=0;
\\
x=mitte\dash{}g$\ast$PC(:,1);
\\
plot(x(1),x(2),color)\hspace{6pt}
\\
x=mitte+g$\ast$PC(:,1);
\\
plot(x(1),x(2),color)\hspace{6pt}
\end{tabbing}}}

\subsection*{Fdistribution.m}
{{\def\dash{\raise2.1pt\hbox{\rule{5pt}{0.3pt}}\hspace{1pt}}\begin{tabbing}
{\textbf{function}}\hspace{6pt}ausgabe=Fdistribution(n)
\\
{\it{\%ausgabe=Fdistribution(n)}}\\
{\it{\%finds\hspace{6pt}the\hspace{6pt}upper\hspace{6pt}limit\hspace{6pt}of\hspace{6pt}the\hspace{6pt}95\%\hspace{6pt}confidence\hspace{6pt}interval}}\\
{\it{\%of\hspace{6pt}an\hspace{6pt}F\hspace{6pt}distribution\hspace{6pt}for\hspace{6pt}2\hspace{6pt}variables\hspace{6pt}and\hspace{6pt}n\hspace{6pt}measurements;
}}\\
{\it{\%values\hspace{6pt}are\hspace{6pt}from\hspace{6pt}the\hspace{6pt}table\hspace{6pt}in\hspace{6pt}J.E.\hspace{6pt}Jackson,\hspace{6pt}A\hspace{6pt}user{'}s\hspace{6pt}guide\hspace{6pt}}}\\
{\it{\%to\hspace{6pt}principal\hspace{6pt}components,\hspace{6pt}John\hspace{6pt}Wiley\hspace{6pt}and\hspace{6pt}sons,\hspace{6pt}New\hspace{6pt}York,1991;}}\\
{\it{\%intermedial\hspace{6pt}values\hspace{6pt}linearly\hspace{6pt}approximated
}}\\

\\
{\it{\%tabulated\hspace{6pt}values:
}}\\
F=\hspace{6pt}{[}1.9950000e+002
\\
\hspace{12pt}1.9000000e+001
\\
\hspace{12pt}9.5500000e+000
\\
\hspace{12pt}6.9400000e+000
\\
\hspace{12pt}5.7900000e+000
\\
\hspace{12pt}5.1400000e+000
\\
\hspace{12pt}4.7400000e+000
\\
\hspace{12pt}4.4600000e+000
\\
\hspace{12pt}4.2600000e+000
\\
\hspace{12pt}4.1000000e+000
\\
\hspace{12pt}3.9800000e+000
\\
\hspace{12pt}3.8900000e+000
\\
\hspace{12pt}3.8100000e+000
\\
\hspace{12pt}3.7400000e+000
\\
\hspace{12pt}3.6800000e+000
\\
\hspace{12pt}3.6300000e+000
\\
\hspace{12pt}3.5900000e+000
\\
\hspace{12pt}3.5500000e+000
\\
\hspace{12pt}3.5200000e+000
\\
\hspace{12pt}3.4900000e+000
\\
\hspace{12pt}3.4700000e+000
\\
\hspace{12pt}3.4400000e+000
\\
\hspace{12pt}3.4200000e+000
\\
\hspace{12pt}3.4000000e+000
\\
\hspace{12pt}3.3900000e+000
\\
\hspace{12pt}3.3700000e+000
\\
\hspace{12pt}3.3500000e+000
\\
\hspace{12pt}3.3400000e+000
\\
\hspace{12pt}3.3300000e+000
\\
\hspace{12pt}3.3200000e+000
\\
\hspace{12pt}3.2300000e+000
\\
\hspace{12pt}3.1500000e+000
\\
\hspace{12pt}3.0700000e+000
\\
\hspace{12pt}3.0000000e+000{]};
\\
\\
{\it{\%to\hspace{6pt}find\hspace{6pt}the\hspace{6pt}right\hspace{6pt}value\hspace{6pt}the\hspace{6pt}number\hspace{6pt}of\hspace{6pt}measurements\hspace{6pt}has}}\\
{\it{\%to\hspace{6pt}be\hspace{6pt}reduced\hspace{6pt}by\hspace{6pt}the\hspace{6pt}number\hspace{6pt}of\hspace{6pt}variables:
}}\\
n=n\dash{}2;
\\

\\
{\textbf{if}}\hspace{6pt}n\hspace{6pt}$<$1
\\
\hspace{36pt}disp({\texttt{{'}Error:\hspace{6pt}Number\hspace{6pt}of\hspace{6pt}measurements\hspace{6pt}must\hspace{6pt}be\hspace{6pt}greater\hspace{6pt}2!{'}}})
\\
\hspace{36pt}
\\
\hspace{12pt}{\textbf{elseif}}\hspace{6pt}n$<$31
\\
\hspace{48pt}
\\
\hspace{48pt}{\it{\%the\hspace{6pt}first\hspace{6pt}30\hspace{6pt}values\hspace{6pt}correspond\hspace{6pt}to\hspace{6pt}1\hspace{6pt}to\hspace{6pt}30\hspace{6pt}measurements:
}}\\
\hspace{48pt}ausgabe=F(n);
\\
\hspace{48pt}
\\
\hspace{24pt}{\textbf{elseif}}\hspace{6pt}n$<$40
\\
\hspace{48pt}
\\
\hspace{48pt}{\it{\%linear\hspace{6pt}interpolation:
}}\\
\hspace{48pt}ausgabe=3.32\dash{}(n\dash{}30)$\ast$0.009;
\\
\hspace{48pt}
\\
\hspace{24pt}{\textbf{elseif}}\hspace{6pt}n$<$60
\\
\hspace{48pt}ausgabe=3.23\dash{}(n\dash{}40)$\ast$0.004;
\\
\hspace{48pt}
\\
\hspace{24pt}{\textbf{elseif}}\hspace{6pt}n$<$121
\\
\hspace{48pt}ausgabe=3.15\dash{}(n\dash{}60)$\ast$8/6000;
\\
\hspace{48pt}
\\
\hspace{24pt}{\textbf{elseif}}\hspace{6pt}n$>$120
\\
\hspace{36pt}disp({\texttt{{'}Warning:\hspace{6pt}Using\hspace{6pt}approximation\hspace{6pt}for}}\\
{\texttt{(number\hspace{6pt}of\hspace{6pt}measurements)\hspace{6pt}$>$$>$\hspace{6pt}120{'}}})
\\
\hspace{36pt}ausgabe=3;
\\
\hspace{30pt}
\\
{\textbf{end}}
\\

\\
\hspace{36pt}
\end{tabbing}}}

\subsection*{loo.m}
{{\def\dash{\raise2.1pt\hbox{\rule{5pt}{0.3pt}}\hspace{1pt}}\begin{tabbing}
{\textbf{function}}\hspace{6pt}loo(data,anz,rep)
\\
{\it{\%loo(data,anz,rep)}}\\
{\it{\%performs\hspace{6pt}DPCA,\hspace{6pt}leaves\hspace{6pt}each\hspace{6pt}measurement\hspace{6pt}out\hspace{6pt}once}}\\
{\it{\%and\hspace{6pt}projects\hspace{6pt}it\hspace{6pt}into\hspace{6pt}the\hspace{6pt}scores\hspace{6pt}plot}}\\
{\it{\%data:\hspace{6pt}(mxn)\hspace{6pt}matrix:\hspace{6pt}values\hspace{6pt}of\hspace{6pt}m\hspace{6pt}variables\hspace{6pt}in\hspace{6pt}n\hspace{6pt}measurements}}\\
{\it{\%anz:\hspace{6pt}number\hspace{6pt}of\hspace{6pt}groups}}\\
{\it{\%rep:\hspace{6pt}(anz)\dash{}vector:\hspace{6pt}number\hspace{6pt}of\hspace{6pt}measurements\hspace{6pt}per\hspace{6pt}group\hspace{6pt}
}}\\

\\
{[}m,n{]}=size(data);
\\

\\
{[}sizerepx,sizerepy{]}=size(rep);
\\
{\textbf{if}}\hspace{6pt}sizerepy==1
\\
\hspace{24pt}rep=repmat(rep,1,anz);
\\
{\textbf{end}}
\\

\\
{\it{\%loop\hspace{6pt}to\hspace{6pt}leave\hspace{6pt}out\hspace{6pt}one\hspace{6pt}measurement\hspace{6pt}after\hspace{6pt}another:
}}\\
column=0;
\\
{\textbf{for}}\hspace{6pt}k=1:anz
\\
\hspace{24pt}
\\
\hspace{24pt}{\textbf{for}}\hspace{6pt}j=1:rep(k)
\\
\hspace{48pt}
\\
\hspace{48pt}{\it{\%reduceddata:\hspace{6pt}matrix\hspace{6pt}without\hspace{6pt}{'}column{'}th\hspace{6pt}measurement
}}\\
\hspace{48pt}column=column+1;
\\
\hspace{48pt}{\textbf{if}}\hspace{6pt}column==1
\\
\hspace{72pt}reduceddata=data(:,2:n);
\\
\hspace{48pt}{\textbf{elseif}}\hspace{6pt}column==n
\\
\hspace{72pt}reduceddata=data(:,1:n\dash{}1);
\\
\hspace{48pt}{\textbf{else}}
\\
\hspace{72pt}reduceddata={[}data(:,1:column\dash{}1),data(:,column+1:n){]};
\\
\hspace{48pt}{\textbf{end}}
\\
\hspace{48pt}
\\
\hspace{48pt}{\it{\%reducing\hspace{6pt}number\hspace{6pt}of\hspace{6pt}measurements\hspace{6pt}in\hspace{6pt}concerned\hspace{6pt}group:
}}\\
\hspace{48pt}reducedrep=rep;
\\
\hspace{48pt}reducedrep(k)=reducedrep(k)\dash{}1;\\
\\
\hspace{48pt}{\it{\%DPCA:
}}\\
\hspace{48pt}{[}dloading,dscore,dvariance,dintvar{]}=dpca(reduceddata,anz,reducedrep);
\\
\hspace{48pt}
\\
\hspace{48pt}{\it{\%scores\hspace{6pt}plot:
}}\\
\hspace{48pt}plotdpc2dmitellipse(dscore,dvariance,anz,reducedrep,2);
\\
\hspace{48pt}
\\
\hspace{48pt}{\it{\%projection\hspace{6pt}of\hspace{6pt}the\hspace{6pt}left\hspace{6pt}out\hspace{6pt}measurement:
}}\\
\hspace{48pt}dpcaprojektion(dloading,dintvar,data(:,column),1,{\texttt{{'}ko{'}}});
\\
\hspace{42pt}\\
\hspace{48pt}{\it{\%halting\hspace{6pt}program\hspace{6pt}before\hspace{6pt}the\hspace{6pt}next\hspace{6pt}measurement\hspace{6pt}is\hspace{6pt}left\hspace{6pt}out:
}}\\
\hspace{48pt}fprintf(1,{\texttt{{'}Press\hspace{6pt}any\hspace{6pt}key\hspace{6pt}to\hspace{6pt}continue!\hspace{6pt}$\backslash$n{'}}})
\\
\hspace{48pt}pause
\\
\hspace{48pt}
\\
\hspace{24pt}{\textbf{end}}
\\
{\textbf{end}}
\\

\\
\hspace{96pt}
\\
\hspace{48pt}
\end{tabbing}}}

\subsection*{loo3.m}
{{\def\dash{\raise2.1pt\hbox{\rule{5pt}{0.3pt}}\hspace{1pt}}\begin{tabbing}
{\textbf{function}}\hspace{6pt}assignment=loo3(data,anz,rep,dimmax)
\\
{\it{\%assignment=loo(data,anz,rep,dimmax)}}\\
{\it{\%performs\hspace{6pt}DPCA,\hspace{6pt}leaves\hspace{6pt}each\hspace{6pt}measurement\hspace{6pt}out\hspace{6pt}once}}\\
{\it{\%and\hspace{6pt}calculates\hspace{6pt}its\hspace{6pt}scores\hspace{6pt}to\hspace{6pt}assign\hspace{6pt}it\hspace{6pt}to\hspace{6pt}the\hspace{6pt}closest\hspace{6pt}group.}}\\
{\it{\%Closeness\hspace{6pt}is\hspace{6pt}defined\hspace{6pt}by\hspace{6pt}the\hspace{6pt}euclidean\hspace{6pt}distances\hspace{6pt}to\hspace{6pt}the}}\\
{\it{\%center\hspace{6pt}of\hspace{6pt}gravity\hspace{6pt}of\hspace{6pt}a\hspace{6pt}groups\hspace{6pt}data\hspace{6pt}points.}}\\
{\it{\%data:\hspace{6pt}(mxn)\hspace{6pt}matrix:\hspace{6pt}values\hspace{6pt}of\hspace{6pt}m\hspace{6pt}variables\hspace{6pt}in\hspace{6pt}n\hspace{6pt}measurements}}\\
{\it{\%anz:\hspace{6pt}number\hspace{6pt}of\hspace{6pt}groups}}\\
{\it{\%rep:\hspace{6pt}(anz)\dash{}vector:\hspace{6pt}number\hspace{6pt}of\hspace{6pt}measurements\hspace{6pt}per\hspace{6pt}group}}\\
{\it{\%dimmax:\hspace{6pt}highest\hspace{6pt}considered\hspace{6pt}DPC}}\\
{\it{\%assignment:\hspace{6pt}(nx(anz+1))\dash{}matrix:\hspace{6pt}first\hspace{6pt}column:\hspace{6pt}number\hspace{6pt}of\hspace{6pt}the\hspace{6pt}}}\\
{\it{\%\hspace{66pt}closest\hspace{6pt}group\hspace{6pt}to\hspace{6pt}a\hspace{6pt}measurement;\hspace{6pt}2.,3.,...\hspace{6pt}column:\hspace{6pt}}}\\
{\it{\%\hspace{66pt}relative\hspace{6pt}distance\hspace{6pt}to\hspace{6pt}the\hspace{6pt}1.,2.,...\hspace{6pt}group
}}\\

\\
{[}m,n{]}=size(data);
\\

\\
{[}sizerepx,sizerepy{]}=size(rep);
\\
{\textbf{if}}\hspace{6pt}sizerepy==1
\\
\hspace{24pt}rep=repmat(rep,1,anz);
\\
{\textbf{end}}
\\

\\
{\it{\%loop\hspace{6pt}to\hspace{6pt}leave\hspace{6pt}out\hspace{6pt}one\hspace{6pt}measurement\hspace{6pt}after\hspace{6pt}another:
}}\\
column=0;
\\
{\textbf{for}}\hspace{6pt}k=1:anz
\\
\hspace{24pt}
\\
\hspace{24pt}{\textbf{for}}\hspace{6pt}j=1:rep(k)
\\
\hspace{48pt}
\\
\hspace{48pt}{\it{\%reduceddata:\hspace{6pt}matrix\hspace{6pt}without\hspace{6pt}{'}column{'}th\hspace{6pt}measurement:
}}\\
\hspace{48pt}column=column+1;
\\
\hspace{48pt}{\textbf{if}}\hspace{6pt}column==1
\\
\hspace{72pt}reduceddata=data(:,2:n);
\\
\hspace{48pt}{\textbf{elseif}}\hspace{6pt}column==n
\\
\hspace{72pt}reduceddata=data(:,1:n\dash{}1);
\\
\hspace{48pt}{\textbf{else}}
\\
\hspace{72pt}reduceddata={[}data(:,1:column\dash{}1),data(:,column+1:n){]};
\\
\hspace{48pt}{\textbf{end}}
\\
\hspace{48pt}
\\
\hspace{48pt}{\it{\%reducing\hspace{6pt}number\hspace{6pt}of\hspace{6pt}measurements\hspace{6pt}in\hspace{6pt}concerned\hspace{6pt}group:
}}\\
\hspace{48pt}reducedrep=rep;
\\
\hspace{48pt}reducedrep(k)=reducedrep(k)\dash{}1;\\
\\
\hspace{48pt}{\it{\%DPCA:
}}\\
\hspace{48pt}{[}dloading,dscore,dvariance,dintvar{]}=dpca(reduceddata,anz,reducedrep);
\\
\hspace{48pt}
\\
\hspace{48pt}{\it{\%projection\hspace{6pt}of\hspace{6pt}left\hspace{6pt}out\hspace{6pt}measurement:
}}\\
\hspace{48pt}score=dpcaprojektion(dloading,dintvar,data(:,column),1,{\texttt{{'}no{'}}});
\\
\hspace{48pt}\\
\hspace{48pt}{\it{\%loop\hspace{6pt}to\hspace{6pt}calculate\hspace{6pt}distances\hspace{6pt}to\hspace{6pt}all\hspace{6pt}groupmeans:\hspace{48pt}
}}\\
\hspace{48pt}min=1;
\\
\hspace{48pt}max=0;
\\
\hspace{48pt}{\textbf{for}}\hspace{6pt}i=1:anz\\
\hspace{72pt}{\it{\%calculate\hspace{6pt}center\hspace{6pt}of\hspace{6pt}gravity:
}}\\
\hspace{72pt}max=max+reducedrep(i);
\\
\hspace{72pt}dpcmean=mean(dscore(1:dimmax,min:max),2);\\
\hspace{72pt}{\it{\%calculate\hspace{6pt}euclidean\hspace{6pt}distance:
}}\\
\hspace{72pt}distance(i)=0;
\\
\hspace{72pt}{\textbf{for}}\hspace{6pt}h=1:dimmax
\\
\hspace{96pt}distance(i)=distance(i)+(score(h)\dash{}dpcmean(h))\^{}2;
\\
\hspace{72pt}{\textbf{end}}
\\
\hspace{72pt}distance(i)=sqrt(distance(i));
\\
\hspace{72pt}min=min+reducedrep(i);
\\
\hspace{48pt}{\textbf{end}}\\

\\
\hspace{48pt}{\it{\%assign\hspace{6pt}measurement\hspace{6pt}to\hspace{6pt}closest\hspace{6pt}group:
}}\\
\hspace{48pt}{[}minimum,minimumindex{]}=min(distance);
\\
\hspace{48pt}assignment(column,1)=minimumindex;\\

\\
\hspace{48pt}{\it{\%calculate\hspace{6pt}relative\hspace{6pt}distances:
}}\\
\hspace{48pt}{\textbf{for}}\hspace{6pt}i=1:anz
\\
\hspace{72pt}assignment(column,1+i)=distance(i)/minimum;
\\
\hspace{48pt}{\textbf{end}}\hspace{48pt}
\\
\hspace{24pt}{\textbf{end}}
\\
{\textbf{end}}\hspace{36pt}
\end{tabbing}}}

\subsection*{pcaprojektion.m}
{{\def\dash{\raise2.1pt\hbox{\rule{5pt}{0.3pt}}\hspace{1pt}}\begin{tabbing}
{\textbf{function}}\hspace{6pt}score=pcaprojektion(loading,mitte,data,index,stil)\\
{\it{\%dpcaprojektion(loading,mitte,data,index,stil)}}\\
{\it{\%projects\hspace{6pt}a\hspace{6pt}data\hspace{6pt}matrix\hspace{6pt}using\hspace{6pt}PC\hspace{6pt}loadings\hspace{6pt}into\hspace{6pt}a\hspace{6pt}scores\hspace{6pt}plot}}\\
{\it{\%loading:\hspace{6pt}(mxm)\dash{}matrix:\hspace{6pt}loadings\hspace{6pt}of\hspace{6pt}m\hspace{6pt}PC\hspace{6pt}as\hspace{6pt}m\hspace{6pt}columns}}\\
{\it{\%mitte:\hspace{6pt}m\dash{}vector:\hspace{6pt}mean\hspace{6pt}values\hspace{6pt}of\hspace{6pt}m\hspace{6pt}variables\hspace{6pt}for\hspace{6pt}the\hspace{6pt}original\hspace{6pt}data}}\\
{\it{\%data:\hspace{6pt}(mxn)\dash{}matrix:\hspace{6pt}values\hspace{6pt}of\hspace{6pt}m\hspace{6pt}variables\hspace{6pt}for\hspace{6pt}n\hspace{6pt}new\hspace{6pt}measurements}}\\
{\it{\%index:\hspace{6pt}number\hspace{6pt}of\hspace{6pt}first\hspace{6pt}displayed\hspace{6pt}PC\hspace{6pt}in\hspace{6pt}the\hspace{6pt}scores\hspace{6pt}plot}}\\
{\it{\%stil:\hspace{6pt}style\hspace{6pt}of\hspace{6pt}the\hspace{6pt}projected\hspace{6pt}data\hspace{6pt}points}}\\
{\it{\%score:\hspace{6pt}(mxn)\dash{}matrix:\hspace{6pt}projected\hspace{6pt}scores\hspace{6pt}of\hspace{6pt}m\hspace{6pt}PC\hspace{6pt}for\hspace{6pt}n\hspace{6pt}new\hspace{6pt}measurements
}}\\

\\
{[}m,n{]}=size(data);
\\

\\
{\it{\%scaling\hspace{6pt}of\hspace{6pt}measurements\hspace{6pt}by\hspace{6pt}total\hspace{6pt}intensity:
}}\\
{\textbf{for}}\hspace{6pt}k=1:n
\\
\hspace{24pt}data(:,k)=data(:,k)/sum(data(:,k));
\\
{\textbf{end}}
\\
\\
{\it{\%mean\hspace{6pt}centering:
}}\\
centered=data\dash{}repmat(mitte,1,n);\\
\\
{\it{\%projection:
}}\\
score=loading{'}$\ast$centered;
\\
\\
{\it{\%plotting\hspace{6pt}can\hspace{6pt}be\hspace{6pt}prevented\hspace{6pt}with\hspace{6pt}style\hspace{6pt}{'}no{'}:
}}\\
{\textbf{if}}\hspace{6pt}stil=={\texttt{{'}no{'}}}
\\
\hspace{24pt}{\textbf{return}}
\\
{\textbf{end}}
\\
\\
{\it{\%plotting:
}}\\
hold\hspace{6pt}on
\\
plot(score(index,:),score(index+1,:),stil);
\end{tabbing}}}

\subsection*{dpcaprojektion.m}
{{\def\dash{\raise2.1pt\hbox{\rule{5pt}{0.3pt}}\hspace{1pt}}\begin{tabbing}
{\textbf{function}}\hspace{6pt}score=dpcaprojektion(loading,varianceintra,data,index,stil)\\
{\it{\%dpcaprojektion(loading,varianceintra,data,index,stil)}}\\
{\it{\%projects\hspace{6pt}a\hspace{6pt}data\hspace{6pt}matrix\hspace{6pt}using\hspace{6pt}DPC\hspace{6pt}loadings\hspace{6pt}into\hspace{6pt}a\hspace{6pt}scores\hspace{6pt}plot}}\\
{\it{\%loading:\hspace{6pt}(mxm)\dash{}matrix:\hspace{6pt}loadings\hspace{6pt}of\hspace{6pt}m\hspace{6pt}DPC\hspace{6pt}as\hspace{6pt}m\hspace{6pt}columns}}\\
{\it{\%varianceintra:\hspace{6pt}(mx1)\dash{}vector:\hspace{6pt}pooled\hspace{6pt}intra\hspace{6pt}group\hspace{6pt}variances\hspace{6pt}of\hspace{6pt}m\hspace{6pt}variables}}\\
{\it{\%data:\hspace{6pt}(mxn)\dash{}matrix:\hspace{6pt}values\hspace{6pt}of\hspace{6pt}m\hspace{6pt}variables\hspace{6pt}for\hspace{6pt}n\hspace{6pt}measurements}}\\
{\it{\%index:\hspace{6pt}number\hspace{6pt}of\hspace{6pt}first\hspace{6pt}displayed\hspace{6pt}DPC\hspace{6pt}in\hspace{6pt}the\hspace{6pt}scores\hspace{6pt}plot}}\\
{\it{\%stil:\hspace{6pt}style\hspace{6pt}of\hspace{6pt}the\hspace{6pt}projected\hspace{6pt}data\hspace{6pt}points}}\\
{\it{\%score:\hspace{6pt}(mxn)\dash{}matrix:\hspace{6pt}projected\hspace{6pt}scores\hspace{6pt}of\hspace{6pt}m\hspace{6pt}DPC\hspace{6pt}for\hspace{6pt}n\hspace{6pt}measurements}}\\
{\it{\%for\hspace{6pt}details\hspace{6pt}on\hspace{6pt}DPCA\hspace{6pt}see\hspace{6pt}P.W.\hspace{6pt}Yendle,\hspace{6pt}H.J.M.\hspace{6pt}Macfie,}}\\
{\it{\%Discriminant\hspace{6pt}principal\hspace{6pt}components\hspace{6pt}analysis,}}\\
{\it{\%J.\hspace{6pt}of\hspace{6pt}Chemometrics\hspace{6pt}3,\hspace{6pt}589\dash{}600,\hspace{6pt}1989
}}\\

\\
{[}m,n{]}=size(data);
\\

\\
{\it{\%scaling\hspace{6pt}of\hspace{6pt}measurements\hspace{6pt}by\hspace{6pt}total\hspace{6pt}intensity:
}}\\
{\textbf{for}}\hspace{6pt}k=1:n
\\
\hspace{24pt}data(:,k)=data(:,k)/sum(data(:,k));
\\
{\textbf{end}}
\\
\\
{\it{\%scaling\hspace{6pt}by\hspace{6pt}pooled\hspace{6pt}intra\hspace{6pt}group\hspace{6pt}variance:
}}\\
{\textbf{for}}\hspace{6pt}k=1:m
\\
\hspace{24pt}scaled(k,:)=data(k,:)/sqrt(varianceintra(k));
\\
{\textbf{end}}\\
\\
{\it{\%projecting:
}}\\
score=loading{'}$\ast$scaled;
\\
\\
{\it{\%plotting\hspace{6pt}can\hspace{6pt}be\hspace{6pt}prevented\hspace{6pt}with\hspace{6pt}style\hspace{6pt}{'}no{'}:
}}\\
{\textbf{if}}\hspace{6pt}stil=={\texttt{{'}no{'}}}
\\
\hspace{24pt}{\textbf{return}}
\\
{\textbf{end}}
\\
\\
{\it{\%plotting:
}}\\
hold\hspace{6pt}on
\\
plot(score(index,:),score(index+1,:),stil);
\end{tabbing}}}

\subsection*{zuordnen.m}
{{\def\dash{\raise2.1pt\hbox{\rule{5pt}{0.3pt}}\hspace{1pt}}\begin{tabbing}
{\textbf{function}}\hspace{6pt}zuordnung=zuordnen(score,loading,intvar,anz,rep,data,dimmax)\\
{\it{\%zuordnung=zuordnen(score,loading,intvar,anz,rep,data,dimmax)}}\\
{\it{\%assigns\hspace{6pt}new\hspace{6pt}measurements\hspace{6pt}to\hspace{6pt}a\hspace{6pt}group\hspace{6pt}of\hspace{6pt}measurements}}\\
{\it{\%by\hspace{6pt}the\hspace{6pt}euclidean\hspace{6pt}distance\hspace{6pt}of\hspace{6pt}the\hspace{6pt}projection\hspace{6pt}into\hspace{6pt}the}}\\
{\it{\%DPC\hspace{6pt}space\hspace{6pt}calculated\hspace{6pt}with\hspace{6pt}the\hspace{6pt}original\hspace{6pt}measurements\hspace{6pt}to\hspace{6pt}the\hspace{6pt}}}\\
{\it{\%centers\hspace{6pt}of\hspace{6pt}gravity\hspace{6pt}of\hspace{6pt}the\hspace{6pt}original\hspace{6pt}groups}}\\
{\it{\%anz:\hspace{6pt}number\hspace{6pt}of\hspace{6pt}groups}}\\
{\it{\%rep:\hspace{6pt}(1xanz)\dash{}vector:\hspace{6pt}number\hspace{6pt}of\hspace{6pt}measurements\hspace{6pt}per\hspace{6pt}group}}\\
{\it{\%loading:\hspace{6pt}(mxm)\dash{}matrix:\hspace{6pt}loadings\hspace{6pt}of\hspace{6pt}m\hspace{6pt}DPC\hspace{6pt}as\hspace{6pt}m\hspace{6pt}columns}}\\
{\it{\%score:\hspace{6pt}(mxn)\dash{}matrix:\hspace{6pt}scores\hspace{6pt}of\hspace{6pt}m\hspace{6pt}DPC\hspace{6pt}for\hspace{6pt}n\hspace{6pt}measurements}}\\
{\it{\%intvar:\hspace{6pt}(mx1)\dash{}vector:\hspace{6pt}pooled\hspace{6pt}intra\hspace{6pt}group\hspace{6pt}variances\hspace{6pt}of\hspace{6pt}m\hspace{6pt}variables}}\\
{\it{\%dimmax:\hspace{6pt}dimensionality\hspace{6pt}of\hspace{6pt}the\hspace{6pt}DPC\hspace{6pt}space}}\\
{\it{\%zuordnung:\hspace{6pt}(nx(anz+1))\dash{}matrix:\hspace{6pt}first\hspace{6pt}column:\hspace{6pt}number\hspace{6pt}of\hspace{6pt}the\hspace{6pt}}}\\
{\it{\%\hspace{66pt}closest\hspace{6pt}group\hspace{6pt}to\hspace{6pt}a\hspace{6pt}measurement;\hspace{6pt}2.,3.,...\hspace{6pt}column:\hspace{6pt}}}\\
{\it{\%\hspace{66pt}relative\hspace{6pt}distance\hspace{6pt}to\hspace{6pt}the\hspace{6pt}1.,2.,...\hspace{6pt}group}}\\
\\
{[}m,n{]}=size(data);\\
{[}sizerepx,sizerepy{]}=size(rep);\\
{\textbf{if}}\hspace{6pt}sizerepy==1\\
\hspace{24pt}rep=repmat(rep,1,anz);\\
{\textbf{end}}\\
\\
{\it{\%DPCA\hspace{6pt}projection\hspace{6pt}of\hspace{6pt}the\hspace{6pt}new\hspace{6pt}data:}}\\
newscore=dpcaprojektion(loading,intvar,data,1,{\texttt{{'}no{'}}});\\
\\
{\it{\%loop\hspace{6pt}for\hspace{6pt}different\hspace{6pt}new\hspace{6pt}measurements\hspace{24pt}}}\\
\hspace{6pt}{\textbf{for}}\hspace{6pt}j=1:n\\
\hspace{30pt}\\
\hspace{30pt}{\it{\%select\hspace{6pt}one\hspace{6pt}measurement:}}\\
\hspace{30pt}newscorej=newscore(:,j);\\
\hspace{30pt}\\
\hspace{48pt}min=1;\\
\hspace{48pt}max=0;\\
\hspace{48pt}\\
\hspace{48pt}{\it{\%loop\hspace{6pt}for\hspace{6pt}different\hspace{6pt}groups:}}\\
\hspace{48pt}{\textbf{for}}\hspace{6pt}i=1:anz\\
\hspace{72pt}max=max+rep(i);\\
\\
\hspace{72pt}{\it{\%center\hspace{6pt}of\hspace{6pt}gravity\hspace{6pt}of\hspace{6pt}the\hspace{6pt}group:}}\\
\hspace{72pt}groupmean=mean(score(1:dimmax,min:max),2);\\
\\
\hspace{72pt}{\it{\%calculate\hspace{6pt}euclidean\hspace{6pt}distance:}}\\
\hspace{72pt}distance(i)=0;\\
\hspace{72pt}{\textbf{for}}\hspace{6pt}h=1:dimmax\\
\hspace{96pt}distance(i)=distance(i)+(newscorej(h)\dash{}groupmean(h))\^{}2;\\
\hspace{72pt}{\textbf{end}}\\
\hspace{72pt}distance(i)=sqrt(distance(i));\\
\hspace{72pt}min=min+rep(i);\\
\hspace{48pt}{\textbf{end}}\\
\hspace{48pt}\\
\hspace{48pt}{\it{\%assign\hspace{6pt}new\hspace{6pt}measurement\hspace{6pt}to\hspace{6pt}closest\hspace{6pt}group:}}\\
\hspace{48pt}{[}minimum,minimumindex{]}=min(distance);\\
\hspace{48pt}zuordnung(j,1)=minimumindex;\\
\\
\hspace{48pt}{\it{\%calculate\hspace{6pt}relative\hspace{6pt}distances\hspace{6pt}to\hspace{6pt}other\hspace{6pt}groups:}}\\
\hspace{48pt}{\textbf{for}}\hspace{6pt}i=1:anz\\
\hspace{72pt}zuordnung(j,1+i)=distance(i)/minimum;\\
\hspace{48pt}{\textbf{end}}\\
\hspace{24pt}{\textbf{end}}
\end{tabbing}}}
\end{appendix}

\newpage
\normalsize
\bibliography{Falk}
\bibliographystyle{plain}

\pagebreak
\section*{Acknowledgements}
It would have been impossible for me to create this work without of the help of several people. I would especially like to thank the following:
\begin{description}
\item[Prof. Dr. Christiane Ziegler] for giving me the possibility to do my diploma thesis on an interesting subject in her work group as well as for the freedom and helpful advice she gave to me for dealing with it.
\item[apl. Prof. Dr. habil. Hubert Gnaser] for an introduction to the ToF-SIMS technique and for fruitful discussions about the use of it. 
\item[Dr. Wolfgang Bock] for solving several technical problems with the TRIFT apparatus.
\item[Manfred Strack] for help with all the other technical problems of everyday laboratory life.
\item[Nicole Lawrence] for all kinds of help throughout my first months in the scientific world, for many discussions about the mysteries of protein physics as well as for an introduction to scanning force microscopy. 
\item[Adam Orendorz] for being helpful on various subjects and for sharing office and laboratory with me.
\item[Dr. Stefan Trellenkamp and Dr. Bert L\"agel,] Nano+Bio Center Kai\-sers\-lau\-tern,  for taking scanning electron microscopy images of my samples and for help on their interpretation.
\item[The whole group ``Interfaces, Nano Materials and Biophysics''] for their warm \linebreak welcome and the pleasant atmosphere. 
\item[My parents] for their moral and financial support throughout my studies. 
\end{description}


\end{document}